\documentclass{article}
\usepackage{times}
\usepackage{a4wide}

\usepackage{bm}

\usepackage{pdfpages}
\usepackage{lineno}

\usepackage{tikz}
\usepackage{tikz-cd}
\usepackage{pgfplots}
\usetikzlibrary{decorations.pathreplacing,calligraphy,shapes,arrows}

\usepackage{latexsym}
\usepackage{amsmath}
\usepackage{amssymb}
\usepackage{ifthen}
\usepackage{mathrsfs}
\usepackage{graphics}

\usepackage{stmaryrd}
\usepackage{amsthm}
\usepackage{arcs}

\usepackage{subcaption}

\newcommand{\nc}{\newcommand}
\nc{\rnc}{\renewcommand} \nc{\nev}{\newenvironment}
\rnc{\subsection}{\secdef\ssa\ssb}
\nc{\ssa}[2][default]{\par\vspace{1ex}\refstepcounter{subsection}\noindent\textbf{\thesubsection.
#1. }} \nc{\ssb}[1]{\par\vspace{2ex}\noindent\textbf{#1. }}

\rnc{\subsubsection}{\secdef\sssa\sssb}
\nc{\sssa}[2][default]{\par\vspace{1ex}\refstepcounter{subsubsection}\noindent\textit{\thesubsubsection.
#1. }} \nc{\sssb}[1]{\par\vspace{1ex}\noindent\textit{#1. }}

\makeatletter \rnc{\@seccntformat}[1]{{\normalfont\bfseries{\csname 
the#1\endcsname}\hspace{1pt}.\hspace{0.4em}}} \rnc{\section}{\@startsection 
        {section}%
        {1}%
        {0mm}%
        {-\baselineskip}%
        {0.5\baselineskip}%
        {\normalfont\normalsize\bfseries\centering}%
} 
\renewcommand{\@makecaption}[2]{\begin{center}#1. #2\end{center}}
\makeatother

\newtheorem{theo}{Theorem}[section]
\newtheorem{lem}[theo]{Lemma}
\newtheorem{cor}[theo]{Corollary}
\newtheorem{prop}[theo]{Proposition}

\theoremstyle{definition} 
\newtheorem{defn}[theo]{Definition}
\newtheorem{rem}[theo]{Remark}
\newtheorem{exa}[theo]{Example}

\numberwithin{equation}{section}

\numberwithin{figure}{section}

\rnc{\proof}{\smallskip\noindent\textit{Proof: }} 
\nc{\proofend}{\hfill$\Box$\vspace{\topsep}\par}

\rnc{\labelenumi}{(\arabic{enumi})} \rnc{\labelitemi}{\text{--}} 
\rnc{\phi}{\varphi} \rnc{\epsilon}{\varepsilon} \nc{\bigmid}{\;\big|\;} 
\nc{\Bigmid}{\;\Big|\;} \rnc{\max}{\textup{max}} \rnc{\min}{\textup{min}} 
\rnc{\log}{\textup{log}\;}

\newlength{\probwidth}
\nc{\prob}[3][9]{ 
\begin{center}
  \normalfont\fbox{
   \begin{tabular}[t]{
     rp{#1cm}}\textit{Instance:}&#2. \\
     \textit{Problem:}&#3
   \end{tabular}}
\end{center}}

\nc{\pprob}[4][9]{ 
\begin{center}
   \normalfont\fbox{
    \begin{tabular}[t]{
     rp{#1cm}}\textit{Instance:}&#2. \\
     \textit{Parameter:}&#3. \\
     \textit{Problem:}&#4
   \end{tabular}}
\end{center}}

\nc{\nprob}[4][9]{ 
\begin{center}
  \normalfont\fbox{

\addtolength{\probwidth}{#1cm}\parbox{\probwidth}{\textsc{#2}\\\hspace*{1.5em} 
     \begin{tabular}[t]{
      rp{#1cm}}\textit{Instance:}&#3. \\
      \textit{Problem:}&#4
     \end{tabular}}}
\end{center}}

\nc{\npprob}[5][9]{ 
\begin{center}
  \normalfont\fbox{

\addtolength{\probwidth}{#1cm}\parbox{\probwidth}{\textsc{#2}\\\hspace*{1.5em} 
    \begin{tabular}[t]{
     rp{#1cm}}\textit{Instance:}&#3. \\
     \textit{Parameter:}&#4. \\
     \textit{Problem:}&#5
    \end{tabular}}}
\end{center}}

\nc{\nppxrob}[5][9]{ \normalfont\fbox{ 

\addtolength{\probwidth}{#1cm}\parbox{\probwidth}{\textsc{#2}\\\hspace*{1.5em} 
   \begin{tabular}[t]{
    rp{#1cm}}\textit{Instance:}&#3. \\
    \textit{Parameter:}&#4. \\
    \textit{Problem:}&#5
   \end{tabular}}}}

\nc{\nppprob}[5][4]{ 
\begin{center}
  \normalfont\fbox{

\addtolength{\probwidth}{#1cm}\parbox{\probwidth}{\textsc{#2}\\\hspace*{1.5em} 
    \begin{tabular}[t]{
     rp{#1cm}}\textit{Instance:}&#3. \\
     \textit{Parameter:}&#4. \\
     \textit{Problem:}&#5
    \end{tabular}}}
\end{center}}

\nc{\FOR}{\textbf{for}} \nc{\FORALL}{\textbf{for all}} \nc{\TO}{\textbf{to}} 
\nc{\DO}{\textbf{do}} \nc{\OD}{\textbf{od}} \nc{\IF}{\textbf{if}} 
\nc{\FI}{\textbf{fi}} \nc{\THEN}{\textbf{then}} \nc{\ELSE}{\textbf{else}} 
\nc{\WHILE}{\textbf{while}} \nc{\REPEAT}{\textbf{repeat}} 
\nc{\UNTIL}{\textbf{until}} \nc{\OR}{\textbf{or}} \nc{\AND}{\textbf{and}} 
\nc{\PRINT}{\textbf{print}} 

\nc{\im}[1]{\item\hspace{#1cm}} 
\nev{algorithm}{\begin{enumerate}\rnc{\labelenumi}{\textit{\small 
\arabic{enumi}.}}\rnc{\itemsep}{0ex}}{\end{enumerate}}

\nc{\fpcl}[1]{\left[#1\right]_{\text{\upshape fp}}} 
\nc{\pr}{\le^{\text{\normalfont fp}}_m} \nc{\FPT}{\textup{FPT}} 
\nc{\EPT}{\textup{EPT}} \nc{\SUBEPT}{\textup{SUBEPT}} 

\rnc{\S}[1]{\text{$\textup{S}[#1]$}} \nc{\SP}{\textup{S[P]}} 
\nc{\MP}{\textup{M[P]}} 

\nc{\PTIME}{\textup{PTIME}} \nc{\PSPACE}{\textup{PSPACE}} 
\nc{\NP}{\textup{NP}} 

\nc{\DTIME}{\textup{DTIME}} 

\nc{\se}{\subseteq} \nc{\re}{\rightarrow}

\nc{\LOEFF}[1]{{o}^{\rm eff}(#1)}

\nc{\algo}[1]{{\mathbb #1}}

\nc{\NAT}{{\mathbb N}} 

\nc{\VC}{\textsc{Vertex-Cover}} \nc{\IS}{\textsc{Independent-Set}} 
\nc{\clique}{\textsc{Clique}} 

\nc{\DS}{\textsc{Dominating-Set}} \nc{\TSAT}{\textsc{3-Sat}}

\nc{\CNF}{\textup{CNF}} \nc{\WSAT}{\textsc{WSat}} \nc{\AWSAT}{\textsc{AWSat}} 
\nc{\SAT}{\textsc{Sat}} \nc{\ASAT}{\textsc{ASat}} \nc{\CIRC}{\textsc{Circ}} 
\nc{\PROP}{\textsc{Prop}} 

\nc{\nva}{\textup{nv}} \nc{\ncl}{\textup{nc}} 

\nc{\gap}{\textit{gap-}} 

\nc{\ETH}{\textup{ETH}} \nc{\gapETH}{\gap\textup{ETH}}

\nc{\serf}{\textup{serf}} \nc{\serfT}{\textup{serf-T}} 
\nc{\var}{\textup{var}} 

\nc{\m}[1]{\textup{$\textsc{Mini}$$#1$}}

\nc{\ceil}[1]{\left\lceil#1\right\rceil} 
\nc{\floor}[1]{\left\lfloor#1\right\rfloor} 

\nc{\bende}{\eqno$\Box$} \nc{\benda}{\tag*{$\Box$}} 

\nc{\pa}{\kappa}

\nc{\MSO}{\textup{MSO}} \nc{\EMSO}{\textup{EMSO}}

\nc{\ESO}{\textup{ESO}} 

\nc{\coNP}{\textup{co-NP}} 

\nc{\ZFC}{\textup{ZFC}}

\nc{\XNP}{\textup{XNP}} \nc{\XNPu}{\ensuremath{\textup{XNP}_{\rm uni}}}

\nc{\XNL}{\textup{XNL}} \nc{\XNLu}{\ensuremath{\textup{XNL}_{\rm uni}}} 

\nc{\XL}{\textup{XL}} \nc{\XLu}{\textup{XL}_{\rm uni}} 

\nc{\co}{\textup{co-}} 

\rnc{\L}{\textup{LOGSPACE}} 

\nc{\NL}{\textup{NLOGSPACE}} 

\nc{\TC}{\textup{TC}} 

\nc{\DTC}{\textup{DTC}} 

\nc{\FO}{\textup{FO}} 

\nc{\mtc}{\models_{\FO[\TC]}} \nc{\mtcy}{\models_{L^{\TC}_\le}} 

\nc{\INV}{\textsc{Inv}} 

\nc{\TAUT}{\textsc{TAUT}} 

\rnc{\P}{\textup{P}} 

\rnc{\angle}[1]{\langle #1\rangle} 

\nc{\opt}{\textup{opt}} 

\nc{\LFP}{\textup{LFP}} 

\nc{\rand}[1]{\marginpar{\raggedright\footnotesize #1}} 
\nc{\yrand}[1]{\rand{\textbf{Yijia: }#1}} 
\nc{\mrand}[1]{\rand{\textbf{Mingjun: }#1}} 
\nc{\jrand}[1]{\rand{\textbf{J\"org: }#1}} 

\nc{\Ppoly}{\textup{P}/\text{\small poly}} 

\nc{\f}{\mathbf f} 

\nc{\s}{\mathbf s} 

\nc{\tim}{\textup{time}} 

\nc{\sat}{\textup{sat}} \nc{\rank}{\textup{rank}} \nc{\PC}{\mathbf{PC}} 
\nc{\bin}{{\rm in}} \nc{\bout}{{\rm out}} \nc{\Mix}{\textsc{Mix}} 
\nc{\MIX}{\mathbf{MIX}} 

\nc{\tdeg}{\textup{tdeg}} \nc{\lspan}{\textup{span}}

\nc{\ma}[1]{\mathbb #1} \nc{\bet}[1]{\| #1\|} \nc{\EFM}[2]{(#1,#2)} 
\nc{\EF}{\text{Ehren\-feucht\--Fra\"i\-ss\'e}} \nc{\AF}{\textup{AF}} 
\nc{\supp}{\textup{supp}} \nc{\ar}{\textup{ar}} \nc{\pol}{\textup{pol-}} 
\nc{\equivg}{\equiv\rangle} 

\nc{\ER}{Erd\H{o}s-R\'enyi} 

\nc{\PCC}{\textup{PCC}} 

\nc{\TCOL}{3\textsc{-Colorability}} 

\nc{\GI}{\textup{GI}} 

\nc{\PCP}{\textup{PCP}} 

\nc{\LPC}{\textup{LPC}} 

\nc{\AC}{\textup{AC}} 

\nc{\paraAC}{\textup{para-$\AC^0$}} 

\nc{\ACC}{\textup{ACC}} 

\nc{\paraACC}{\textup{para-$\ACC^0$}} 

\nc{\MOD}{\textup{MOD}} 

\nc{\phalt}{\ensuremath{p\textsc{-Halt}}} 

\nc{\pclique}{\ensuremath{p\textsc{-Clique}}} 
\nc{\pds}{\ensuremath{p\textsc{-Dominating-Set}}} 
\nc{\pgapclique}[1]{\ensuremath{p\textit{-{#1}-gap}\textsc{-Clique}}} 
\nc{\pgapds}[1]{\ensuremath{p\textit{-{#1}-gap}\textsc{-Dominating-Set}}}

\nc{\para}{\textup{para-}} 

\nc{\vc}{\textit{vc}} \nc{\ds}{\textit{ds}} \nc{\cli}{\omega} 

\nc{\kg}{\textup{kg}}

\nc{\entry}{\textup{entry}}

\renewcommand{\mod}[1]{\ \textup{mod}\; #1}

\newcommand{\dotcup}{\;\dot\cup\;}

\newcommand{\tw}{\textup{tw}}
\newcommand{\CFI}{\textup{CFI}}

\newcommand{\Aut}{\textup{Aut}}
\newcommand{\aut}{\textup{aut}}
\newcommand{\base}{\textit{base}}
\newcommand{\Hom}{\textup{Hom}}
\newcommand{\Per}{\textup{Per}}
\newcommand{\id}{\textup{id}}

\title{Some remarks on the uncolored versions of the original CFI-graphs}

\author{Yijia Chen\\\normalsize School of Computer Science\\
 \normalsize Shanghai Jiao Tong University\\
 \normalsize yijia.chen@cs.sjtu.edu.cn\\
 \and
 J\"{o}rg Flum\\\normalsize Mathematisches Institut \\
 \normalsize Universit\"{a}t Freiburg i.Br.\\
 \normalsize flum@uni-freiburg.de
 \and
 Mingjun Liu\\\normalsize School of Computer Science\\
 \normalsize Shanghai Jiao Tong University\\
 \normalsize liumingjun@sjtu.edu.cn\\
 }

\date{}

\begin{document}

\maketitle

%\linenumbers

%
%
%

% abstract and introduction
\begin{abstract}  
The CFI-graphs, named after Cai, Fürer, and Immerman, are central to the study of the graph iso-
morphism testing and of first-order logic with counting. They are  colored graphs, and the coloring
plays a  role in many of their applications. As usual, it is not hard to remove the coloring by some
extra graph gadgets, but at the cost of blowing up the size of the graphs and changing some parameters of them as well. This might lead to suboptimal combinatorial bounds important to their applications.

Since then for some uncolored variants
of the CFI-graphs it has been shown that they serve the same purposes. We show that this already applies
to the graphs obtained from the original CFI-graphs by forgetting the colors. Moreover,
we will see that there is a first-order formula $\varphi(x,y)$ expressing in almost all uncolored CFI-graphs that $x$ and $y$ have the same color in the corresponding colored graphs.

\end{abstract}

\section{Introduction}\label{sec:int}

In 1992 Cai, F\"urer, and Immerman~\cite{caifurimm92} introduced an ingenious 
graph-theoretic construction, whose resulting graphs are now widely known as 
\emph{CFI-graphs}. Using such graphs Cai, F\"urer, and Immerman resolved two 
well-known open problems in theoretical computer science. The first is the 
question of whether, for any fixed $k\ge 1$, the $k$-dimensional 
Weisfeiler-Leman algorithm is a complete graph isomorphism test, that is, 
distinguishes all pairs of non-isomorphic graphs. The second is the question 
of whether fixed-point logic with counting captures PTIME.

Specifically, for all $k\ge 1$, Cai, F\"urer, and Immerman presented 
a pair of \emph{colored} graphs, we denote in this Introduction by~$X^k$ 
and $\tilde X^k$, on $\Theta(k)$ vertices with the following properties. 
\begin{itemize}
\item[(a)] $X^k$ and $\tilde X^k$ are not isomorphic.

\item[(b)] $X^k$ and $\tilde X^k$ satisfy the same sentences of $C^k$, the 
    fragment of first-order logic with counting consisting of the sentences 
    with at most $k$ variables.\label{page:(b)(c)}
     
\item[(c)] There is a polynomial time algorithm that accepts all $ X^{k} $'s
    and rejects all $\tilde X^k$'s. 
\end{itemize}
To get their result on fixed-point logic with counting Cai, F\"urer, and 
Immerman used the relationship of this logic to the logics 
$C^k$~(see~\cite{graott92}). Particularly, $X^k$ and $\tilde X^k$ in Part~(b) 
cannot be distinguished by fixed-point logic with counting using at most $k$ 
variables. To get the result on the Weisfeiler-Leman algorithms they used the 
fact that Part~(b) can be restated by saying that the $k-1$-dimensional 
Weisfeiler-Leman algorithm cannot distinguish $X^k$ and $\tilde X^k$ 
(see~\cite{ caifurimm92,immlan}). 
It implies that to distinguish $X^k$ and $\tilde X^k$ two non-isomorphic 
graphs on $k$ vertices, we need an $\Omega(k)$-dimensional Weisfeiler-Leman 
algorithm, an algorithm that runs in time super-polynomial in $k$ and hence,  
by (c), is not optimal. 

\begin{figure} 	
\centering
\input{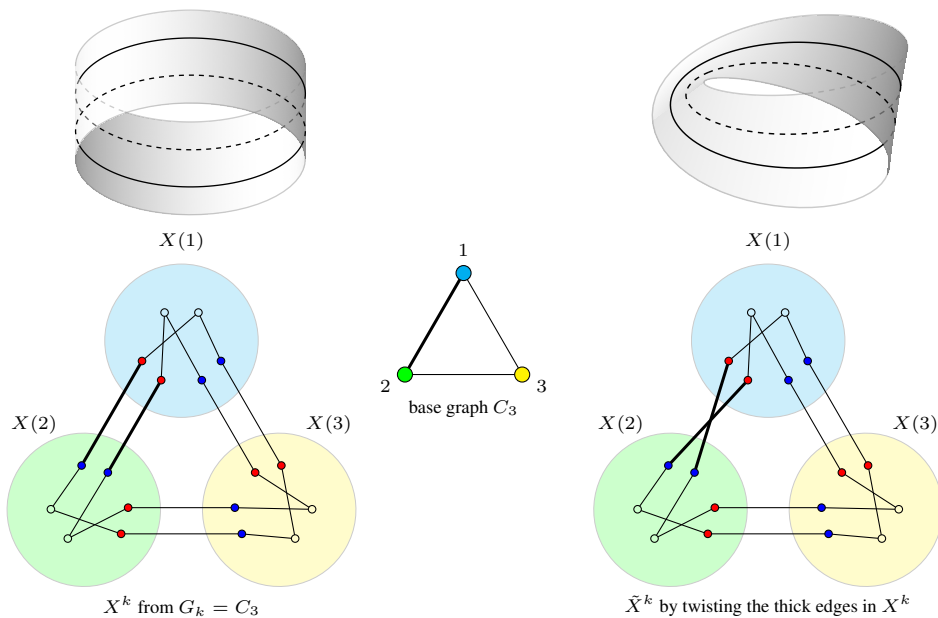}
\caption{A pair of $X^k$ and $\tilde X^k$ is constructed from a colored triangle $C_3$. Each colored circle represents a CFI-gadget replacing a vertex in $C_3$ with the same color, e.g., two $X(1)$'s correspond to the vertex $1$.
Observe that two disjoint cycles of $X^k$ can be embedded on a strip, while the twist in
$\tilde X^k$ joins the two cycles on a M\"obius strip.
}\label{fig:CFIC3}  
\end{figure}

The construction of a pair of CFI-graphs $X^k$ and $\tilde X^k$ starts with a 
base graph $G_k$. In a first step, we replace each vertex in $G_k$ by a 
so-called colored \emph{CFI-gadget}, whose automorphism properties induce a 
notion of orientation. Then we add edges between all 
the gadgets in the same way as how the original vertices are connected in the 
base graph $G_k$. This already gives us the graph $X^k$. Finally, by 
introducing a single ``twist'' between an arbitrary pair of adjacent gadgets 
in $X^k$ we obtain $\tilde X^k$. Figure~\ref{fig:CFIC3} exemplifies the 
construction where the base graph is a colored triangle $C_3$. In $\tilde X^k$ we have twisted two (thick) edges between $X(1)$ and $X(2)$, which correspond to the (thick) edge between vertices $1$ and $2$ in $C_3$.
By the properties of the automorphism group 
of the CFI-gadget, this single ``twist'' can be ``propagated'' globally to 
anywhere in the graph (see 
Corollary~\ref{cor:welldefinedtildeXG} for a precise statement), making $X^k$ and $\tilde X^k$ locally 
indistinguishable, while they can be distinguished by their orientation. 
In Figure~\ref{fig:CFIC3}, $X^k$ consists of two disjoint cycles which can be 
visualized as ``staying'' on the two opposite faces of a strip. And in 
$\tilde X^k$, because of the twist, the two cycles are merged into one, turning 
the underlying strip into a M\"obius strip. 
It is demonstrated in~\cite{caifurimm92} that the local indistinguishability of 
$X^k$ and~$\tilde X^k$ witnesses the weakness of the $k-1$-dimensional 
Weisfeiler-Leman algorithm. 
\medskip

\noindent
Besides the above group-theoretic viewpoint, it has been realized that the CFI-graphs 
can also be understood as systems of linear equations~\cite{DGHL09,tin91} and by means of 
Tseitin tautologies~\cite{bergro15,BG17}. And over time, the original construction of the 
CFI-graphs has been tweaked or adapted to address a variety of 
applications (see, e.g. \cite{AMRSSV19,Rad24,grolicneusch23,neu24,Roberson}).  All  these diverse perspectives reveal the multifaceted nature of CFI graphs as a foundational tool for logic, algorithms, and beyond.

\medskip

\noindent
As already mentioned (see also Figure~\ref{fig:CFIC3}), in~\cite{caifurimm92}~both the CFI-graphs and their 
building blocks -- the CFI-gadgets -- are colored. More precisely, each 
CFI-gadget is colored by a distinct set of colors, and thus two different 
gadgets can be readily separated by their colors. As a result, any partial 
isomorphism from $X^k$ to $\tilde X^k$ has to be an automorphism on any 
gadget within its domain. This observation is often the first, sometimes 
implicit, step in the proofs in~\cite{caifurimm92} and also frequently used 
in many of the subsequent works relying on CFI-graphs. Of course, 
such coloring makes the graph isomorphism problem easier and the lower bounds 
stronger. And as already commented in~\cite[Page~7, footnote~3]{caifurimm92}, 
coloring can be easily replaced by adding some extra graph gadgets into $X^k$ 
and $\tilde X^k$. But this might come with some price.   

Recall that each pair of $X^k$ and $\tilde X^k$ is constructed from a base 
graph $G_k$, which is chosen as a $3$-regular expander graph on $\Theta(k)$ 
vertices in~\cite{caifurimm92}. As discussed above, this leads to $X^k$ and 
$\tilde X^k$ containing $\Theta(k)$ many CFI-gadgets, thus with $\Theta(k)$ 
different colors. Thereby, we need at least $\Theta(k)$ many new gadgets to 
replace these colors, blowing up the size of the graphs $X^k$ and $\tilde X^k$ from $\Theta(k)$ to $\Omega(k^2)$. As a consequence, we can only deduce 
from such a construction that an $\Omega(\sqrt{n})$-dimensional
Weisfeiler-Leman algorithm is necessary to distinguish two non-isomorphic 
\emph{uncolored} graphs with $n$ vertices, suboptimal to the desired $\Omega(n)$\ lower bound. 
Another aspect of the construction that will experience a similar degradation is the 
degree of the CFI-graphs. By choosing a $3$-regular graph as the base graph 
$G_k$ we get the pair $X^k$ and $\tilde X^k$ both $3$-regular as well. This 
property is leveraged in~\cite{caifurimm92} to show that the Weisfeiler-Leman 
algorithm fails even for \emph{colored} graphs of degree $3$.
However, once we 
replace colors by additional gadgets, the degree of CFI-graphs has to 
increase. Therefore, the original construction in~\cite{caifurimm92} does not 
immediately witness the failure of the Weisfeiler-Leman algorithm on 
\emph{uncolored} graphs of degree $3$.

The use of an unbounded number of colors also introduces complications for 
logic-related results. In that context, we typically work with classes of 
structures over a \emph{fixed} (finite) vocabulary. For instance, graphs are 
identified with structures over the vocabulary $\{E\}$, where $E$ is a binary 
relation symbol. Coloring then can be translated to unary relations (for 
subsets of vertices of the same color) on these structures. However, when the 
number of the colors is unbounded, as in~\cite{caifurimm92}, the vocabulary 
is no longer fixed, i.e., containing a fixed finite number of unary relation 
symbols. So strictly speaking, the polynomial time algorithm in Part~(c) 
cannot be understood as deciding a class of structures of a fixed vocabulary. 
Moreover, we cannot deduce from (a)--(c) that for a fixed vocabulary there 
are $C^{O(k)}$-sentences that are not equivalent to any $C^k$-sentence. 
\medskip

\noindent
In this article, we systematically investigate the uncolored version of the 
CFI-graphs. Among others, we show that already the graphs $Y^k$ and $\tilde 
Y^k$ obtained from $X^k$ and $\tilde X^k$ by forgetting their colors satisfy 
{(a)--(c)}, too. Moreover, these uncolored graphs share various other 
properties with their colored counterparts, as well as with other known 
uncolored variants of the CFI-graphs. This is not as straightforward as it 
might look at first glance. For example, while Part~(b) clearly holds for 
$Y^k$ and $\tilde Y^k$, it is not completely obvious why $Y^k$ and $\tilde 
Y^k$ are not isomorphic as required in Part~(a). As previously discussed, any 
partial isomorphism from $X^k$ to $\tilde X^k$ must act as an automorphism on 
any CFI-gadget within its domain. We will see that this is not the case for 
the uncolored version. That is, there exist partial isomorphisms from $Y^k$ 
to $\tilde Y^k$ that map a CFI-gadget in $Y^k$ to fragments of different 
CFI-gadgets in $\tilde Y^k$. For the similar reasons designing a polynomial 
time algorithm to distinguish $Y^k$ and $\tilde Y^k$, as in Part~(c), becomes 
more intricate. To establish the corresponding results it is very helpful to 
relate the automorphisms of the uncolored CFI-graphs to the automorphisms of 
the colored CFI-graphs.  

Once having a good understanding of the uncolored CFI-graphs, we will prove 
the uncolored version of many results that have been shown using the colored 
CFI graphs. Admittedly, many of them are already known. But our proof is 
based on a unified treatment of CFI-graphs, both colored and uncolored. 
Sometimes, such a treatment leads to better bounds that could be important 
for the desired applications. For example, while Otto~\cite{ott17} witnessed the strictness of the $C^k$ hierarchy by graphs on $2^{O(k)}$ vertices, we obtain this result for graphs of cardinality $O(k)$.
 That is, for every $k\ge 1$ there exists a pair of 
uncolored graphs $Y^k$ and $\tilde Y^k$ on $O(k)$ vertices which satisfy the 
same sentences of $C^k$ but not of $C^{k+1}$. 
To prove this, we use 
CFI-graphs that are completely different from~\cite{ott17}. The key 
observation is that the maximum number $k$ such that $C^k$ cannot distinguish 
$Y^k$ and $\tilde Y^k$ is precisely the \emph{tree-width} of the base graph 
$G_k$. This remarkable connection between the tree-width of a base graph and 
the expressive power of $C^k$ on $Y^k$ and~$\tilde Y^k$ was first observed by 
Dawar and Richerby on ``colored''  CFI-graphs~\cite{dawric}. Thus, to obtain the improved version of Otto's 
result, all we need are base graphs $G_k$ of tree-width $k$ on $O(k)$ 
vertices. Interestingly, such graphs can be constructed from expander graphs 
on $O(k)$ vertices as used in the original article~\cite{caifurimm92}, albeit without 
colors.

\paragraph{Results and organization of this paper.}
As  already mentioned, the main purpose of this paper is to show that the central results obtained via the original CFI-graphs can already be obtained by the corresponding uncolored  CFI-graphs.

In Section~\ref{sec:pre} we fix some notation and recall some simple facts that
we need in this paper. We introduce the CFI-gadgets of the (colored and 
uncolored) 
CFI-graphs in Section~\ref{sec:unc+col}. For $d\ge 1$ the colored gadget 
$X(d)$ and the uncolored gadget $Y(d)$ have the same vertices and edges. We 
recapitulate the automorphisms of $X(d)$ in~Section~\ref{sec:autc}. 
In~Section~\ref{sec:autu} we show that in general every automorphism of 
$Y(d)$ is determined by an automorphism of $X(d)$ and a permutation of the 
``central building block'' $A(d)$ of $Y(d)$ that  consists of $d$ vertices. 
This result only holds for $d\notin \{1,2,4 \}$. 

In Section~\ref{sec:ygxg+tilde} for a connected graph $G$ with at least two 
vertices we introduce the original colored CFI-graphs $X(G)$ and  $\tilde 
X(G)$ and the uncolored graphs $Y(G)$ and $\tilde Y(G)$ obtained from the 
previous ones just by deleting the colors. We study the relationship between 
the automorphisms of $X(G)$ and that of $Y(G)$ in Section~\ref{sec:auts}: For 
a graph of degree at least $3$, every automorphism of $Y(G)$ is obtained from 
an automorphism of $X(G)$ by means of an automorphism of $G$. In particular, 
this allows to show that $Y(G)$ and $\tilde Y(G)$ are not isomorphic.  
In Section~\ref{sec:poly}  we present a polynomial time algorithm accepting 
$Y(G)$ for all graphs $G$ and rejecting the corresponding $\tilde Y(G)$'s. 

Using techniques developed so far, we show that there is a first-order formula $\varphi(x,y)$ expressing in  almost all graphs $Y(G)$ and $\tilde Y(G)$  that $x$ and $y$  have the same color in $X(G)$ and 
$\tilde X(G)$, respectively (see Section~\ref{sec:fodef}).

 For 
$ k\ge 1$, let $L^k$ and $C^k$  denote the fragment of first-order logic and 
of first-order logic with counting, respectively, whose formulas contain at 
most $k$ variables. It is known that the $L^k$-equivalence of colored 
CFI-graphs $X(G)$ and  $\tilde X(G)$ already implies their  
$C^k$-equivalence. In Section~\ref{sec:ck} we present a proof of this fact. 
For $k=2$ this result does not hold for the uncolored CFI-graphs, however it 
is not known whether there are counterexamples for $k\ge 3$. 

Section~\ref{sec:tw1} and Section~\ref{sec:tw2} relate the tree-width of the 
graph $G$ with the $C^k$-equivalence of the CFI-graphs; more precisely, 
$Y(G)$ and $\tilde Y(G)$ (and $X(G)$ and $\tilde X(G)$) are $C^k$-equivalent 
if and only if the tree-width of $G$ is at least $k$. Finally, in 
Section~\ref{sec:cclke},  we use the CFI-graphs to show that the problem to 
decide  whether graphs $H_1$ and~$H_2$ are $L^{k}$-equivalent is 
\textup{coNP}-hard under polynomial-time many-one reductions (here the input 
consists of the graphs $H _1$ and $H_2$ and of the natural number $k$). The corresponding result for the   $C^{k}$-equivalence was shown in~\cite{licrassch25,sepp24}.

% Preliminaries
\section{Preliminaries}\label{sec:pre}

For a natural number $d\ge 1$ let $[d]:= \{1, 2, \ldots, d\}$. For a set $A$ 
we denote by $|A|$ its cardinality. For sets~$A$ and~$ B$ we denote by 
$A\dotcup B$ their disjoint union; if the sets are not disjoint, we tacitly 
pass to disjoint copies. 

We denote the \emph{symmetric difference of sets $A$ and $B$} by $A\,\Delta\, 
B$, i.e., $A\,\Delta\, B= (A\setminus B)\cup (B\setminus A)$. Later we will 
use the following well-known equalities and facts. 
\begin{itemize}
\item $A\,\Delta\, B= B\,\Delta\, A$, \ \ $(A\,\Delta\, B)\,\Delta\, C= 
    A\,\Delta\, (B\,\Delta\, C)$, \ \ $A\,\Delta\, A=\emptyset$, \ \ 
    $A\Delta\emptyset= A$, \ \ $A\cap B=(A\cup B)\setminus (A\,\Delta\, 
    B)$.

\item If $A$ and $B$ are finite sets of even cardinality, then $A\,\Delta\, 
    B$ is of even cardinality, too. 
\end{itemize}

\medskip
\noindent Sometimes we consider functions $f:A\to B$, where $A$ contains some 
of its subsets, say $X\subseteq A$, also as an element (i.e., $X\in A$). To 
avoid misunderstandings we write $f(X)$ for the value of $X$ under the  
function~$f$ but $\bar f(X)$ for the set $\{f(a)\mid a\in X\}$. 

\medskip
\noindent For graphs we use the notation $G= (V(G), E(G))$ common in graph 
theory. Here $V(G)$ is the non-empty set of vertices of $G$ and $E(G)$ is the 
set of edges. We only consider finite, simple, and undirected graphs and 
briefly speak of graphs. To express that there is an edge connecting the 
vertices $u$ and~$v$ of the graph~$G$, we use one of the notations $uv\in 
E(G)$ and $\{u,v\}\in E(G)$ (depending on which one is more readable in the 
given context).

For $u\in V(G)$ we denote by $\deg(u)$ its \emph{degree}, i.e., $\deg(u)= 
|\{v\in V(G) \mid uv\in E(G)\}|$. The \emph{degree of $G$}, $\deg(G)$ (often 
denoted by $\Delta(G)$ in the literature), is the \emph{maximum} degree of 
its vertices, i.e., $\deg(G)= \max\{\deg(u) \mid u\in V(G)\}$. 

If $X$ is a set of vertices of the graph $G$ we denote by $G\setminus X$ the 
graph induced by $G$ on $V(G)\setminus X$. For graphs $G$ and~$H$ we denote 
by $G\dotcup H$ the \emph{disjoint union of $G$ and $H$}, i.e., the graph 
with vertex set $V(G)\dotcup V(H)$ and edge set $E(G)\dotcup E(H)$. 

In examples or counterexamples, graphs that are paths or cycles will play an
important role. For $n\ge 1$ we denote by $P_n$ a path with $n$ edges (and 
$n+1$ vertices) and for $n\ge 3$ by $C_n$ a cycle of $n$ edges (and $n$ 
vertices). We also say that $P_n$ and $C_n$ have length $n$. If we write ``let $P_n$ be the path $u_1,\ldots,u_{n+1}$", we mean  $V(P_n)=\{u_i\mid i \in [n+1]\}$ and $E(P_n)=\{u_iu_{i+1}\mid i \in [n]\}$.

Let $G$ and $H$ be graphs and $f:V(G)\to V(H)$. The function~$f$ is a 
\emph{homomorphism from $G$ to $H$} if it preserves edges, i.e., if $uv\in 
E(G)$ implies $f(u)f(v)\in E(H)$ for all $u,v\in V(G)$. The function $f$ is 
an \emph{isomorphism between $G$ and $H$}, shortly $f:G\cong H$, if $f$ is a 
bijection between $V(G)$ and $V(H)$ and $f$ preserves edges and non-edges, 
i.e., if for $u$ and $v$ in $V(G)$, 
\[
uv\in E(G)\iff f(u)f(v)\in E(H).
\]
An isomorphism from $G$ to $G$ is an \emph{automorphism of $G$}. By 
$\Hom(G,H)$ and $\Aut(G)$ we denote the set of isomorphisms from $G$ to $H$ 
and the set of automorphisms of $G$, respectively. We set $\hom(G,H)= 
|\Hom(G,H)|$ and $\aut(G)= |\Aut(G)|$. 

We also consider colored graphs (see the main text). Homomorphisms and 
isomorphisms between colored graphs must preserve the colors too.

% The CFI-gadgets 
\section{The CFI-gadgets}\label{sec:unc+col}

In this section we first introduce the uncolored CFI-gadgets. They are 
obtained from the colored ones introduced by Cai, F\"urer, and Immeman 
in~\cite{caifurimm92} by just deleting the colors. We present the colored 
ones in the second part of this section. 

\paragraph{The uncolored CFI-gadget.} 
For $d\ge 1$ the (uncolored) \emph{\CFI-gadget} $Y(d)$ is determined by 
disjoint sets $A(d)$ and $B(d)$ with $|A(d)|= |B(d)|=d$ and a bijection 
$':A(d)\to B(d)$ from $A(d)$ onto $B(d)$. We set 
\[
M(d):=\{m\subseteq A(d) \mid \text{$|m|$ is even}\}
\]
(recall that by $|m|$ we denote the cardinality of $m$).

\medskip
\noindent The CFI-gadget $Y(d)$ is the graph with vertex set 
\[
V(Y(d)):= (A(d)\cup B(d))\dotcup M(d)
\]
and edge set 
\begin{equation}\label{eq:m-edges}
E(Y(d)):= \big\{am \mid a\in A(d), \ m\in M(d), \ a\in m\big\} 
 \cup \big\{a'm \mid a\in A(d), \ m\in M(d), \ a\notin m\big\}.
\end{equation}
We call the vertices in $M(d)$ \emph{middle vertices} and the vertices in 
$L(d):= A(d)\cup B(d)$ \emph{link vertices}.

 For later purposes, for every $a\in A(d)$ we call $a'\in B(d)$ 
the \emph{twin of $a$}, and vice versa. Thereby we extend the bijection 
$':A(d)\to B(d)$ to a bijection $': L(d)\to L(d)$ by setting for $b\in B(d)$, 
say $b= a'$ with $a\in A(d)$, 
\begin{equation}\label{eq:sibling}
b':=a.
\end{equation}
Thus, every $x\in L(d)$ has its unique twin $x'$.

\medskip
\noindent To get familiar with these gadgets we prove the following results we  
will use later.
\begin{lem}\label{lem:cig}
The graphs $Y(1)$ and $Y(2)$ contain no cycle. For $d\ge 3$, every two 
vertices in $Y(d)$ are on at least one cycle of length at most $8$.
\end{lem}

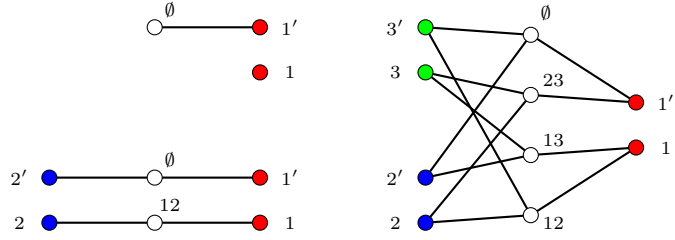
\begin{figure} 	
\centering
\begin{tikzpicture}[scale=0.2,vertex style/.style={draw,
                                   circle,
                                   minimum size=2mm,
                                   inner sep=0pt,
                                   outer sep=0pt,
                                   %shade
                                   }]

{\scriptsize
\begin{scope}[xshift=0cm,yshift=0cm]
  \path \foreach \i in {1}{%
   (11,19) coordinate[vertex style, fill=white] (m\i)};

  \draw (12,20.2) node{{%\small
		\textcolor{black}{$\emptyset$}}};

  \path \foreach \i in {1,2}{%
	(18,13+3*\i) coordinate[vertex style, fill=red] (one\i)};
  \draw (20,16) node{{%\small
		\textcolor{black}{$1$}}};
  \draw (20,19) node{{%\small
		\textcolor{black}{$1'$}}};

  \draw[thick] (m1)--(one2);
  
\end{scope}}

{\scriptsize                                   
\begin{scope}[xshift=0cm,yshift=-10cm]
  \path \foreach \i in {1,...,2}{%
   (11,13+3*\i) coordinate[vertex style, fill=white] (m\i)};

%  \draw (11,17.2) node{{%\small
%		\textcolor{black}{${\{1,2\}}$}}};
%  \draw (11,20.2) node{{%\small
%		\textcolor{black}{$\emptyset$}}};

  \draw (12,17.2) node{{%\small
		\textcolor{black}{$12$}}};
  \draw (12,20.2) node{{%\small
		\textcolor{black}{$\emptyset$}}};

  \path \foreach \i in {1,2}{%
	(18,13+3*\i) coordinate[vertex style, fill=red] (one\i)};
  \draw (20,16) node{{%\small
		\textcolor{black}{$1$}}};
  \draw (20,19) node{{%\small
		\textcolor{black}{$1'$}}};

  \path \foreach \i in {1,2}{%
	(4,13+3*\i) coordinate[vertex style, fill=blue] (two\i)};
  \draw (2,16) node{{%\small
		\textcolor{black}{$2$}}};
  \draw (2,19) node{{%\small
		\textcolor{black}{$2'$}}};

%  \path \foreach \i in {1,2}{%
%	(4,18+3*\i) coordinate[vertex style, fill=green] (three\i)};
%  \draw (2,21) node{{%\small
%		\textcolor{black}{$3$}}};
%  \draw (2,24) node{{%\small
%		\textcolor{black}{$3'$}}};

  \draw[thick] (m1)--(one1);
  \draw[thick] (m1)--(two1); 
%  \draw[thick] (m1)--(three2); 

%  \draw[thick] (m2)--(one1);
%  \draw[thick] (m2)--(two2); 
%  \draw[thick] (m2)--(three1); 

%  \draw[thick] (m3)--(one2);
%  \draw[thick] (m3)--(two1); 
%  \draw[thick] (m3)--(three1); 

  \draw[thick] (m2)--(one2);
  \draw[thick] (m2)--(two2); 
%  \draw[thick] (m4)--(three2); 
  
\end{scope}}

{\scriptsize
\begin{scope}[xshift=25cm,yshift=-5cm]
  \path \foreach \i in {1,...,4}{%
    (11,7.5+4*\i) coordinate[vertex style, fill=white] (m\i)};
%
%  \draw (12,10) node{{%\small
%		\textcolor{black}{${\{1,2\}}$}}};
%  \draw (12,16.8) node{{%\small
%		\textcolor{black}{${\{1,3\}}$}}};
%  \draw (12,20.5) node{{%\small
%		\textcolor{black}{${\{2,3\}}$}}};
%  \draw (11,25) node{{%\small
%		\textcolor{black}{${\emptyset}$}}};

  \draw (12.5,11) node{{%\small
		\textcolor{black}{$12$}}};
  \draw (12.5,16.5) node{{%\small
		\textcolor{black}{$13$}}};
  \draw (12.5,20.5) node{{%\small
		\textcolor{black}{$23$}}};
  \draw (12,25) node{{%\small
		\textcolor{black}{$\emptyset$}}};

  \path \foreach \i in {1,2}{%
	(18,13+3*\i) coordinate[vertex style, fill=red] (one\i)};
  \draw (20,16) node{{%\small
		\textcolor{black}{$1$}}};
  \draw (20,19) node{{%\small
		\textcolor{black}{$1'$}}};

  \path \foreach \i in {1,2}{%
	(4,8+3*\i) coordinate[vertex style, fill=blue] (two\i)};
  \draw (2,11) node{{%\small
		\textcolor{black}{$2$}}};
  \draw (2,14) node{{%\small
		\textcolor{black}{$2'$}}};

  \path \foreach \i in {1,2}{%
	(4,18+3*\i) coordinate[vertex style, fill=green] (three\i)};
  \draw (2,21) node{{%\small
		\textcolor{black}{$3$}}};
  \draw (2,24) node{{%\small
		\textcolor{black}{$3'$}}};

  \draw[thick] (m1)--(one1);
  \draw[thick] (m1)--(two1); 
  \draw[thick] (m1)--(three2); 

  \draw[thick] (m2)--(one1);
  \draw[thick] (m2)--(two2); 
  \draw[thick] (m2)--(three1); 

  \draw[thick] (m3)--(one2);
  \draw[thick] (m3)--(two1); 
  \draw[thick] (m3)--(three1); 

  \draw[thick] (m4)--(one2);
  \draw[thick] (m4)--(two2); 
  \draw[thick] (m4)--(three2); 

\end{scope}}

r\end{tikzpicture}
\caption{The gadgets $X(1)$ with $A(1)= [1]$, $X(2)$ with $A(2)=[2]$, and $X(3)$ 
with $A(3)=[3] $. For the middle vertices we write, say $ 12$ instead of $ \{1,2 \}$ to declutter the picture.}\label{fig:X1X2X3}
\end{figure}

\proof For the first statement see Figure~\ref{fig:X1X2X3}: the gadgets 
$Y(1)$ and $Y(2)$ are obtained from the gadgets $X(1)$ and $X(2)$ by 
forgetting the colors. For the second statement it is easy to see that it 
suffices to show the claim for $d= 3$. Assume $A(3):= [3]$. For example, a 
cycle containing the vertices 1 and~$1'$ is 
\[
1\to \{1,2 \} \to 3'\to \emptyset\to 1'\to \{2,3 \}\to 3\to \{1,3 \}\to 1. \benda
\]
By~\eqref{eq:m-edges} neighbors of link vertices in $Y(d)$ are middle 
vertices and neighbors of middle vertices  are link vertices. Moreover, for 
every link vertex $x\in L(d)$ and every middle vertex $m\in Y(d)$ we have (again by~\eqref{eq:m-edges}), 
\begin{equation}\label{eq:cmeg}
x m\in E(Y(d))\iff x' m\notin E(Y(d)).
\end{equation}
We can show a little bit more. 
\begin{lem}\label{lem:li'uni}
Assume $d\ge 3$ and $x\in L(d)$. Then the twin $x'$ of $ x$ is the only vertex $y$ 
of $Y(d)$ satisfying 
\[
\text{for all $m\in M(d)$}: x m\in  E(Y(d))\iff y m\notin  E(Y(d)).
\]
\end{lem}

\proof For a contradiction assume that $y\ne x'$ is a vertex satisfying the 
equivalence 
\[
x m\in  E(Y(d))\iff y m\notin  E(Y(d))
\]
for all $m\in M(d)$. Clearly, $y\ne x$, too. 

We first prove that $y$ must be a link vertex. So assume that $y$ is not a 
link vertex. As $d\ge 3$ we get in case that $x\in A(d)$ distinct vertices 
$a_1, a_2\in A(d)$ with $a_1\ne x$ and $a_2\ne x$. Then, for $m:= \{a_1, 
a_2\}$ we have $y\{a_1, a_2\}\notin E(G)$ (as $y$ is not a link vertex) and 
thus by the equivalence above, $x\{a_1,a_2\}\in E(G)$, i.e., $x\in \{a_1, 
a_2\}$, a contradiction. In case that  $x\in B(d)$ we 
have for $a\in A(d)$ with $a\ne x'$ that $y \{a,x'\}\notin E(G)$ and thus by 
the equivalence above, $x\{a,x' \}\in E(G)$, again a contradiction. 
 
Hence $y$ is a link vertex, i.e., $y\in L(d)$. In the next table the 
first line  lists all possible scenarios for $x,y$ and the second line 
contains middle vertices $m\in M(d)$ that lead to a contradiction in the 
corresponding case. Here $a$ is a vertex in $A(d)$ such that $a\notin 
\{x,x',y,y'\}$ (recall that $d\ge 3$). 
\[
\begin{array}{c|c|c|c|c}
& x, y\in A(d) &x, y\in B(d)& \text{$x\in A(d)$ and $y\in B(d)$} 
 & \text{$x\in B(d)$ and $y\in A(d)$} \\\hline \\[-3mm] 
 m  &\{x, y\} &\{x', y'\} &\{y', a\} &\{x', a\}
\end{array}. \benda
\]

\medskip
\begin{rem}\label{rem:neighbors}
In $Y(1)$ with $A(1)= [1]$ we get a counterexample to the displayed statement 
of the lemma. In fact, for $x:=1'$ not only $x'$ but also $y:= \emptyset$ 
satisfies the displayed equivalence in Lemma~\ref{lem:li'uni} (as $1'\emptyset \in  E(Y(1))$ and  $\emptyset 
\emptyset \notin  E(Y(1))$). 

In $Y(2)$ with $A(2)=[2]$ we get a counterexample for $x:=1$ and $y:=2'$ (as 
$1\{1,2 \}\in E(Y(2))$, $2'\{1,2 \}\notin E(Y(2))$ and $1\emptyset\notin 
E(Y(2))$, $2'\emptyset\in E(Y(2))$). $\hfill{\dashv}$
\end{rem}

\paragraph{The colored CFI-gadgets.}
The \emph{colored \CFI-gadget $X(d)$} is the pair $(Y(d),c)$, where $c$ is a 
coloring of the link vertices in $L(d)= A(d)\cup B(d)$ with the property 
\begin{eqnarray}\label{eq:ccg}
\text{$c(a)= c(a')$ for $a\in A(d)$}
 & \text{and} &
\text{$c(a_1)\ne c(a_2)$ for $a_1,a_2\in A(d)$ with $a_1\ne a_2$}. 
\end{eqnarray}
 Thus, $c(x)= c(x')$ for every $x\in L(d)$, i.e., twins have the same color. See 
Figure~\ref{fig:X1X2X3}. 

% Automorphisms
\section{Automorphisms of the colored CFI-gadgets}\label{sec:autc}

 Recall that by $\Aut(G)$ we denote the set of automorphisms of the graph~$G$. 
 If the graph~$G$ is colored, then $\Aut(G)$ denotes 
the set of automorphisms that preserve the color of the colored vertices.

In this section we study the automorphisms of the colored CFI-gadget $X(d)$. 
Note that by the coloring of $X(d)$ we have for every $f\in \Aut(X(d))$,
\begin{eqnarray}\label{eq:autad}
\text{for $a\in A(d)$, \ \ $\{f(a), f(a')\}= \{a, a'\}$} 
 & \text{and thus}, &
\text{for $m\in M(d)$, \ \ $f(m)\in M(d)$.}
\end{eqnarray}
We use these facts to characterize the functions in $\Aut(X(d))$.
\begin{prop}\label{pro:autcolgad}
For every $m\in M(d)$ there is exactly one automorphism in $\Aut(X(d))$, 
which we denote by $f_m$, with the property 
\[
f_m(\emptyset)= m
\]
(here $\emptyset$ refers to the empty set as an \emph{element} of $M(d)$). 
Hence, 
\begin{equation}
\Aut(X(d))=\{f_m \mid m\in M(d)\} .
\end{equation}
The values of $f_m$ are given by
\begin{equation}\label{eq:auta2d}
\text{for $a\in A(d)$}, \quad
f_m(a)= 
\begin{cases}
a & \text{if $a\notin m$} \\
a' & \text{if $a\in m$} 
\end{cases}
\end{equation}
and thus by \eqref{eq:autad},
\begin{equation}\label{eq:auta2db}
\text{for $a'\in B(d)$}, \quad
f_m(a'):= 
\begin{cases}
a' & \text{if $a\notin m$} \\
a & \text{if $a\in m$}. 
\end{cases}
\end{equation}
Furthermore for $m^*\in M(d)$,
\begin{equation}\label{eq:autmd}
\text{for $m^*\in M(d)$}, \ \ f_m(m^*)= m^*\,\Delta\, m 
\end{equation}
(note that $(m^*\,\Delta\, m) \in M(d) $, as $ m^*$ and $ m$ are of even cardinality). In particular, as $\emptyset\, \Delta\, m =m$ and by~\eqref{eq:auta2d},
\begin{equation}\label{eq:resemp} 
f_m(\emptyset)= m= \{a\in A(d) \mid f_m(a)= a'\}.
\end{equation}
The automorphism $f_m$ is self-inverse, that is, $f_m^{-1}= f_m$.
\end{prop}

\smallskip 
\begin{exa}\label{exa:autcolgad}
Let $d= 3$, $A(3):= [3]$, and $B(3):= \{1',2',3'\}$. 
Figure~\ref{fig:exaautoX3} illustrates the automorphism $f:= f_{\{1, 2\}}\in 
\Aut(X(3))$. \hfill{$ \dashv$}
\begin{figure} 	
\centering
\begin{tikzpicture}[scale=0.2,vertex style/.style={draw,
                                   circle,
                                   minimum size=2mm,
                                   inner sep=0pt,
                                   outer sep=0pt,
                                   %shade
                                   }]

{\scriptsize
\begin{scope}[]
  \path \foreach \i in {1,...,4}{%
    (11,7.5+4*\i) coordinate[vertex style, fill=white] (m\i)};
%
%  \draw (12,10) node{{%\small
%		\textcolor{black}{${\{1,2\}}$}}};
%  \draw (12,16.8) node{{%\small
%		\textcolor{black}{${\{1,3\}}$}}};
%  \draw (12,20.5) node{{%\small
%		\textcolor{black}{${\{2,3\}}$}}};
%  \draw (11,25) node{{%\small
%		\textcolor{black}{${\emptyset}$}}};

  \draw (12.5,11) node{{%\small
		\textcolor{black}{$12$}}};
  \draw (12.5,16.5) node{{%\small
		\textcolor{black}{$13$}}};
  \draw (12.5,20.5) node{{%\small
		\textcolor{black}{$23$}}};
  \draw (12,25) node{{%\small
		\textcolor{black}{$\emptyset$}}};

  \path \foreach \i in {1,2}{%
	(18,13+3*\i) coordinate[vertex style, fill=red] (one\i)};
  \draw (20,16) node{{%\small
		\textcolor{black}{$1$}}};
  \draw (20,19) node{{%\small
		\textcolor{black}{$1'$}}};

  \path \foreach \i in {1,2}{%
	(4,8+3*\i) coordinate[vertex style, fill=blue] (two\i)};
  \draw (2,11) node{{%\small
		\textcolor{black}{$2$}}};
  \draw (2,14) node{{%\small
		\textcolor{black}{$2'$}}};

  \path \foreach \i in {1,2}{%
	(4,18+3*\i) coordinate[vertex style, fill=green] (three\i)};
  \draw (2,21) node{{%\small
		\textcolor{black}{$3$}}};
  \draw (2,24) node{{%\small
		\textcolor{black}{$3'$}}};

  \draw[thick] (m1)--(one1);
  \draw[thick] (m1)--(two1); 
  \draw[thick] (m1)--(three2); 

  \draw[thick] (m2)--(one1);
  \draw[thick] (m2)--(two2); 
  \draw[thick] (m2)--(three1); 

  \draw[thick] (m3)--(one2);
  \draw[thick] (m3)--(two1); 
  \draw[thick] (m3)--(three1); 

  \draw[thick] (m4)--(one2);
  \draw[thick] (m4)--(two2); 
  \draw[thick] (m4)--(three2); 

\end{scope}}

%{\scriptsize
%\begin{scope}[xshift=23cm]
%  
%  \draw (0,17.5) node{{
%		\textcolor{black}{$\Longrightarrow$}}};
%
%\end{scope}}

{\scriptsize
\begin{scope}[xshift=25cm]
  \path \foreach \i in {1,...,4}{%
   (11,7.5+4*\i) coordinate[vertex style, fill=white] (m\i)};

%  \draw (12,11) node{{%\small
%		\textcolor{black}{$f(\emptyset)$}}}; 
%  \draw (12.5,17.1) node{{%\small
%		\textcolor{black}{$f(\{2,3\})$}}};
%  \draw (11.1,20.8) node{{%\small
%		\textcolor{black}{$f(\{1,3\})$}}};
%  \draw (12.5,25.3) node{{%\small
%		\textcolor{black}{$f(\{1,2\})$}}};

  \draw (12.5,10.5) node{{%\small
		\textcolor{black}{$f(\emptyset)$}}};
  \draw (12.5,17) node{{%\small
		\textcolor{black}{$f(23)$}}};
  \draw (12.5,21) node{{%\small
		\textcolor{black}{$f(13)$}}};
  \draw (12.5,25) node{{%\small
		\textcolor{black}{$f(12)$}}};

  \path \foreach \i in {1,2}{%
	(18,13+3*\i) coordinate[vertex style, fill=red] (one\i)};
  \draw (20.2,16) node{{%\small
		\textcolor{black}{$f(1')$}}};
  \draw (20,19) node{{%\small
		\textcolor{black}{$f(1)$}}};

  \path \foreach \i in {1,2}{%
	(4,8+3*\i) coordinate[vertex style, fill=blue] (two\i)};
  \draw (1.8,11) node{{%\small
		\textcolor{black}{$f(2')$}}};
  \draw (2,14) node{{%\small
		\textcolor{black}{$f(2)$}}};

  \path \foreach \i in {1,2}{%
	(4,18+3*\i) coordinate[vertex style, fill=green] (three\i)};
  \draw (2,21) node{{%\small
		\textcolor{black}{$f(3)$}}};
  \draw (1.8,24) node{{%\small
		\textcolor{black}{$f(3')$}}};

  \draw[thick] (m1)--(one1);
  \draw[thick] (m1)--(two1); 
  \draw[thick] (m1)--(three2); 

  \draw[thick] (m2)--(one1);
  \draw[thick] (m2)--(two2); 
  \draw[thick] (m2)--(three1); 

  \draw[thick] (m3)--(one2);
  \draw[thick] (m3)--(two1); 
  \draw[thick] (m3)--(three1); 

  \draw[thick] (m4)--(one2);
  \draw[thick] (m4)--(two2); 
  \draw[thick] (m4)--(three2); 

\end{scope}}

r\end{tikzpicture}
\caption{We apply the automorphism $f:= f_{\{1, 2\}}$ to the CFI-gadget $X(3)$.}\label{fig:exaautoX3}
\end{figure}
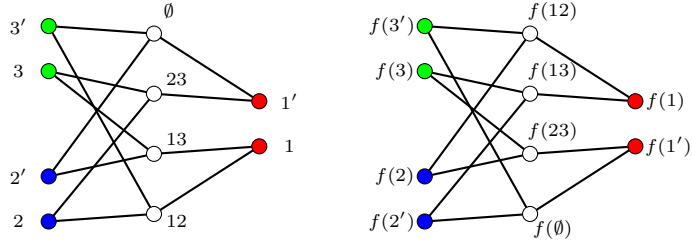
\end{exa}

\noindent \textit{Proof of Proposition~\ref{pro:autcolgad}:} Assume $m\in M$, 
$f\in \Aut(X(d))$, and $f(\emptyset)= m$. We show that $f=f_m$, i.e., $f$ 
satisfies the equalities~\eqref{eq:auta2d} and \eqref{eq:autmd} (for $f$ 
instead of $f_m$). 

Let $a\in A(d)$. Then $f(a)\in \{a, a'\}$ by~\eqref{eq:autad}. As $a 
\emptyset\notin E(X(d))$ (by~\eqref{eq:m-edges}), we get
$f(a)f(\emptyset)\notin E(X(d))$, i.e, $f(a)m\notin E(X(d))$. Therefore 
(cf.~\eqref{eq:m-edges}), 
\begin{eqnarray*}
\text{if $a\in m$, then $f(a)=a'$}
 & \text{and} &
\text{if $a\notin m$, then $f(a)=a$.}
\end{eqnarray*}
This yields~\eqref{eq:auta2d} and the equality $m= \{a\in A(d) \mid f(a)= a' 
\}$. We turn to~\eqref{eq:autmd}. Let $m^*\in M(d)$ and $a\in A(d)$. Then 
\begin{align*} 
 & am^* \ \in E(X(d)) \\ 
 & \iff f(a)f(m^*)\in E(X(d)) 
 & \text{\big(as $f\in \Aut(X(d))$\big)} \\
 & \iff \text{\big($f(a)=a$ and $a\in f(m^*)$\big) \ or \ \big($f(a)=a'$ and $a\notin f(m^*)$\big)}
 & \text{\big(by~\eqref{eq:m-edges}\big)} \\
 & \iff \text{$a\in (f(m^*)\setminus m)$ \ or \ $a\in (m\setminus f(m^*))$} 
 & \text{\big(as $m= \{a\in A(d) \mid f(a)= a'\}$\big)} \\
 & \iff a\in f(m^*)\,\Delta\, m. 
\end{align*}
Thus $f(m^*)\,\Delta\, m= m^*$ and hence, $f(m^*)= \big((f(m^*)\,\Delta\, 
m)\, \Delta\, m\big)= m^*\Delta m$. Therefore, $f= f_m$. 

\medskip 
\noindent We still have to show that $f_m\in \Aut(X(d))$ and that $f_m$ is 
self-inverse. By~\eqref{eq:autad} we have $f_m(f_m(a))= a$ and $f_m(f_m(a'))= a'$ for $a\in A(d)$. 
 By~\eqref{eq:autmd} we get for $m^*\in M(d)$, 
\[
f_m(f_m(m^*))= f_m(m^*)\, \Delta\, m= (m^*\, \Delta\, m)\, \Delta\, m= m^*.
\]
So $f_m$ is self-inverse and therefore bijective. Arguing similarly as in the 
chain of equivalences above one verifies that $f_m$ preserves edges. E.g., 
for $a\in A(d)$ and $m^*\in M(d)$, 
\begin{align*} 
am^* \in E(X(d)) 
 & \iff 
 \text{\big($f_m(a)=a$ and $f_m(a)\in m^*$\big) 
  or \big($f_m(a)=a'$ and $f_m(a)\notin m^*$\big)} \\
 & \iff f_m(a)\in (m^*\setminus m)\cup (m\setminus m^*) 
  \ \ (= m^*\,\Delta \,m  = f_m(m^*)) \\
 & \iff f_m(a)f_m(m^*)\in E(X(G)). \benda
\end{align*}

\begin{cor}\label{cor:uniquefm} 
For every $m_1, m_2\in M(d)$ there is a unique automorphism $f\in \Aut(X(d))$ 
with $f(m_1)= m_2$, namely $f= f_{m_1\Delta m_2}$.
\end{cor}

\proof Assume $f\in \Aut(X(d))$ with $f(m_1)=m_2$. By the previous result we 
know $f= f_m$ for some $m\in M(d)$. By~\eqref{eq:autmd}, $f_m(m_1)= 
m_1\,\Delta\, m= m_2$. Thus, $m_1\,\Delta\, m_2= (m_1\,\Delta\,(m_1\,\Delta\, 
m))= m$. Clearly, $f_{m_1\Delta m_2}(m_1)= (m_1\Delta m_2)\Delta 
m_1=m_2$.\proofend 

\medskip
\noindent Later we will make use of the following reformulation of parts of 
Proposition~\ref{pro:autcolgad}. 
\begin{cor}\label{cor:uniquefm2} 
For every $a\in A(d)$ choose $u(a)\in \{a,a'\}$ such that 
\[
m:= \{a\in A(d) \mid u(a) =a'\}
\]
has an even cardinality. Then there is a unique $f\in \Aut(X(d))$ with $f(a)= 
u(a)$ for all $a\in A(d)$, namely~$f_m$. In particular, $f_m(\emptyset)= m$.  
\end{cor}

\noindent
In the following for $f\in \Aut(X(d))$ we set 
\begin{equation}\label{eq:deffm}
m_f:= \{a\in A(d) \mid f(a)=a'\} \ (=f(\emptyset)).
\end{equation}
Hence, by \eqref{eq:autmd} and \eqref{eq:resemp} we have for all $ m^*\in M(d)$
\begin{equation}\label{eq:deffm*}f(m^*)=m^*\Delta m_f=m^*\Delta \{a\in A(d) \mid f(a)=a'\}.
  \end{equation}

  \section{Automorphisms of the uncolored CFI-gadgets}\label{sec:autu}

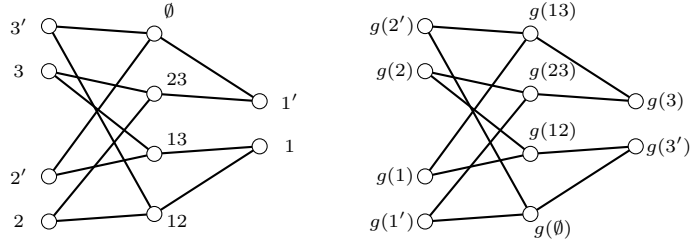
\begin{figure} 	
\centering
\begin{tikzpicture}[scale=0.2,vertex style/.style={draw,
                                   circle,
                                   minimum size=2mm,
                                   inner sep=0pt,
                                   outer sep=0pt,
                                   %shade
                                   }]

{\scriptsize
\begin{scope}[]
  \path \foreach \i in {1,...,4}{%
   (11,7.5+4*\i) coordinate[vertex style, fill=white] (m\i)};

%  \draw (12,10) node{{%\small
%		\textcolor{black}{${\{1,2\}}$}}};
%  \draw (12,16.8) node{{%\small
%		\textcolor{black}{${\{1,3\}}$}}};
%  \draw (12,20.5) node{{%\small
%		\textcolor{black}{${\{2,3\}}$}}};
%  \draw (11,25) node{{%\small
%		\textcolor{black}{${\emptyset}$}}};

  \draw (12.5,11) node{{%\small
		\textcolor{black}{$12$}}};
  \draw (12.5,16.5) node{{%\small
		\textcolor{black}{$13$}}};
  \draw (12.5,20.5) node{{%\small
		\textcolor{black}{$23$}}};
  \draw (12,25) node{{%\small
		\textcolor{black}{$\emptyset$}}};

  \path \foreach \i in {1,2}{%
	(18,13+3*\i) coordinate[vertex style, fill=white] (one\i)};
  \draw (20,16) node{{%\small
		\textcolor{black}{$1$}}};
  \draw (20,19) node{{%\small
		\textcolor{black}{$1'$}}};

  \path \foreach \i in {1,2}{%
	(4,8+3*\i) coordinate[vertex style, fill=white] (two\i)};
  \draw (2,11) node{{%\small
		\textcolor{black}{$2$}}};
  \draw (2,14) node{{%\small
		\textcolor{black}{$2'$}}};

  \path \foreach \i in {1,2}{%
	(4,18+3*\i) coordinate[vertex style, fill=white] (three\i)};
  \draw (2,21) node{{%\small
		\textcolor{black}{$3$}}};
  \draw (2,24) node{{%\small
		\textcolor{black}{$3'$}}};

  \draw[thick] (m1)--(one1);
  \draw[thick] (m1)--(two1); 
  \draw[thick] (m1)--(three2); 

  \draw[thick] (m2)--(one1);
  \draw[thick] (m2)--(two2); 
  \draw[thick] (m2)--(three1); 

  \draw[thick] (m3)--(one2);
  \draw[thick] (m3)--(two1); 
  \draw[thick] (m3)--(three1); 

  \draw[thick] (m4)--(one2);
  \draw[thick] (m4)--(two2); 
  \draw[thick] (m4)--(three2); 

\end{scope}}

%
%{\scriptsize
%\begin{scope}[xshift=23cm]
%  
%  \draw (0,17.5) node{{
%		\textcolor{black}{$\Longrightarrow$}}};
%
%\end{scope}}

{\scriptsize
\begin{scope}[xshift=25cm]
  \path \foreach \i in {1,...,4}{%
   (11,7.5+4*\i) coordinate[vertex style, fill=white] (m\i)};

%  \draw (12,11) node{{%\small
%		\textcolor{black}{$g(\emptyset)$}}}; 
%  \draw (12.5,17.1) node{{%\small
%		\textcolor{black}{$g(\{1,2\})$}}};
%  \draw (11.1,20.8) node{{%\small
%		\textcolor{black}{$g(\{2,3\})$}}};
%  \draw (12.5,25.3) node{{%\small
%		\textcolor{black}{$g(\{1,3\})$}}};

  \draw (12.5,10.5) node{{%\small
		\textcolor{black}{$g(\emptyset)$}}};
  \draw (12.5,17) node{{%\small
		\textcolor{black}{$g(12)$}}};
  \draw (12.5,21) node{{%\small
		\textcolor{black}{$g(23)$}}};
  \draw (12.5,25) node{{%\small
		\textcolor{black}{$g(13)$}}};

  \path \foreach \i in {1,2}{%
	(18,13+3*\i) coordinate[vertex style, fill=white] (one\i)};
  \draw (20.2,16) node{{%\small
		\textcolor{black}{$g(3')$}}};
  \draw (20,19) node{{%\small
		\textcolor{black}{$g(3)$}}};

  \path \foreach \i in {1,2}{%
	(4,8+3*\i) coordinate[vertex style, fill=white] (two\i)};
  \draw (1.8,11) node{{%\small
		\textcolor{black}{$g(1')$}}};
  \draw (2,14) node{{%\small
		\textcolor{black}{$g(1)$}}};

  \path \foreach \i in {1,2}{%
	(4,18+3*\i) coordinate[vertex style, fill=white] (three\i)};
  \draw (2,21) node{{%\small
		\textcolor{black}{$g(2)$}}};
  \draw (1.8,24) node{{%\small
		\textcolor{black}{$g(2')$}}};

  \draw[thick] (m1)--(one1);
  \draw[thick] (m1)--(two1); 
  \draw[thick] (m1)--(three2); 

  \draw[thick] (m2)--(one1);
  \draw[thick] (m2)--(two2); 
  \draw[thick] (m2)--(three1); 

  \draw[thick] (m3)--(one2);
  \draw[thick] (m3)--(two1); 
  \draw[thick] (m3)--(three1); 

  \draw[thick] (m4)--(one2);
  \draw[thick] (m4)--(two2); 
  \draw[thick] (m4)--(three2); 

\end{scope}}

r\end{tikzpicture}
\caption{The automorphism $g\in \Aut(Y(3))$ maps $\{1,1'\}$ to $\{2,2'\}$, 
$\{2,2'\}$ to $\{3,3'\}$, and $\{3,3'\}$ to $\{1,1'\}$.}\label{fig:autoY3}
\end{figure}

Throughout this section, we assume that $ d\ge 1$. Of course, every $f$ in $\Aut(X(d))$ is in $\Aut(Y(d))$. The coloring forces 
$f\in \Aut(X(d))$ to satisfy $f(a)\in \{a,a'\}$ for $a\in A(d)$.  As $ X(d)$ and $ Y(d)$ have the same edges, by \eqref{eq:ccg} we see that  this is the only additional condition that an automorphism of $Y(d)$ has to satisfy in order to be in $\Aut(X(d))$. For future discussions we record this in a lemma.
\begin{lem}\label{lem:AutoYdbeAutoXd}
Let $g\in \Aut(Y(d))$. Then $g\in \Aut(X(d))$ if and only if for every $a\in 
A(d)$ we have
\[
\{g(a), g(a')\}= \{a, a'\}.
\] 
\end{lem}
\noindent
But 
in $Y(d)$ for $a_1,a_2 \in A(d)$ we cannot 
distinguish the role of the link vertices $\{a_1, a_1'\}$ from the role of 
the link vertices $\{a_2, a_2'\}$.
For example, the automorphism $g$ of  $Y(3)$ given by the following table   and illustrated in Figure~\ref{fig:autoY3} completely  shuffles the link vertices in $Y(3)$.
\[
\begin{array}{c|c|c|c|c|c|c|c|c|c|c}
 x &1 &1' &2 &2' &3 &3' &\emptyset &\{1,2\} &\{1,3\} &\{2,3\} \\\hline \\[-3mm] 
g(x) &2' &2 &3 &3' &1' &1 &\{1,2\} &\{1,3\} & \emptyset  & \{2,3\}
\end{array}
\]
However, this $g$ still retains the following property, which is trivially true
for any $g\in \Aut(X(d))$ viewed as an automorphism of $Y(d)$.

\begin{figure} 	
\centering
\begin{tikzpicture}[scale=0.2,vertex style/.style={draw,
                                   circle,
                                   minimum size=2mm,
                                   inner sep=0pt,
                                   outer sep=0pt,
                                   %shade
                                   }]

{\scriptsize                                   
\begin{scope}[]
  \path \foreach \i in {1,...,2}{%
   (11,13+3*\i) coordinate[vertex style, fill=white] (m\i)};

  \draw (12,17.2) node{{%\small
		\textcolor{black}{$12$}}};
  \draw (12,20.2) node{{%\small
		\textcolor{black}{$\emptyset$}}};

  \path \foreach \i in {1,2}{%
	(18,13+3*\i) coordinate[vertex style, fill=white] (one\i)};
  \draw (20,16) node{{%\small
		\textcolor{black}{$1$}}};
  \draw (20,19) node{{%\small
		\textcolor{black}{$1'$}}};

  \path \foreach \i in {1,2}{%
	(4,13+3*\i) coordinate[vertex style, fill=white] (two\i)};
  \draw (2,16) node{{%\small
		\textcolor{black}{$2$}}};
  \draw (2,19) node{{%\small
		\textcolor{black}{$2'$}}};

  \draw[thick] (m1)--(one1);
  \draw[thick] (m1)--(two1);

  \draw[thick] (m2)--(one2);
  \draw[thick] (m2)--(two2); 
  
\end{scope}}

{\scriptsize                                   
\begin{scope}[xshift=30cm,yshift=0cm]
  \path \foreach \i in {1,...,2}{%
   (11,13+3*\i) coordinate[vertex style, fill=white] (m\i)};

  \draw (12,17.3) node{{%\small
		\textcolor{black}{$g(\emptyset)$}}};
  \draw (12,20.3) node{{%\small
		\textcolor{black}{$g(12)$}}};

  \path \foreach \i in {1,2}{%
	(18,13+3*\i) coordinate[vertex style, fill=white] (one\i)};
  \draw (20.2,16) node{{%\small
		\textcolor{black}{$g(2')$}}};
  \draw (20.2,19) node{{%\small
		\textcolor{black}{$g(1)$}}};

  \path \foreach \i in {1,2}{%
	(4,13+3*\i) coordinate[vertex style, fill=white] (two\i)};
  \draw (1.8,16) node{{%\small
		\textcolor{black}{$g(1')$}}};
  \draw (1.8,19) node{{%\small
		\textcolor{black}{$g(2)$}}};

  \draw[thick] (m1)--(one1);
  \draw[thick] (m1)--(two1);

  \draw[thick] (m2)--(one2);
  \draw[thick] (m2)--(two2); 
  
\end{scope}}

r\end{tikzpicture}
\caption{The automorphism $g\in \Aut(Y(2))$ sends $1$ to $1'$ and $1'$ to $2$, 
thus is not twin-preserving.}\label{fig:autoY2}
\end{figure}
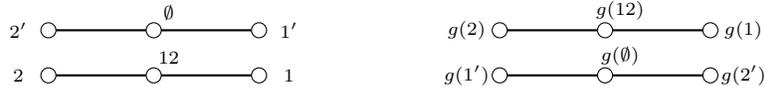

\begin{defn}\label{def:twinpreserving}
An automorphism $g\in \Aut(Y(d))$ is \emph{twin-preserving} if for all $x\in 
L(d)\ (= A(d)\cup B(d))$ we have $g(x)\in L(d)$ and 
$
g(x')= g(x)'.
$\hfill{$\dashv$}
\end{defn}
\noindent
The automorphism of Figure~\ref{fig:autoY3} is twin-preserving and  Figure~\ref{fig:autoY2} gives an automorphism $g\in \Aut(Y(2))$ which is 
\emph{not} twin-preserving.
\medskip

\noindent
Recall  that a function $\pi$ is a \emph{permutation of a 
    set $Z$} if $\pi: Z\to Z$ is bijective. We use $\Per(Z)$ to denote the 
    set of permutations of $Z$.

\begin{rem}\label{rem:twinpreserving}
It is straightforward to verify that for every $g\in \Aut(Y(d))$ all the 
following are equivalent. 
\begin{enumerate}
\item[(i)] $g$ is twin-preserving.

\item[(ii)] For every $a\in A(d)$ we have $ g(a)\in L(d)$ and  $g(a')= g(a)'$.

\item[(iii)] There is a permutation $\pi\in \Per(A(d))$  such 
    that $\{g(a), g(a')\}= \{\pi(a), \pi(a)'\}$ for all $a\in A(d)$.\hfill{$\dashv$} 
    
\end{enumerate}
\end{rem}
\noindent
It turns out that the notion of twin-preserving plays a crucial role in connecting 
$\Aut(Y(d))$ with $\Aut(X(d))$. First, in Proposition~\ref{pro:yfpitog} we 
will see how every $f\in \Aut(X(d))$ and every permutation $\pi$ of $A(d)$ 
induces a twin-preserving automorphism of $Y(d)$.

The following lemma  will be useful in the proof of this  proposition.
\begin{lem}\label{lem:3ver} For every $ \pi\in \Per(A(d))$ there is a unique $ \rho\in\Aut(Y(d))$ with
  \begin{equation}\label{eq:1}\mbox{ for all $ a\in A(d)$: \quad $ \rho(a)=\pi(a)$ \ \ and \ \ $ \rho(a')=\pi(a)'$},
  \end{equation}
   in particular, $ \rho$ is an extension of $ \pi$ and it is  twin-preserving.
  Moreover, $ \rho$ satisfies
  \begin{equation}\label{eq:2}\mbox{ for all $ m\in M(d)$}:\quad \rho(m)=\{\pi(a) \mid a\in m \}.
      \end{equation}
    \end{lem}
    \proof First note that for a function $ \rho$ satisfying \eqref{eq:1} we have
    \[
\bar\rho(A(d))=A(d)\quad \mbox{ and}\quad \bar\rho(B(d))=B(d).
\]
Assume now that $ \rho\in\Aut(Y(d))$ satisfies \eqref{eq:1}. We prove \eqref{eq:2}.
Let $ a\in A(d)$ and $ m\in M(d)$. Then
\begin{align*}a\in\rho(m) & \iff \{a, \rho(m)\}\in E(Y(d))\\
                          &\iff \{\rho^{-1}(a), m\}\in E(Y(d))& (\mbox{as $ \rho\in\Aut(Y(d))$})\\
                          &\iff  \{\pi^{-1}(a), m\}\in E(Y(d))& (\mbox{by $\bar\rho(A(d))=A(d)$ and \eqref{eq:1} })\\
  &\iff \pi^{-1}(a)\in m.
\end{align*}
Therefore, $\rho(m)=\{a \mid a\in A(d) \mbox{ and\ } \pi^{-1}(a)\in m\} = \{\pi(a) \mid a\in m \}$.

As by \eqref{eq:1} and \eqref{eq:2} all values of $ \rho$ are fixed, we have seen so far that at most one $\rho\in \Aut(d) $ can sastify~\eqref{eq:1}.
We still have to show that a function $ \rho$ defined on $ V(Y(d))$ by  \eqref{eq:1} and  \eqref{eq:2} preserves edges (and non-edges) and thus is  an automorphism of $ Y(d)$. Note that
for $ a\in A(d)$ and $ m\in M(d)$,
\begin{align*}\{a, m\}\in E(Y(d)) &  \iff a\in m\\
                                 &  \iff  \pi(a)\in \rho(m) & (\mbox{by \eqref{eq:2}})\\
                                 &   \iff  \rho(a)\in \rho(m) & (\mbox{by \eqref{eq:1}})\\
 &   \iff \{ \rho(a),\rho(m) \}\in E(Y(d)) & (\mbox{as $\bar\rho (A(d))=A(d)$}).
\end{align*}
And thus,
\begin{align*}\{a', m\}\in E(Y(d))&   \iff \{a, m\}\notin E(Y(d)  \\
                                 &   \iff \{ \rho(a),\rho(m) \}\notin E(Y(d)) &(\mbox{as just shown}) \\
 &  \iff  \{ \rho(a'),\rho(m) \}\in E(Y(d)) & (\mbox{by \eqref{eq:1}}).\benda
\end{align*}

\noindent
In the following, for $ \pi\in \Per(A(d))$ we denote its unique extension according to the previous lemma by~$ \rho_\pi$; in particular, $ \rho_\pi\in\Aut(Y(d))$.

\begin{prop}\label{pro:yfpitog}
For every $f\in \Aut(X(d))$ and every $\pi\in \Per(A(d))$ 
there is a unique twin-preserving $g\in \Aut(Y(d))$ such that for all $a\in A(d)$,
\begin{equation*}g(a)=f(\pi(a)).
\end{equation*}
For this $ g$ we have
\[
g=f\circ \rho_\pi.
\]
Moreover, for $m\in M(d)$, 
\begin{equation}\label{eq:fgm}
g(m)=\{\pi(a)\in A(d) \mid a\in m \} \, \Delta \, m_f
\end{equation}
 (recall that $m_f= \{a\in A(d)\mid f(a) =a'\}$).
\end{prop}
\noindent
Observe that the automorphism of $ Y(3)$ shown in Figure~\ref{fig:autoY3} corresponds to the $ g$
obtained according to the proposition for $ f=f_{\{1,2  \}}$ and the permutation $ \pi$ of $A(3)$ with $\pi(1)= 2$, $\pi(2)= 3$, and $\pi(3)= 1$.
\medskip

\medskip

\noindent \textit{Proof of Proposition~\ref{pro:yfpitog}:} Let $ f\in \Aut(X(d))$ and 
 $ \pi\in \Per(A(d))$. Clearly $f$ can be understood as an automorphism of $ Y(d)$ as 
well. We know that $ \rho_\pi\in\Aut(Y(d))$.
           So
\[
g:= f\circ \rho_\pi, \quad \text{as depicted by}\ \ 
\begin{tikzcd}
Y(d) \arrow[r, "\rho_\pi"] \arrow[rd,"g"] 
 & Y(d) \arrow[d, "f"]\\%
 & Y(d)
\end{tikzcd},
\]
is an automorphism of $Y(d)$. Observe that both $ f$ and $ \rho_\pi$ are twin-preserving, which implies that  $ g  $ is twin-preserving too. Moreover
 for $ a\in A(d)$ by~\eqref{eq:1},
\[
g(a)=f(\rho_\pi(a))=f(\pi(a)).
\]
To see~\eqref{eq:fgm}, for every $m\in M(d)$ we have 
\begin{align*}
g(m) = f(\rho_\pi(m))
 & = f(\{\pi(a)\mid a\in m\})& \text{(by~\eqref{eq:2})} \\
  & = \{\pi(a)\mid a\in m\}\, \Delta \, m_f & \text{(by~\eqref{eq:deffm})}.
\end{align*} 
We still need to establish the uniqueness of $g$. To that end, let $g_1\in 
\Aut(Y(d))$ be twin-preserving with $ g_1(a)=f(\pi(a))$ for all $ a\in A(d)$. Define
\[
\rho_1:= f^{-1}\circ g_1
\quad \text{i.e.,}\ \ 
\begin{tikzcd}
Y(d) \arrow[r, "\rho_1"] \arrow[rd,"g_1"] 
 & Y(d) \\%
 & Y(d) \arrow[u, "f^{-1}"swap]
\end{tikzcd}.
\]
Since $f, g_1\in \Aut(Y(d))$, we deduce $\rho_1\in \Aut(Y(d))$ as well. 
 Furthermore, for every $ a\in A(d)$,
\begin{equation}\label{eq:pi'a}
\rho_1(a)= f^{-1}(g_1(a))= f^{-1}(f(\pi(a)))= \pi(a)= \rho_\pi(a),
\end{equation}
and (recall that $ \rho_\pi$ satisfies \eqref{eq:1}, and both $ g_1$ and $ f$ are twin-preserving)
\begin{equation}\label{eq:pi'b}
\rho_1(a')= f^{-1}(g_1(a'))= f^{-1}(f(\pi(a)'))= \pi(a)'=\rho_\pi(a').
\end{equation}
Thus, $ \rho_1$ satisfies~\eqref{eq:1}, too. Hence, $ \rho_1=\rho_\pi$ by the previous lemma.  Thus, $ g_1=f\circ\rho_1=f\circ\rho_\pi=g$.\proofend
\medskip

\noindent
Now we prove that  the following converse of Proposition~\ref{pro:yfpitog} is true 
as well. 
\begin{prop}\label{pro:Ydtwinpreservingdecompose}
Let $g\in \Aut(Y(d))$ be twin-preserving. Then there are a unique $f\in 
\Aut(X(d))$ and a unique $\pi\in \Per(A(d))$ such that $g(a)= f(\pi(a))$ for 
all $a\in A(d)$.
\end{prop}
\medskip

\noindent
Thus, an automorphism $g\in \Aut(Y(d))$ is twin-preserving if and 
only if $g(a)= f(\pi(a))$ for 
all $a\in A(d)$ holds for  an $f\in \Aut(X(d))$ and $\pi\in \Per(A(d))$. Or equivalently, $g= f\circ \rho_\pi$ for an $f\in \Aut(X(d))$ and $\pi\in \Per(A(d))$.
\medskip

\noindent
\textit{Proof of Proposition~\ref{pro:Ydtwinpreservingdecompose}}. As $g$ is twin-preserving, Remark~\ref{rem:twinpreserving}~(iii) 
implies that there is a permutation $\pi$ of $A(d)$ such that for all $a\in 
A(d)$, 
\begin{equation}\label{eq:gpi}
\{g(a), g(a')\}= \{\pi(a), \pi(a)'\}.
\end{equation}
Using the twin-preserving extension $ \rho_\pi\in \Aut(Y(d)) $ of $ \pi $  we can 
rewrite the previous 
equality by $\{g(a), g(a')\}= \{\rho_\pi(a), \rho_\pi(a')\}$. Therefore, %
\begin{equation}\label{eq:grhoinverse}
\hspace{-1mm}\{g(\rho_\pi^{-1}(a)), g(\rho_\pi^{-1}(a)')\}
 = \{\rho_\pi(\rho_\pi^{-1}(a)), \rho_\pi(\rho_\pi^{-1}(a)')\}
 = \{\rho_\pi(\rho_\pi^{-1}(a)), \rho_\pi(\rho_\pi^{-1}(a'))\}
  = \{a,a'\},
\end{equation}
where the second equality is by the fact that $\rho_\pi$, and hence $\rho_\pi^{-1}$, 
are  twin-preserving. Then we define 
\[
f:= g\circ \rho_\pi^{-1} \quad \ \ 
\begin{tikzcd}
Y(d) \arrow[rd,"g"] 
 & Y(d) \arrow[l, "\rho_\pi^{-1}"swap] \arrow[d, "f"]\\%
 & Y(d)
\end{tikzcd}.
\]
Thus, $f\in \Aut(Y(d))$ and $g(a)=f(\rho_\pi(a))= f(\pi(a))$ for all $a\in A(d)$. 
To show $f\in \Aut(X(d))$,  we use Lemma~\ref{lem:AutoYdbeAutoXd} 
and deduce $\{f(a), f(a')\}= \{a,a'\}$ for every $a\in A(d)$ by 
\begin{align*}
\{f(a), f(a')\}
 & = \{g(\rho_\pi^{-1}(a)), g(\rho_\pi^{-1}(a'))\} \\
  & = \{g(\rho_\pi^{-1}(a)), g(\rho_\pi^{-1}(a)')\} & \text{(as $\rho_\pi^{-1} $ is twin-preserving)} \\
 & = \{a, a'\} &  \text{(by~\eqref{eq:grhoinverse})}.
\end{align*}
To prove the uniqueness of $f$ and $\pi$, let $f_1\in \Aut(X(d))$ and 
$\pi_1\in \Per(A(d))$ such that for all $a\in A(d)$ 
\[
g(a)= f_1(\pi_1(a)).
\]
Hence, for $a\in A(d)$, 
\begin{align*}
\{\pi_1(a), \pi_1(a)'\} 
 & = \{f_1(\pi_1(a)), f_1(\pi_1(a))'\} & \text{(as $f_1\in \Aut(X(d))$ and $\pi_1(a)\in A(d)$)} \\
  & = \{g(a), g(a)'\} & \text{(by $g(a)= f_1(\pi_1(a))$)} \\
   & = \{g(a), g(a')\} & \text{($g$ is twin-preserving)} \\
 & = \{\pi(a), \pi(a)'\} & \text{(by~\eqref{eq:gpi})}.
\end{align*}
Thus $\pi_1(a)= \pi(a)$ for $a\in A(d)$, i.e., $\pi_1= \pi$. To show $f_1= 
f$, by Corollary~\ref{cor:uniquefm2} it suffices to verify that $f_1(a)= 
f(a)$ for every $a\in A(d)$. This can be seen by 
\begin{align*}
f_1(a) & = g(\pi_1^{-1}(a))
 & \text{(by $\pi_1\in \Per(A(d))$ and $g(a)= f_1(\pi_1(a))$ for all $a\in A(d)$)} \\
 & = g(\pi^{-1}(a)) & \text{(by $\pi=\pi_1$)} \\
 & = f(a) & \text{(as $\rho_\pi$ extends $\pi$ and  $f= g\circ \rho_\pi^{-1}$)}. \benda
\end{align*} 
\medskip

\noindent
As already illustrated by~Figure~\ref{fig:autoY2}, there is an automorphism 
$g\in \Aut(Y(2))$ that is not twin-preserving. Thus, not all automorphisms of 
$Y(2)$ can be obtained from automorphisms of $X(2)$ by means of a permutation 
of $A(2)$ as in~Proposition~\ref{pro:yfpitog}. The next two examples 
demonstrate that this can also happen with the gadgets $Y(1)$  and $Y(4)$.  

\begin{exa}\label{exa:gY1}
Assume $d=1$ and let $A(d):= \{1\}$. Then $B(d)= \{1'\}$, $M(d)= 
\{\emptyset\}$, and $E(Y(d))$ only contains the edge $1'\emptyset$ (see 
Figure~\ref{fig:X1X2X3}). We have $\Aut(Y(d))= \{\id_{Y(d)}, g\}$, where~$\id_{Y(d)}$ denotes the 
identity on $V(Y(d))$ and $ g$ is the automorphism 
 with $g(1)= 1$, 
$g(1')= \emptyset$, and $g(\emptyset)= 1'$,  which is not twin-preserving. Moreover, note that $\bar g(A(d))\subseteq A(d)$, 
$\bar g(L(d))\not\subseteq L(d)$ and equivalently, $\bar 
g(M(d))\not\subseteq M(d)$. Clearly, $\id_{Y(d)}$ is  twin-preserving.
\hfill{$ \dashv$}
\end{exa}

\begin{exa}\label{exa:gY4} 
Assume $d= 4$ and let $Y(d)$ be given by $A(4)=[4]$. We present a $g\in 
\Aut(Y(d))$ with 
\[
g(4)= \emptyset \ (\in M(4)) \ \ \text{and} \ \ g^{-1}=g.
\]
Thus, $g(\emptyset)= 4$. As $1', 2',3',4'$ are the neighbors of $\emptyset$ 
in $Y(4)$, the vertices $g(1'), \ldots, g(4')$ must be the neighbors of~$4$ 
in $Y(4)$. We set 
\[
g(1')= \{1,4\}, \ \ g(2')= \{2,4\}, \ \ g(3')=\{3, 4\}, \ \ g(4')=\{1, 2, 3, 4\}.
\]
The vertices 
\[
1, \ 2, \ 3, \ \{1, 2\}, \{1 ,3\}, \  \{2, 3\}
\]
are precisely the vertices that so far are neither in the domain nor in the 
range of $g$. 
We set 
\[
g(1)= \{2, 3\}, \ \ g(2)= \{1, 3\}, \ \ g(3)= \{1, 2\}.
\]
Then $\bar g(L(d))= M(d)$. As we want to have $g^{-1}= g$, the function $g$ 
is already defined for all vertices as in Figure~\ref{fig:f4}.  We leave it to 
the reader to verify that $g\in \Aut(Y(d))$. Note that $g$ is clearly not 
twin-preserving. \hfill{$ \dashv$}
\begin{figure} 	
\centering
\begin{tikzpicture}[scale=0.2,vertex style/.style={draw,
                                   circle,
                                   minimum size=2mm,
                                   inner sep=0pt,
                                   outer sep=0pt,
                                   %shade
                                   }]

{\scriptsize
\begin{scope}[]
  \path \foreach \i in {1,...,8}{%
   (10,7.5+4*\i) coordinate[vertex style, fill=white] (m\i)};

  \draw (10,10) node{{%\small
		\textcolor{black}{$1234$}}};
  \draw (10,14) node{{%\small
		\textcolor{black}{$34$}}};
  \draw (10,18) node{{%\small
		\textcolor{black}{$24$}}};
  \draw (10,22) node{{%\small
		\textcolor{black}{$23$}}};
  \draw (10,29) node{{%\small
		\textcolor{black}{$14$}}};
  \draw (10,33) node{{%\small
		\textcolor{black}{$13$}}};
  \draw (10,37) node{{%\small
		\textcolor{black}{$12$}}};
  \draw (10,41) node{{%\small
		\textcolor{black}{$\emptyset$}}};

  \path \foreach \i in {1,2}{%
	(20,13+3*\i) coordinate[vertex style, fill=white] (one\i)};
  \draw (22,16) node{{%\small
		\textcolor{black}{$1$}}};
  \draw (22,19) node{{%\small
		\textcolor{black}{$1'$}}};

  \path \foreach \i in {1,2}{%
	(20,28+3*\i) coordinate[vertex style, fill=white] (two\i)};
  \draw (22,31) node{{%\small
		\textcolor{black}{$2$}}};
  \draw (22,34) node{{%\small
		\textcolor{black}{$2'$}}};

  \path \foreach \i in {1,2}{%
	(-0,13+3*\i) coordinate[vertex style, fill=white] (three\i)};
  \draw (-2,16) node{{%\small
		\textcolor{black}{$3$}}};
  \draw (-2,19) node{{%\small
		\textcolor{black}{$3'$}}};

  \path \foreach \i in {1,2}{%
	(-0,28+3*\i) coordinate[vertex style, fill=white] (four\i)};
  \draw (-2,31) node{{%\small
		\textcolor{black}{$4$}}};
  \draw (-2,34) node{{%\small
		\textcolor{black}{$4'$}}};

  \draw[thick] (m1)--(one1);
  \draw[thick] (m1)--(two1); 
  \draw[thick] (m1)--(three1);
  \draw[thick] (m1)--(four1); 

  \draw[thick] (m2)--(one2);
  \draw[thick] (m2)--(two2); 
  \draw[thick] (m2)--(three1);
  \draw[thick] (m2)--(four1); 

  \draw[thick] (m3)--(one2);
  \draw[thick] (m3)--(two1); 
  \draw[thick] (m3)--(three2); 
  \draw[thick] (m3)--(four1); 

  \draw[thick] (m4)--(one2);
  \draw[thick] (m4)--(two1); 
  \draw[thick] (m4)--(three1);
  \draw[thick] (m4)--(four2);
  
  \draw[thick] (m5)--(one1);
  \draw[thick] (m5)--(two2); 
  \draw[thick] (m5)--(three2);
  \draw[thick] (m5)--(four1); 

  \draw[thick] (m6)--(one1);
  \draw[thick] (m6)--(two2); 
  \draw[thick] (m6)--(three1);
  \draw[thick] (m6)--(four2); 

  \draw[thick] (m7)--(one1);
  \draw[thick] (m7)--(two1); 
  \draw[thick] (m7)--(three2);
  \draw[thick] (m7)--(four2);
  
  \draw[thick] (m8)--(one2);
  \draw[thick] (m8)--(two2); 
  \draw[thick] (m8)--(three2);
  \draw[thick] (m8)--(four2);  
  
\end{scope}}

% image of g

{\scriptsize
\begin{scope}[xshift=32cm]
  \path \foreach \i in {1,...,8}{%
   (10,7.5+4*\i) coordinate[vertex style, fill=white] (m\i)};

  \draw (10,10) node{{
		\textcolor{black}{$g(4')$}}};
  \draw (10,14.3) node{{
		\textcolor{black}{$g(3')$}}};
  \draw (10,18.2) node{{
		\textcolor{black}{$g(2')$}}};
  \draw (10,22) node{{
		\textcolor{black}{$g(1)$}}};
  \draw (10,29) node{{
		\textcolor{black}{$g(1')$}}};
  \draw (10,33) node{{
		\textcolor{black}{$g(2)$}}};
  \draw (10,37) node{{
		\textcolor{black}{$g(3)$}}};
  \draw (10,41) node{{%\small
		\textcolor{black}{$g(4)$}}};

  \path \foreach \i in {1,2}{%
	(20,13+3*\i) coordinate[vertex style, fill=white] (one\i)};
  \draw (22.2,16) node{{
		\textcolor{black}{$g(23)$}}};
  \draw (22.2,19) node{{
		\textcolor{black}{$g(14)$}}};

  \path \foreach \i in {1,2}{%
	(20,28+3*\i) coordinate[vertex style, fill=white] (two\i)};
  \draw (22.2,31) node{{
		\textcolor{black}{$g(13)$}}};
  \draw (22.2,34) node{{
		\textcolor{black}{$g(24)$}}};

  \path \foreach \i in {1,2}{%
	(-0,13+3*\i) coordinate[vertex style, fill=white] (three\i)};
  \draw (-2.4,16) node{{
		\textcolor{black}{$g(12)$}}};
  \draw (-2.4,19) node{{
		\textcolor{black}{$g(34)$}}};

  \path \foreach \i in {1,2}{%
	(-0,28+3*\i) coordinate[vertex style, fill=white] (four\i)};
  \draw (-2.2,31) node{{
		\textcolor{black}{$g(\emptyset)$}}};
  \draw (-3.2,34) node{{
		\textcolor{black}{$g(1234)$}}};

  \draw[thick] (m1)--(one1);
  \draw[thick] (m1)--(two1); 
  \draw[thick] (m1)--(three1);
  \draw[thick] (m1)--(four1); 

  \draw[thick] (m2)--(one2);
  \draw[thick] (m2)--(two2); 
  \draw[thick] (m2)--(three1);
  \draw[thick] (m2)--(four1); 

  \draw[thick] (m3)--(one2);
  \draw[thick] (m3)--(two1); 
  \draw[thick] (m3)--(three2); 
  \draw[thick] (m3)--(four1); 

  \draw[thick] (m4)--(one2);
  \draw[thick] (m4)--(two1); 
  \draw[thick] (m4)--(three1);
  \draw[thick] (m4)--(four2);
  
  \draw[thick] (m5)--(one1);
  \draw[thick] (m5)--(two2); 
  \draw[thick] (m5)--(three2);
  \draw[thick] (m5)--(four1); 

  \draw[thick] (m6)--(one1);
  \draw[thick] (m6)--(two2); 
  \draw[thick] (m6)--(three1);
  \draw[thick] (m6)--(four2); 

  \draw[thick] (m7)--(one1);
  \draw[thick] (m7)--(two1); 
  \draw[thick] (m7)--(three2);
  \draw[thick] (m7)--(four2);
  
  \draw[thick] (m8)--(one2);
  \draw[thick] (m8)--(two2); 
  \draw[thick] (m8)--(three2);
  \draw[thick] (m8)--(four2);  
  
\end{scope}}

r\end{tikzpicture}
\caption{The above $g\in \Aut(Y(4))$ interchanges the link vertices and middle vertices, 
thus is not link-preserving.}\label{fig:f4}
\end{figure}
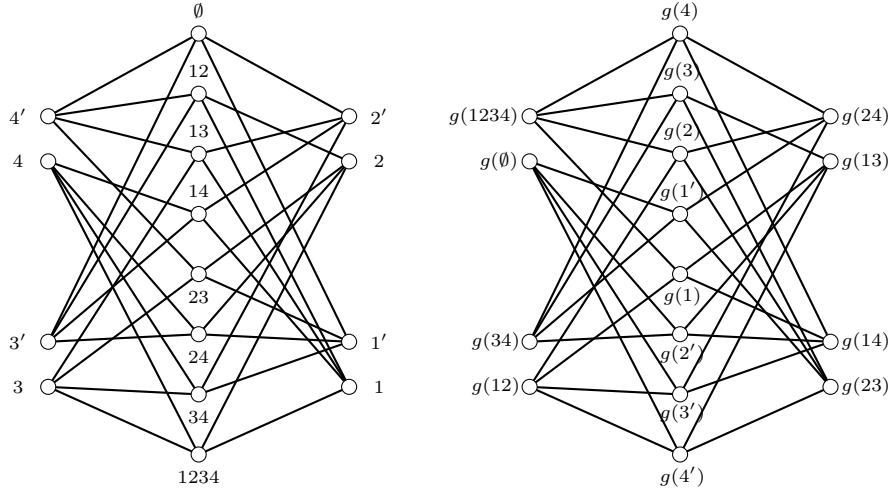
\end{exa}

\noindent
Interestingly, Figure~\ref{fig:autoY2}, Example~\ref{exa:gY1}, and 
Example~\ref{exa:gY4} are the only  cases for non-twin-preserving 
automorphisms of $Y(d)$.

\begin{prop}\label{pro:none124}
Assume $d\notin \{1,2,4\}$ and $g\in \Aut(Y(d))$. Then $g$ is 
twin-preserving. 
\end{prop}

Then combining Propositions~\ref{pro:yfpitog}, 
\ref{pro:Ydtwinpreservingdecompose}, and~\ref{pro:none124} we obtain 
immediately: 
\begin{theo}\label{thm:124}
For every $d\notin \{1,2,4\}$
\[
\Aut(Y(d))= \big\{f\circ \rho_\pi\bigmid 
 \text{$f\in \Aut(X(d))$ and $\pi\in \Per(A(d))$}\big\}.
\]
Thus by the uniqueness stated in Propositions~\ref{pro:yfpitog} and
\ref{pro:Ydtwinpreservingdecompose} we get 
$
\aut(Y(d))= 2^d\cdot \aut(X(d))
$ (recall that  $ \aut(\ldots )$  denotes the number of automorphisms of $ \ldots $). 
\end{theo}

\noindent
In the remaining part of the section, we prove Proposition~\ref{pro:none124}, 
for which we need a further notion concerning automorphisms of  $Y(d)$. 
\begin{defn}\label{def:linkpreserving}
An automorphism $g\in \Aut(Y(d))$ is \emph{link-preserving} if 
\[
\bar g(L(d))= L(d).
\]
Or 
equivalently, $\bar g(M(d))= M(d)$. 
\end{defn} 

\begin{rem}\label{rem:linkpreserving}
 Clearly, twin-preserving implies link-preserving. But the 
converse is not true as witnessed by the automorphism $g\in Y(2)$ in 
Figure~\ref{fig:autoY2}. %
In addition, observe that the automorphisms in Example~\ref{exa:gY1} and 
Example~\ref{exa:gY4} are both non-link-preserving.\hfill{$ \dashv$}
\end{rem}
\begin{lem}\label{lem:no2linktotwinpreserving}
Assume that $d\ne 2$  and $g\in \Aut(Y(d))$ is 
link-preserving. 
Then $g$ is twin-preserving.
\end{lem}

\proof The case of $d=1$ can be easily seen by the  discussion in 
Example~\ref{exa:gY1}, so we assume $d\ge 3$. Let $ x\in L(d)$. By~\eqref{eq:cmeg} and $g\in 
\Aut(Y(d))$, 
\[
\text{for all $m\in M(d)$}: \ \ \
 g(x)g(m)\in E(Y(d))\iff g(x') g(m)\notin  E(Y(d)).
\]
Since $ g$ is link-preserving we know that $ g(x)\in L(d)$ and  $\bar g(M(d))= M(d)$. Hence,
\[
\text{for all $m\in M(d)$}: \ \ \
 g(x)m\in E(Y(d))\iff g(x') m\notin  E(Y(d)).
\]
Thus, 
by~Lemma~\ref{lem:li'uni} we get $ g(x')\in L(d)$ and
$g(x)'= g(x')$. \proofend 
\medskip

\noindent The assumption ``$ g$ link-preserving'' of 
the lemma  is necessary as shown by the automorphism $g\in \Aut(Y(4))$ 
exhibited in Example~\ref{exa:gY4} (and also see Figure~\ref{fig:f4}).

\medskip
\begin{lem}\label{lem:dno14linkpreserving}
Assume $d\notin \{1,4\}$.  Then every $g\in 
\Aut(Y(d))$ is link-preserving.
\end{lem}

\proof Let $ d\ge 2$. Since the set $M(d)$ has cardinality $2^{d-1}$, the vertices in $L(d)$  have degree $2^{d-2}$ in $Y(d)$. The vertices in $M(d)$ have degree $d$.
As $2^{d-2}\ne d$ for $d\ne 
4$, the function $g$ must map link 
vertices to link vertices and middle vertices to middle vertices, i.e., $g$ 
is link-preserving. \proofend 

\medskip
\noindent Now Proposition~\ref{pro:none124} follows directly from 
Lemma~\ref{lem:dno14linkpreserving} and 
Lemma~\ref{lem:no2linktotwinpreserving}.

% The CFI-graphs 

\medskip
\begin{center}
\fbox{\fbox{\parbox{10cm} {\emph{In the remaining sections, unless mentioned 
otherwise, $G$ will always denote a connected graph with more than one 
vertex.}}}} 
\end{center}

\section{The uncolored CFI-graphs $Y(G)$ and $\tilde Y(G)$ 
and the colored CFI-graphs $X(G)$ and $\tilde X(G)$}\label{sec:ygxg+tilde}

\paragraph{The graphs $Y(G)$ and $X(G)$.} 
Let $u$ be a vertex of degree $d$ of the graph $G$. Then $Y(u)$ is the gadget 
$Y(d)$ determined by $A(d):= A(u)$ and $B(d):= B(u)$ where 
\begin{itemize}
\item $A(u):= \big\{a(u,v) \bigmid v\in V(G), \ uv\in E(G)\big\}$ \ \ and \ 
    \ $B(u):= \big\{b(u,v) \bigmid v\in V(G), \ uv\in E(G)\big\}$, 

\item $':A(u)\to B(u)$ is given by $a(u,v)\mapsto b(u,v)$. 
\end{itemize}
Thus, the set of link vertices in $Y(u)$ is $L(u):= A(u)\cup B(u)$. 
Similarly, the set of middle vertices $M(d)$ is denoted by $M(u)$ and in 
particular the middle vertex $\emptyset$ by $\emptyset_{u}$.

\medskip
\noindent The \emph{uncolored \CFI-graph $Y(G)$ of $G$} is defined by 
\begin{itemize}
\item $V(Y(G)):= \dot{\bigcup}_{u\in V(G)} V(Y(u))$, i.e., the disjoint 
    union of all vertices in $Y(u)$'s. To aid discussions, we let 
    \begin{eqnarray*}
    L(Y(G)):= \bigcup_{u\in V(G)} L(u)
     & \text{and} & 
    M(Y(G)):= \bigcup_{u\in V(G)} M(u),
    \end{eqnarray*}
    i.e., the sets of link vertices and middle vertices in $Y(G)$, 
    respectively. Thus, $V(Y(G))= L(Y(G))\cup M(Y(G))$.

\item $E(Y(G))$ contains the edges of $Y(u)$ for all $u\in V(G)$ and for 
    each $uv\in E(G)$ the edges 
    \begin{eqnarray}\label{eq:edyg}
    \{a(u,v),a(v,u)\} & \text{and} & \{b(u,v),b(v,u)\}.
    \end{eqnarray}
     Hence, these are the only edges between the gadgets $Y(u)$ and $Y(v)$ 
    if $uv\in E(G)$ (and this is the reason why in the gadget $Y(d)$ we 
    called \emph{link} vertices the vertices in $L(d)= A(d)\cup B(d)$). If 
    $uv\notin E(G)$, there are no edges between $Y(u)$ and $Y(v)$. 

  \end{itemize}

\noindent
As in the gadgets $ Y(d)$ (cf.~\eqref{eq:sibling}), now we can view $'$ as a 
mapping 
\begin{equation*}%
': L(Y(G))\to L(Y(G)),
\end{equation*}
which maps every link vertex $x\in L(Y(G))$ to its \emph{twin} $x'$. Note 
that $x$ and $x'$ are twins in a  gadget $Y(u)$ for some $u\in V(G)$.

As for $ u $ in $ G$ and $ m\in M(u)$, the degree of $ m$ in $ Y(G)$ equals the degree of $ u$ in $ G$, we have
\begin{equation}\label{eq:degcfi}\deg(Y(G))\ge \deg(G).
\end{equation}

\medskip
\noindent We turn to the definition of the colored \CFI-graph $X(G)$ of $G$.
By definition (see~\eqref{eq:ccg}) the colored gadget $X(d)$ is the pair 
$(Y(d),c)$ where $c$ is a coloring of $L(d)$ with the property 
\begin{eqnarray*}
\text{$c(a)=c(a')$ \ for $a\in A(d)$}
 & \text{and} & 
\text{$c(a_1)\ne c(a_2)$ \ for $a_1,a_2\in A(d)$ with $a_1\ne a_2$}.
\end{eqnarray*}
The 
\begin{equation}\label{eq:defxg}
\text{\emph{colored \CFI-graph $X(G)$ of~$G$} is the structure $(Y(G),c)$},
\end{equation}
where $c= (c_1, c_2)$. Here, $c_1$ is a coloring defined on $\bigcup_{u\in 
V(G)} L(u)$ such that for every $u\in V(G)$ the gadget $Y(u)$ together with 
the restriction of the coloring $c_1$ to $L(u)$ is a colored gadget $X(u)$. 
Furthermore, $c_2$ is a coloring of $V(Y(G))$ obtained from an 
\emph{injective} coloring\footnote{That is, every vertex in $V(G)$ receives a 
distinct color.} of $V(G)$ by assigning, for every vertex $u\in V(G)$, to all 
elements of $X(u)$ the color of $u$ in $G$. 

Hence, for every $f\in \Aut(X(G))$ and $u\in V(G)$ the restriction 
$f\upharpoonright_{V(X(u))}$ of $f$ to $V(X(u))$ is in $\Aut(X(u))$. In 
particular, for $a(u,v)\in V(X(G))$ (where $u\in V(G)$ and $uv\in E(G)$) we 
have (cf. \eqref{eq:autad})
\[
\{f(a(u,v)), f(b(u,v))\}= \{a(u,v), b(u,v)\}.
\]

\paragraph{The graphs $\tilde Y(G)$ and $\tilde X(G)$.} 
We fix a graph $G$. Let $t\ge 0$ and let 
\[
\bar e = v_1w_1, v_2w_2, \ldots, v_tw_t
\]
be a sequence of $t$ edges of $G$ (possibly with repetitions). We define the 
colored graph $X^{\bar e}(G)$ by induction on $t$. After each induction step 
for all edges $uv$ of $G$ we will have in the corresponding $X^{\bar e}(G)$ 
exactly two of the edges $\{a(u,v), a(v,u)\}$, $\{b(u,v), b(v,u)\}$, 
$\{a(u,v), b(v,u)\}$, and $\{b(u,v), a(v,u)\}$, namely either the edges 
\begin{equation}\label{eq:eun} 
\{a(u,v), a(v,u)\}
 \ \  \text{and} \ \ 
\{b(u,v), b(v,u)\}
\end{equation}
or the edges 
\begin{equation}\label{eq:etw} 
\{a(u,v), b(v,u)\}
 \ \ \text{and} \ \ 
\{b(u,v), a(v,u)\}.
\end{equation}
Let $t= 0$, i.e., $\bar e= \varepsilon$, the empty sequence. Then 
$X^{\varepsilon}(G):= X(G)$. By the definition of $X(G)$ we 
know that for all edges $uv$ of $G$ the edges in~\eqref{eq:eun} are present 
in $X(G)$ and none of the edges in~\eqref{eq:etw}. 
\medskip

\noindent
For the induction step assume that $X^{\bar e}(G)$ has already been defined 
for $\bar e$ with $|\bar e|= t$ (by $|\bar e|$ we denote the length of $\bar 
e$). Let $uv\in E(G)$. Then $X^{\bar e, \, uv}(G)$ is obtained by 
\emph{twisting the edge $uv$ in $X^{\bar e}(G)$}. Here, twisting the edge 
$uv$ in $X^{\bar e}(G)$ means the following. If the edges in~\eqref{eq:eun} 
are present in $X^{\bar e}(G)$, we obtain $X^{\bar e,\, uv}(G)$ by deleting 
them and adding the edges in~\eqref{eq:etw}. Otherwise, if the edges 
in~\eqref{eq:etw} are present in $X^{\bar e}(G)$, then we obtain $X^{\bar 
e,\, uv}(G)$ by deleting them and adding the edges in~\eqref{eq:eun}. 
\medskip

\noindent
The graphs $Y^{\bar e}(G)$ are obtained in the same way by
replacing always $X(G)$ by $Y(G)$. Or equivalently, we obtain $Y^{\bar e}(G)$ 
by deleting the colors in $X^{\bar e}(G)$. 
\medskip

\noindent
Observe that for $uv\in E(G)$, the edges in~\eqref{eq:etw} are present in 
$X^{\bar e}(G)$ if $uv$ appears an odd number of times in $\bar e$, i.e., if 
we twist $uv$ an odd number of times. Similarly, the edges in~\eqref{eq:eun} 
are present in $X^{\bar e}(G)$ if $uv$ appears an even number of times in 
$\bar e$. In other words, for $\bar e= e_1,\ldots, e_t$ the graph 
\begin{equation}\label{eq:multiset}
\text{$X^{\bar e}(G)$ is determined by the function 
 $e\ (\in E(G)) \mapsto \big|\{e=e_i \mid i\in [t]\}\big| \mod 2$}.
\end{equation}
In particular we see that 
\begin{equation}\label{eq:zwgl}
X^{\bar e}(G)=X^{\bar e, \, uv, \, uv}(G).
\end{equation}
This observation together with the next lemma will allow us to derive a much 
stronger statement. 
\begin{lem}\label{lem:twist2adjacentedges}
Let $e,e'\in E(G)$ be \emph{adjacent} edges, i.e., they share exactly one 
vertex $u\in V(G)$. Then for any sequence of edges $\bar e$ in $G$ there is 
an isomorphism 
\[
f: X^{\bar e}(G)\cong X^{\bar e,\, e,\, e'}(G).
\]
Moreover, for every $v'\in V(G)\setminus \{u\}$ the restriction of $f$ to 
$X(v')$ is the identity mapping. 
\end{lem}

\proof Let $e= uv$ and $e'= uw$ with $v\ne w$. As $V(X^{\bar e}(G))= 
V(X^{\bar e,\, e,\, e'}(G))= V(X(G))$ and this set is the disjoint union of all 
the sets $V(X(v'))$ where $v'$ ranges over all vertices of $G$, it suffices 
to define the restriction of $f$ to each $V(X(v'))$. As already mentioned the 
restriction of $f$ to $X(v')$ with $v'\ne u$ is the identity mapping. The 
restriction of $f$ to $V(X(u))$ is the function $f_{\{a(u,v), a(u,w)\}}$ (cf.~Proposition~\ref{pro:autcolgad}). 
More explicitly, $f$ is given by 
\[
f(x):=
\begin{cases}
 x & \text{if $x\notin X(u)$} \\
 b(u,v) & \text{if $x= a(u,v)$} \\
 a(u,v) & \text{if $x= b(u,v)$} \\
 b(u,w) & \text{if $x= a(u,w)$} \\
 a(u,w) & \text{if $x= b(u,w)$} \\
m\,\Delta\,\{ a(u,v), a(u,w)\} & \text{if $x= m\in M(u)$}.
\end{cases}
\]
As the restriction of $f$ to every gadget is an isomorphism, it suffices to 
prove that $f$ preserves edges between gadgets and again here we only have to 
consider edges between the relevant gadgets, i.e., edges between $X(u)$ and 
$X(v)$ and edges between $X(u)$ and $X(w)$. For example, if $a(u,w)b(w,u)$ is 
present in $X^{\bar e}(G)$, then as $e\ne uw$ the edge 
$a(u,w)b(w,u)$ is present in 
$X^{\bar e, \, e}(G)$, too. However, by its definition the graph $X^{\bar e,\, 
e,\, e'}(G)= X^{\bar e,\, e, \, uw}(G)$ contains the edge $b(u,w)b(w,u)$, i.e., 
$f(a(u,w))f(b(w,u))$, since $f$ is the identity on $X(w)$. \proofend

\medskip
\begin{lem}\label{lem:twist2nonadjacentedges}
Let $e,e'\in E(G)$. Then for any sequence of edges $\bar e$ in $G$, 
\[
X^{\bar e}(G) \cong X^{\bar e,\, e,\, e'}(G).
\]
\end{lem}

\proof By~\eqref{eq:zwgl} and the previous lemma, we can assume that $e$ and
$e'$ have no vertex in common. Recall that we always assume that $G$ is a 
connected graph. Hence, there are for some $k\ge 4$ pairwise distinct 
vertices $v_1, \ldots, v_k$ such that 
\[
\underbrace{v_1v_2}_e,\, v_2v_3,\ \ldots ,\, v_{k-2}v_{k-1},
 \, \underbrace{v_{k-1} v_k}_{e'} \in E(G).
\]
We obtain, by repeatedly invoking Lemma~\ref{lem:twist2adjacentedges} 
and~\eqref{eq:multiset}, 
\begin{align*}
X^{\bar e}(G) 
& \cong X^{\bar e, \, v_1v_2,\, v_2v_3} (G) \\
 & \cong X^{\bar e, \, v_1v_2,\, v_2v_3,\, v_2v_3,\, v_3v_4}(G) \\
 & \qquad \vdots \\
 & \cong X^{\bar e, \, v_1v_2,\, v_2v_3,\, v_2v_3,\, v_3v_4, 
   \, \cdots,\, v_{k-3} v_{k-2},\, v_{k-2} v_{k-1},\, v_{k-2} v_{k-1},\, v_{k-1} v_{k}}(G) \\
 & = X^{\bar e,\, v_1v_2, \, v_{k-1} v_k}(G) \\
 & = X^{\bar e,\, e,\, e'}(G) \qquad (\text{as $e = v_1v_2$ and $e'= v_{k-1} v_k$} )
\end{align*}
This finishes the proof. \proofend

\begin{cor}\label{cor:welldefinedtildeXG}
For $e,e'\in E(G)$, \ $X^{e}(G)\cong X^{e'}(G)$. 
\end{cor}

\proof By Lemma~\ref{lem:twist2nonadjacentedges} and~\eqref{eq:zwgl} we get 
$X^{e}(G)\cong X^{e,\, e,\, e'}(G)= X^{e'}(G)$. \proofend

\begin{defn}\label{def:tilde} 
For a graph $G$ choose an \emph{arbitrary} edge $e\in E(G)$ and set 
\[
\tilde X(G):= X^e(G).
\]
By the previous corollary the graph $\tilde X(G)$ is well-defined (up to 
isomorphism). 
\end{defn}

\begin{lem}\label{lem:parity}
Let $\bar e$ be a sequence of edges in $G$. Then 
\begin{eqnarray*}
 X^{\bar e}(G) \cong X(G) \ \ \text{if\/ $|\bar e|$ is even}
 \ \ & \text{and} & \ \ 
 X^{\bar e}(G) \cong \tilde X(G) \ \ \text{if\/ $|\bar e|$ is odd}.
\end{eqnarray*}
\end{lem}

\proof A simple induction on $|\bar e|$ using
Lemma~\ref{lem:twist2nonadjacentedges}. \proofend 

\noindent %

\noindent Of course, isomorphisms between $X^{\bar e}(G)$ and $X^{\bar 
e'}(G)$ are still isomorphisms between $Y^{\bar e}(G)$ and $Y^{\bar e'}(G)$. 
Therefore,  Lemma~\ref{lem:twist2nonadjacentedges} and Corollary~\ref{cor:welldefinedtildeXG}
hold for the corresponding uncolored graphs. In particular, we can define 
$\tilde Y(G)$ according to Definition~\ref{def:tilde}: 
\begin{defn}\label{def:ytilde} 
For a graph $G$ choose an \emph{arbitrary} edge $e\in E(G)$ and set 
\[
\tilde Y(G):= Y^e(G).
\]
By the version for $Y(G)$ of Corollary~\ref{cor:welldefinedtildeXG} 
 the graph $\tilde Y(G)$ is well-defined 
(up to isomorphism). 
\end{defn}
\noindent However, in general an isomorphism between some $Y^{\bar e}(G)$ and 
$Y^{\bar e'}(G)$ is not  necessarily an isomorphism between 
$X^{\bar e}(G)$ and $X^{\bar e'}(G)$, as possibly it does not 
preserve the colors of the vertices. In particular, the following proof does 
not guarantee that $Y(G)$ and $\tilde Y(G)$ are not isomorphic. Nevertheless, 
we will see later that $Y(G)\not\cong \tilde Y(G)$. 

\medskip
\begin{theo}\label{thm:XGiso}
Let $\bar e$ be a sequence of edges in $G$. Then 
\begin{description}
\item[(i)] $X^{\bar e}(G)$ is isomorphic to $X(G)$ if and only if\/ $|\bar 
    e|$ is even. 

\item[(ii)] $X^{\bar e}(G)$ is isomorphic to $\tilde X(G)$ if and only if\/ 
    $|\bar e|$ is odd. 
\end{description}
In particular, $X(G)\not\cong \tilde X(G)$. 
\end{theo}

\proof Lemma~\ref{lem:parity} already establishes the direction from right to 
left in both, (i) and (ii). For the converse directions, we show that $X(G)$ 
and $\tilde X(G)$ are not isomorphic by a double counting argument. Towards a 
contradiction, assume that $h$ is an isomorphism from $X(G)$ to $\tilde 
X(G)$. By the coloring of $X(G)$ and $\tilde X(G)$, for $u\in V(G)$ we know 
that the restriction $h\upharpoonright_{V(X(u))}$  of $h$ to $V(X(u))$ is an 
automorphism of $X(u)$ (more precisely, an isomorphism between the copy of 
$X(u)$ in $X(G)$ and the copy of $X(u)$ in $\tilde X(G)$). In particular, the 
set~(cf.~\eqref{eq:deffm}) 
\[
m_{h\upharpoonright_{V(X(u))}}
 \ \big(= \{a(u,v) \mid v\in V(G), \ uv \in E(G), \ h(a(u,v))= b(u,v)\}\big)
\]
has even cardinality and thus also $F_h$ has even cardinality, where 
\[
F_h:= \bigcup_{u\in V(G)}m_{h\upharpoonright_{V(X(u))}}.
\]
Now we count the number of elements of $F_h$ in a different way, i.e., 
\[
F_h= \bigcup_{uv\in E(G)}\big(\{a(u,v) \mid h(a(u,v))= b(u,v)\}\cup
 \{a(v,u) \mid h(a(v,u))= b(v,u)\}\big),
\]
thereby showing that $|F_h|$ must be odd, the desired contradiction.

Note that we have twisted one edge, say $e_0= \{u_0,v_0\}$, to obtain $\tilde 
X(G)$ from $X(G)$; that is, $\tilde X(G)= X^{e_0}(G)$. Consider first an edge 
$e=\{u,v\}$ distinct from $e_0$. Then 
\[
\text{$a(u,v)a(v,u)$ \ and \ $b(u,v)b(v,u)$ 
 are in $E(X(G))$ and in $E(\tilde X(G))$}.
\]
As $\{h(a(u,v)), h(a(v,u))\}\in E(\tilde X(G))$, we see that 
\[
\text{either \ $\{h(a(u,v)), h(a(v,u))\}= \{a(u,v), a(v,u)\}$
 \ or \ $\{h(a(u,v)), h(a(v,u))\}=\{b(u,v), b(v,u)\}$}. 
\]
That is, the edge $e$ contributes to $F_h$ zero vertices or two vertices
(namely, $a(u,v)$ and $a(v,u)$). Now we consider the edge $e_0:= \{u_0,v_0 
\}$. By definition of $X^{e_0}(G) \ (=\tilde X(G))$ we know that 
\begin{align*}
\{a(u_0,v_0),a(v_0,u_0)\}, \ \{b(u_0,v_0),&b(v_0,u_0)\} \in E(X(G))\\
 \text{but } & \{a(u_0,v_0),b(v_0,u_0)\}, \ \{b(u_0,v_0),a(v_0,u_0)\}\in E(\tilde X(G)).
\end{align*}
As again $\{h(a(u_0,v_0)), h(a(v_0,u_0))\}\in E(\tilde X(G))$, we see that 
\begin{align*}
\text{either \ $h(a(u_0,v_0))= a(u_0,v_0)$}
   & \text{ and $h(a(v_0,u_0))=b(v_0,u_0)$} \\
  \text{ or } & h(a(u_0,v_0))= b(u_0,v_0) \text{ and } h(a(v_0,u_0))= a(v_0,u_0).
\end{align*}
So, in both cases the edge $e_0$ contributes to $F_h$ one vertex. Hence, 
$|F_h|$ is odd. \proofend 

\medskip
\noindent Among others in the next section we will see that $Y(G)\not\cong 
\tilde Y(G)$. Here is a first step, the result for graphs of degree two. Note 
that (connected) graphs $G$ with $\deg(G)\in[2]$ are paths $P_k$ with $k$ 
edges, where $k\ge 1$, or cycles~$C_k$ with $k\ge 3$ edges. For graphs $H_1$ 
and $H_2$ we denote by $H_1\dotcup H_2$ the disjoint union of $H_1$ and~$ 
H_2$.

\begin{lem}\label{lem:1-2}
\begin{itemize}
\item[(i)] For $k\ge 1$, $Y(P_k)\cong P_{3(k-1)+1}\dotcup P_{3(k-1)+3}$ and 
    $\tilde Y(P_k)\cong P_{3(k-1)+2}\dotcup P_{3(k-1)+2}$ 
    (cf.~Figure~\ref{fig:gYP3}).\vspace{-0.2cm} 

\item[(ii)] For $k\ge 3$, $Y(C_k)\cong C_{3k}\dotcup C_{3k}$ and $\tilde 
    Y(C_k)\cong C_{6k}$ (cf.~Figure~\ref{fig:gYC3}). 
\end{itemize}%

\smallskip \noindent Therefore, if $\deg(G)\in[2]$, then $Y(G)\not\cong 
\tilde Y(G)$ and $\deg(Y(G)), \deg(\tilde Y(G))\in [2]$. Moreover, if $ G$ 
and $ G'$ are graphs of degree at most 2 and $Y(G)\cong Y(G')$, then $ 
G\cong G'$.
\end{lem}

\begin{figure}[h]
  \centering
\begin{tikzpicture}[scale=0.25,vertex style/.style={draw,
                                   circle,
                                   minimum size=2mm,
                                   inner sep=0pt,
                                   outer sep=0pt,
                                   %shade
                                   }]

{\scriptsize                                  
\begin{scope}[xshift=-10cm,yshift=0cm]
% Y(P3)
% draw vertices

\def\cloudwidth{10}
\def\horwidth{3}

\def \j{1}
\pgfmathsetmacro\jaone{int(\j+1)}
\draw[fill=gray, opacity=.2] (\cloudwidth*\j-15+\horwidth-1,5.5) rectangle (\cloudwidth*\j-15+2*\horwidth+1,12.5);
\draw (\cloudwidth*\j-15+\horwidth*1.5,13.5) node {{\textcolor{black}{$Y(\j)$}}};

\draw (\cloudwidth*\j-15+\horwidth,9) coordinate[vertex style] (middleup\j);
\draw (\cloudwidth*\j-15+\horwidth,10.5) node {{\textcolor{black}{$\emptyset_{\j}$}}};

\draw (\cloudwidth*\j-15+2*\horwidth,6.5) node {{\textcolor{black}{$\jaone$}}};
\draw (\cloudwidth*\j-15+2*\horwidth,8) coordinate[vertex style] (rightdown\j);
\draw (\cloudwidth*\j-15+2*\horwidth,10) coordinate[vertex style] (rightup\j);
\draw (\cloudwidth*\j-15+2*\horwidth,11.5) node {{\textcolor{black}{$\jaone'$}}};

\foreach \j in {2,3} {
    \draw[fill=gray, opacity=.2] (\cloudwidth*\j-15-1,5.5) rectangle (\cloudwidth*\j-15+2*\horwidth+1,12.5);
    \draw (\cloudwidth*\j-15+\horwidth,13.5) node {{\textcolor{black}{$Y(\j)$}}};
    
    \pgfmathsetmacro\jmone{int(\j-1)}
    \pgfmathsetmacro\jaone{int(\j+1)}
    \draw (\cloudwidth*\j-15,6.5) node {{\textcolor{black}{$\jmone$}}};
    \draw (\cloudwidth*\j-15,8) coordinate[vertex style] (leftdown\j);
    \draw (\cloudwidth*\j-15,10) coordinate[vertex style] (leftup\j);
    \draw (\cloudwidth*\j-15,11.5) node {{\textcolor{black}{$\jmone'$}}};

    \draw (\cloudwidth*\j-15+\horwidth,6.5) node {{\textcolor{black}{$\jmone\jaone$}}};
    \draw (\cloudwidth*\j-15+\horwidth,8) coordinate[vertex style] (middledown\j);
    \draw (\cloudwidth*\j-15+\horwidth,10) coordinate[vertex style] (middleup\j);
    \draw (\cloudwidth*\j-15+\horwidth,11.5) node {{\textcolor{black}{$\emptyset_{\j}$}}};
    
    \draw (\cloudwidth*\j-15+2*\horwidth,6.5) node {{\textcolor{black}{$\jaone$}}};
    \draw (\cloudwidth*\j-15+2*\horwidth,8) coordinate[vertex style] (rightdown\j);
    \draw (\cloudwidth*\j-15+2*\horwidth,10) coordinate[vertex style] (rightup\j);
    \draw (\cloudwidth*\j-15+2*\horwidth,11.5) node {{\textcolor{black}{$\jaone'$}}};
}

\def \j{4}
\pgfmathsetmacro\jmone{int(\j-1)}
\draw[fill=gray, opacity=.2] (\cloudwidth*\j-15-1,5.5) rectangle (\cloudwidth*\j-15+\horwidth+1,12.5);
\draw (\cloudwidth*\j-15+\horwidth*0.5,13.5) node {{\textcolor{black}{$Y(\j)$}}};

\draw (\cloudwidth*\j-15,6.5) node {{\textcolor{black}{$\jmone$}}};
\draw (\cloudwidth*\j-15,8) coordinate[vertex style] (leftdown\j);
\draw (\cloudwidth*\j-15,10) coordinate[vertex style] (leftup\j);
\draw (\cloudwidth*\j-15,11.5) node {{\textcolor{black}{$\jmone'$}}};

\draw (\cloudwidth*\j-15+3,9) coordinate[vertex style] (middleup\j);
\draw (\cloudwidth*\j-15+3,10.5) node {{\textcolor{black}{$\emptyset_{\j}$}}};

% draw lines

% lines in a cloud
\foreach \j in {1,...,4} {
    \ifthenelse{\not\equal{\j}{1}}{
        \draw (leftup\j)--(middleup\j)
    }{};
    \ifthenelse{\not\equal{\j}{4}}{
        \draw (middleup\j)--(rightup\j);
    }{};
    \ifthenelse{\not\equal{\j}{1} \and \not\equal{\j}{4}}{
        \draw (leftdown\j)--(middledown\j);
        \draw (middledown\j)--(rightdown\j);
    }{};
}

% lines between clouds
\foreach \j in {1,...,3} {
    \pgfmathsetmacro\jaone{int(\j+1)}
    \draw (rightup\j)--(leftup\jaone);
    \draw (rightdown\j)--(leftdown\jaone);
}
\end{scope}}

{\scriptsize
\draw (-16,5) node {{\textcolor{black}{$Y(P_3)$}}};
\draw[dashed] (-18,4)--(20,4);
\draw (-16,3) node {{\textcolor{black}{$\tilde Y(P_3)$}}};}

{\scriptsize                                   
\begin{scope}[xshift=-10cm,yshift=-10cm]
%\tilde Y(P3)
% draw vertices

\def\cloudwidth{10}
\def\horwidth{3}

\def \j{1}
\pgfmathsetmacro\jaone{int(\j+1)}
\draw[fill=gray, opacity=.2] (\cloudwidth*\j-15+\horwidth-1,5.5) rectangle (\cloudwidth*\j-15+2*\horwidth+1,12.5);
\draw (\cloudwidth*\j-15+\horwidth*1.5,4.5) node {{\textcolor{black}{$Y(\j)$}}};

\draw (\cloudwidth*\j-15+\horwidth,9) coordinate[vertex style] (middleup\j);
\draw (\cloudwidth*\j-15+\horwidth,10.5) node {{\textcolor{black}{$\emptyset_{\j}$}}};

\draw (\cloudwidth*\j-15+2*\horwidth,6.5) node {{\textcolor{black}{$\jaone$}}};
\draw (\cloudwidth*\j-15+2*\horwidth,8) coordinate[vertex style] (rightdown\j);
\draw (\cloudwidth*\j-15+2*\horwidth,10) coordinate[vertex style] (rightup\j);
\draw (\cloudwidth*\j-15+2*\horwidth,11.5) node {{\textcolor{black}{$\jaone'$}}};

\foreach \j in {2,3} {
    \draw[fill=gray, opacity=.2] (\cloudwidth*\j-15-1,5.5) rectangle (\cloudwidth*\j-15+2*\horwidth+1,12.5);
    \draw (\cloudwidth*\j-15+\horwidth,4.5) node {{\textcolor{black}{$Y(\j)$}}};
    
    \pgfmathsetmacro\jmone{int(\j-1)}
    \pgfmathsetmacro\jaone{int(\j+1)}
    \draw (\cloudwidth*\j-15,6.5) node {{\textcolor{black}{$\jmone$}}};
    \draw (\cloudwidth*\j-15,8) coordinate[vertex style] (leftdown\j);
    \draw (\cloudwidth*\j-15,10) coordinate[vertex style] (leftup\j);
    \draw (\cloudwidth*\j-15,11.5) node {{\textcolor{black}{$\jmone'$}}};

    \draw (\cloudwidth*\j-15+\horwidth,6.5) node {{\textcolor{black}{$\jmone\jaone$}}};
    \draw (\cloudwidth*\j-15+\horwidth,8) coordinate[vertex style] (middledown\j);
    \draw (\cloudwidth*\j-15+\horwidth,10) coordinate[vertex style] (middleup\j);
    \draw (\cloudwidth*\j-15+\horwidth,11.5) node {{\textcolor{black}{$\emptyset_{\j}$}}};
    
    \draw (\cloudwidth*\j-15+2*\horwidth,6.5) node {{\textcolor{black}{$\jaone$}}};
    \draw (\cloudwidth*\j-15+2*\horwidth,8) coordinate[vertex style] (rightdown\j);
    \draw (\cloudwidth*\j-15+2*\horwidth,10) coordinate[vertex style] (rightup\j);
    \draw (\cloudwidth*\j-15+2*\horwidth,11.5) node {{\textcolor{black}{$\jaone'$}}};
}

\def \j{4}
\pgfmathsetmacro\jmone{int(\j-1)}
\draw[fill=gray, opacity=.2] (\cloudwidth*\j-15-1,5.5) rectangle (\cloudwidth*\j-15+\horwidth+1,12.5);
\draw (\cloudwidth*\j-15+\horwidth*0.5,4.5) node {{\textcolor{black}{$Y(\j)$}}};

\draw (\cloudwidth*\j-15,6.5) node {{\textcolor{black}{$\jmone$}}};
\draw (\cloudwidth*\j-15,8) coordinate[vertex style] (leftdown\j);
\draw (\cloudwidth*\j-15,10) coordinate[vertex style] (leftup\j);
\draw (\cloudwidth*\j-15,11.5) node {{\textcolor{black}{$\jmone'$}}};

\draw (\cloudwidth*\j-15+3,9) coordinate[vertex style] (middleup\j);
\draw (\cloudwidth*\j-15+3,10.5) node {{\textcolor{black}{$\emptyset_{\j}$}}};

% draw lines

% lines in a cloud
\foreach \j in {1,...,4} {
    \ifthenelse{\not\equal{\j}{1}}{
        \draw (leftup\j)--(middleup\j)
    }{};
    \ifthenelse{\not\equal{\j}{4}}{
        \draw (middleup\j)--(rightup\j);
    }{};
    \ifthenelse{\not\equal{\j}{1} \and \not\equal{\j}{4}}{
        \draw (leftdown\j)--(middledown\j);
        \draw (middledown\j)--(rightdown\j);
    }{};
}

% lines between clouds

% twisted edge
\draw[color=red,line width=0.4mm] (rightup1)--(leftdown2);
\draw[color=red,line width=0.4mm] (rightdown1)--(leftup2);

\foreach \j in {2,...,3} {
    \pgfmathsetmacro\jaone{int(\j+1)}
    \draw (rightup\j)--(leftup\jaone);
    \draw (rightdown\j)--(leftdown\jaone);
}
\end{scope}}

\end{tikzpicture}
\caption{The uncolored CFI-graphs $Y(P_3)$ and $\tilde Y(P_3)$ with $V(P_3)
= \{1,2,3,4\}$ and $E(P_3)=\{\{1,2\},\{2,3\},\{3,4\}\}$. For illustration purpose, $a(u,v)$ 
in $Y(u)$ are abbreviated by $v$ and $b(u,v)$ in $Y(u)$ are abbreviated 
by $v'$. The twisting of $\{1,2\} $ leads to the edges marked red  in $\tilde Y(P_3)$.}\label{fig:gYP3} 
\end{figure}
\begin{figure}[h]
\centering
\begin{tikzpicture}[scale=0.18,vertex style/.style={draw,
                                   circle,
                                   minimum size=2mm,
                                   inner sep=0pt,
                                   outer sep=0pt,
                                   %shade
                                   }]

{\scriptsize                                  
\begin{scope}[xshift=0cm,yshift=0cm]
% Y(C3)

% draw vertices
\foreach \i in {0,1,2} {
\def\cenx{0}
\def\ceny{10}
\pgfmathsetmacro\iplus{int(\i+1)}
\pgfmathsetmacro\prev{int(mod(int(\i-1+3),3)+1)}
\pgfmathsetmacro\succ{int(mod(int(\i+1+3),3)+1)}
  \begin{scope}[rotate around={120*\i:(0,0)}]
    \draw[fill=gray, opacity=.2] (\cenx,\ceny) circle (6.8);
    \draw (\cenx,\ceny+8.8) node {{\textcolor{black}{$Y(\iplus)$}}};
    
    \draw (\cenx-2.8,\ceny+3.8) node {{\textcolor{black}{$\emptyset_{\iplus}$}}};
    \draw (\cenx-1.5,\ceny+2.5) coordinate[vertex style] (emp\i);
    \draw (\cenx+1.5,\ceny+2.5) coordinate[vertex style] (full\i);
    \draw (\cenx+2.8,\ceny+3.8) node {{\textcolor{black}{$\prev \succ$}}};

    \draw (\cenx-3.5,\ceny-1.8) coordinate[vertex style] (succ\i);
    \draw (\cenx-1.8,\ceny-3.5) coordinate[vertex style] (succ\i');
    \draw (\cenx+3.5,\ceny-1.8) coordinate[vertex style] (prev\i);
    \draw (\cenx+1.8,\ceny-3.5) coordinate[vertex style] (prev\i');
    
    \draw (\cenx-4.8,\ceny-1.2) node {{\textcolor{black}{$\succ$}}};
    \draw (\cenx-1.3,\ceny-5.2) node {{\textcolor{black}{$\succ'$}}};
    \draw (\cenx+4.8,\ceny-1.2) node {{\textcolor{black}{$\prev$}}};
    \draw (\cenx+1.3,\ceny-5.2) node {{\textcolor{black}{$\prev'$}}};

    \draw (prev\i')--(emp\i)--(succ\i');
    \draw (prev\i)--(full\i)--(succ\i);
  \end{scope}
}

%draw edges
\foreach \i in {0,1,2} {
    \pgfmathsetmacro\prev{int(mod(int(\i-1+3),3))}
    \pgfmathsetmacro\succ{int(mod(int(\i+1+3),3))}
    \draw (succ\i)--(prev\succ);
    \draw (succ\i')--(prev\succ');
}

\end{scope}}

{\scriptsize
\draw (16,18) node {{\textcolor{black}{$Y(C_3)$}}};
\draw[dashed] (19,-15)--(19,20);
\draw (22,18) node {{\textcolor{black}{$\tilde Y(C_3)$}}};}

{\scriptsize                                  
\begin{scope}[xshift=38cm,yshift=0cm]
% Y(C3)

% draw vertices
\foreach \i in {0,1,2} {
\def\cenx{0}
\def\ceny{10}
\pgfmathsetmacro\iplus{int(\i+1)}
\pgfmathsetmacro\prev{int(mod(int(\i-1+3),3)+1)}
\pgfmathsetmacro\succ{int(mod(int(\i+1+3),3)+1)}
  \begin{scope}[rotate around={120*\i:(0,0)}]
    \draw[fill=gray, opacity=.2] (\cenx,\ceny) circle (6.8);
    \draw (\cenx,\ceny+8.8) node {{\textcolor{black}{$Y(\iplus)$}}};
    
    \draw (\cenx-2.8,\ceny+3.8) node {{\textcolor{black}{$\emptyset_{\iplus}$}}};
    \draw (\cenx-1.5,\ceny+2.5) coordinate[vertex style] (emp\i);
    \draw (\cenx+1.5,\ceny+2.5) coordinate[vertex style] (full\i);
    \draw (\cenx+2.8,\ceny+3.8) node {{\textcolor{black}{$\prev\succ$}}};

    \draw (\cenx-3.5,\ceny-1.8) coordinate[vertex style] (succ\i);
    \draw (\cenx-1.8,\ceny-3.5) coordinate[vertex style] (succ\i');
    \draw (\cenx+3.5,\ceny-1.8) coordinate[vertex style] (prev\i);
    \draw (\cenx+1.8,\ceny-3.5) coordinate[vertex style] (prev\i');
    
    \draw (\cenx-4.8,\ceny-1.2) node {{\textcolor{black}{$\succ$}}};
    \draw (\cenx-1.3,\ceny-5.2) node {{\textcolor{black}{$\succ'$}}};
    \draw (\cenx+4.8,\ceny-1.2) node {{\textcolor{black}{$\prev$}}};
    \draw (\cenx+1.3,\ceny-5.2) node {{\textcolor{black}{$\prev'$}}};

    \draw (prev\i')--(emp\i)--(succ\i');
    \draw (prev\i)--(full\i)--(succ\i);
  \end{scope}
}

%draw edges
\foreach \i in {0,1,2} {
    \pgfmathsetmacro\prev{int(mod(int(\i-1+3),3))}
    \pgfmathsetmacro\succ{int(mod(int(\i+1+3),3))}
    %twisted edge
    \ifthenelse{\not\equal{\i}{0}} {
        \draw (succ\i)--(prev\succ);
        \draw (succ\i')--(prev\succ');
    } {
        \draw[color=red,line width=0.4mm] (succ\i)--(prev\succ');
        \draw[color=red,line width=0.4mm] (succ\i')--(prev\succ);
    }
}
\end{scope}}

\end{tikzpicture}
\caption{The uncolored CFI-graphs $Y(C_3)$ and $\tilde Y(C_3)$ with $V(C_3)
= \{1,2,3\}$ and $E(C_3)= \big\{\{1,2\}, \{2,3\}, \{3,1\}\big\}$. 
For illustration purpose, $a(u,v)$ 
in $Y(u)$ are abbreviated by $v$ and $b(u,v)$ in $Y(u)$ are abbreviated by $v'$. 
The twist of $\{1,2\} $ leads to the edges marked red  in $\tilde Y(C_3)$.}\label{fig:gYC3}
\end{figure}

\noindent We leave the simple proof to the reader. The lemma shows that the 
graphs $Y(P_k)$, $\tilde Y(P_k)$, and $\tilde Y(C_k)$ are not connected. This 
does not happen for (connected) graphs of degree at 
least three.

\begin{prop}\label{pro:3conn}
Assume $\deg(G)\ge 3$. Then both, $Y(G)$ and $\tilde Y(G)$, are connected 
graphs. 
\end{prop}

\proof We look first at $Y(G)$. As $G$ is connected, there must be a 
connected component $C(a)$ of $Y(G)$ containing all $a(u,v)$ for $uv\in E(G)$ 
and the vertices in $M(u)$ distinct from the empty set. By the same reason 
there must be a connected component $C(b)$ of $Y(G)$ containing all $b(u,v)$ 
for $uv\in E(G)$ and for~$u$ in $G$ the vertices in $M(u)$ of cardinality 
less than the degree of $u$. Note that $C(a)\cup C(b)=V(Y(G))$. Let $u\in 
V(G)$ be of degree at least 3, Then there is an $m\in M(u)$ with $0< |m|< 
\deg(u).$ Hence, $m\in C(a)\cap C(b)$ and therefore, $C(a)= C(b)=V(Y(G))$, 
i.e., $Y(G)$ is connected. 

Now we consider $\tilde Y(G)$. Let $u\in V(G)$ have at least three neighbors, 
say, $v_1,v_2, v_3$. By Definition~\ref{def:ytilde} we can assume that $\tilde 
Y(G)$ is obtained from $Y(G)$ by twisting the edge $ v_1u$ of $ G$.
 By the argument leading to the result for $Y(G)$ it suffices to 
show that $a(v_1,u)$ and $a(u,v_1)$ lie in the same connected component of 
$\tilde Y(G)$ and that the same applies to $b(v_1,u)$ and $b(u,v_1)$. We show 
it for $a(v_1,u)$ and $a(u,v_1)$ by presenting the following path from 
$a(v_1,u)$ to $a(u,v_1)$ in $\tilde Y(G)$: 
\[
a(v_1,u),\ b(u,v_1),\ \{a(u,v_2),a(u,v_3)\},\ a(u,v_2),
 \ \{a(u,v_1), a(u,v_2)\},\ a(u,v_1). \benda
\]

% The relationship between $\Aut(X(G))$ and $\Aut(Y(G))$
\section{The relationship between $\Aut(X(G))$ and $\Aut(Y(G))$}\label{sec:auts}

In this section we analyze the relationship between $\Aut(X(G))$ and 
$\Aut(Y(G))$. As a byproduct of the analysis we get 
\[ 
Y(G)\not\cong \tilde Y(G). 
\]
We develop a machinery in the same spirit as the one in 
Section~\ref{sec:autu}.

Since for distinct vertices $ u$ and $ v$ of $ G$ the color of all vertices of $ X(u)$ is different from the color of the vertices of $ X(v)$, for every $ f\in \Aut(X(G))$ the restriction of $ f$ to $ X(u)$ is in $ \Aut(X(u))$, in particular, $ \bar f(V(X(u)))= V(X(u))$. The graph $ Y(G)$ is uncolored. The next lemma shows that any 
automorphism of $G$ can be lifted to an automorphism of $Y(G)$ in a natural 
way. We leave its routine proof to the reader. 
\begin{lem}\label{lem:GautotoYGauto}
For every $\sigma\in \Aut(G)$ there is a unique $\tau\in \Aut(Y(G))$ such that 
for all $u,v\in V(G)$ 
with $uv\in E(G)$, 
\begin{eqnarray*}
\tau(a(u,v))= a(\sigma(u), \sigma(v))
 & \text{and} &  
\tau(b(u,v))= b(\sigma(u), \sigma(v)).
\end{eqnarray*}
Moreover, this $ \tau$ satisfies for every $m\in M(u)$, 
\[
\tau(m)= \big\{a(\sigma(u), \sigma(w))
 \bigmid \text{$w\in V(G)$, $uw\in E(G)$, and $a(u,w)\in m$}\big\}.
\]
In particular, $\bar \tau(V(Y(u)))= V(Y(\sigma(u)))$.
\medskip

\noindent
In the following we denote for $\sigma\in \Aut(G)$ the corresponding $ \tau$ by $ \tau_\sigma$.
\end{lem}

\noindent
Taking  in the following proposition as  $f\in \Aut(X(G))$ the identity, we see that \eqref{eq:YGgtwinpreserving} corresponds to the first statement displayed in the preceding lemma (where $ \tau$ plays the role of $ g$).
\smallskip
\begin{prop}\label{pro:XGautoplusGpitoYGauto}
For every $f\in \Aut(X(G))$ and $\sigma\in \Aut(G)$ there is a unique $g\in 
\Aut(Y(G))$ such that for all $u,v\in V(G)$ with $uv\in E(G)$,
\begin{equation}\label{eq:YGgtwinpreserving}g(a(u,v))= f(a(\sigma(u), \sigma(v)))
\quad\text{and} \quad
    g(b(u,v))= f(b(\sigma(u), \sigma(v))).
  \end{equation}
  Moreover, this $ g$ satisfies $\bar g(V(Y(u)))= V(Y(\sigma(u)))$ and $g= f\circ \tau_\sigma$.
\end{prop}

\proof It is straightforward to verify that $g:= f\circ \tau_\sigma$ satisfies~\eqref{eq:YGgtwinpreserving}. 
To see the uniqueness of $g$, by~\eqref{eq:YGgtwinpreserving} it suffices to 
show that $g$ is uniquely determined by $f$ and $\sigma$ on any middle vertex in 
$Y(G)$. Let $u\in V(G)$ and $ m\in M(u)$. In particular, we will see that $g(m)\in M(\sigma(u))$, thus getting 
$\bar g(V(Y(u)))= V(Y(\sigma(u)))$, too.  For some (even for all) $ v\in V(G) $ with $uv\in E(G)$ we have
\[
a(u,v)\,m\in E(G) \ \ \ \mbox{or} \ \ \ b(u,v)\,m\in E(G).
\]
As $g\in \Aut(Y(G))$, we get
\[
g(a(u,v))\,g(m)\in E(G) \ \ \ \mbox{or} \ \ \ g(b(u,v))\,g(m)\in E(G).
\]
Since $ \{f(a(\sigma(u), \sigma(v))), f(b(\sigma(u), \sigma(v)))\}= \{a(\sigma(u), \sigma(v)),b(\sigma(u), \sigma(v))\}$ (as $ f\in \Aut(X(G))$), we obtain by $ $\eqref{eq:YGgtwinpreserving},
\[
a(\sigma(u),\sigma(v))\,g(m)\in E(G) \ \ \ \mbox{or} \ \ \ b(\sigma(u),\sigma(v))\,g(m)\in E(G).
\]
Therefore, 
$g(m)\in M(\sigma(u))$ and hence, $ g(m)\se A(\sigma(u))$.
Every  vertex $x$ in $A(\sigma(u))$  can be written as $x= a(\sigma(u), v)$ for some 
$v\in V(G)$. Since $\sigma\in \Aut(G)$, we can further 
write 
\[
x= a(\sigma(u), \sigma(w)) 
\]
for some $w\in V(G)$ with $uw\in E(G)$. By~\eqref{eq:YGgtwinpreserving} and 
$f\in \Aut(X(G))$,
\[
\big\{g(a(u,w)), g(b(u,w))\big\} 
= \big\{f(a(\sigma(u), \sigma(w))), f(b(\sigma(u), \sigma(w)))\big\}
=  \big\{a(\sigma(u), \sigma(w)), b(\sigma(u), \sigma(w))\big\}.
\]
We deduce
\begin{align*}
x & \in g(m) \\
 & \iff a(\sigma(u), \sigma(w))\in g(m) \\
 & \iff \big(\text{$g(a(u,w))= a(\sigma(u),\sigma(w))$ and\ $g(a(u,w))\in g(m)$}\big) \\
 & \hspace{2cm} \text{or}\ \big(\text{$g(b(u,w))= a(\sigma(u),\sigma(w))$ and\ $g(b(u,w))\in g(m)$}\big) \qquad  \qquad(\text{as $ g$ is bijective)} \\
 & \iff \big(\text{$g(a(u,w))= a(\sigma(u),\sigma(w))$ and\ $\{g(a(u,w)), g(m)\}\in E(Y(G))$}\big) \\
 & \hspace{2cm} \text{or}\ \big(\text{$g(b(u,w))= a(\sigma(u),\sigma(w))$ and\ $\{g(b(u,w)), g(m)\}\in E(Y(G))$}\big) \\
 & \iff \big(\text{$g(a(u,w))= a(\sigma(u),\sigma(w))$ and\ $\{a(u,w), m\}\in E(Y(G))$}\big) \\
 & \hspace{2cm} \text{or}\ \big(\text{$g(b(u,w))= a(\sigma(u),\sigma(w))$ and\ $\{b(u,w), m\}\in E(Y(G))$}\big)
  \quad (\text{by $g\in \Aut(Y(G))$}) \\
 & \iff \big(\text{$f(a(\sigma(u), \sigma(w)))= a(\sigma(u),\sigma(w))$ and\ $\{a(u,w), m\}\in E(Y(G))$}\big) \\
 & \hspace{2cm} \text{or}\ \big(\text{$f(b(\sigma(u), \sigma(w)))= a(\sigma(u),\sigma(w))$ and\ $\{b(u,w), m\}\in E(Y(G))$}\big).
  \qquad (\text{by~\eqref{eq:YGgtwinpreserving}})
\end{align*}
Note that the last condition in the above chain of equivalences only depends on $f$ 
and $\sigma$. Hence, $g(m)$ is completely determined by $f$ and $\sigma$ as 
desired. \proofend

\noindent
Proposition~\ref{pro:XGautoplusGpitoYGauto} immediately implies
\[
\big\{f\circ \tau_\sigma \bigmid \text{$f\in \Aut(X(G))$ and $\sigma\in \Aut(G)$}\}
 \subseteq \Aut(Y(G)).
\]
Are these two sets equal? In other words,  can all automorphisms of $Y(G)$ be 
obtained in this way, i.e., by an automorphism of $G$ and an automorphism of 
$X(G)$? We get a positive answer for graphs $G$ of degree at least three in 
Theorem~\ref{thm:fifg}. The following example shows that the answer is 
negative for some graphs of degree two.  
\begin{exa}\label{exa:ck} 
For $k\ge 3$ we know that $Y(C_k)$ is the disjoint union of two cycles 
$C_{3k}$ (see Lemma~\ref{lem:1-2} and Figure~\ref{fig:gYC3}). Hence, for any 
two vertices $x$ and $y$ in $Y(C_k)$ there is a $g\in \Aut(Y(C_k))$ with 
$g(x)= y$. For example, for $u,v,w\in V(C_k)$ with $uv\in E(C_k)$ and an 
arbitrary $m\in M(w)$ there is an $g\in \Aut(Y(C_k))$ with $g(a(u,v))=m$. 
Note that for such a $g$  equalities of the form in~\eqref{eq:YGgtwinpreserving} cannot 
hold for any $\sigma\in \Aut(C_k)$ and $f\in \Aut(X(C_k))$. Furthermore, $ \bar g(L(Y(C_k)))\not\se L(Y(C_k))$ and in general, for $ x\in L(Y(d))$ the equality $ g(x')=g(x)'$ does not hold; in particular, it does not hold for $ x:=a(u,v)$ as in this case $g(x)' $ is not even defined. \hfill{$ \dashv$}
\end{exa}  

\noindent
We turn to the positive result for graphs of degree at least three.

\begin{theo}\label{thm:fifg}
Assume $G$ is a graph of degree at least $3$. Let $g\in \Aut(Y(G))$. Then there are a unique $\sigma\in \Aut(G)$ and a 
unique $f\in\Aut(X(G))$  such that for 
all $uv\in E(G)$ the equalities in~\eqref{eq:YGgtwinpreserving} hold, i.e., 
\begin{eqnarray*}
    g(a(u,v))= f(a(\sigma(u), \sigma(v)))
     & \text{and} &  
    g(b(u,v))= f(b(\sigma(u), \sigma(v))). 
\end{eqnarray*}
Together with Proposition~\ref{pro:XGautoplusGpitoYGauto}, we get
\[
\Aut(Y(G))= \{f\circ\tau_\sigma\mid \text{$f\in \Aut(X(G))$ and $\sigma\in \Aut(G)$}\}.
\]
\end{theo}

\noindent
The following notion concerning automorphisms of $ Y(G)$ will play an important  role in our 
proof of Theorem~\ref{thm:fifg}. 
\begin{defn}\label{def:ggadgetpreserving}
    Let $ g\in\Aut(Y(G))$ or $ g:Y(G)\cong \tilde Y(G) $. We say that 
 $g$   is \emph{gadget-preserving} if for every $u\in 
V(G)$ there is a $u^+\in V(G)$ such that $\bar g(V(Y(u)))= V(Y(u^+))$. 
\end{defn}

\noindent Observe that the $g\in \Aut(C_k)$ of
Example~\ref{exa:ck} is not gadget-preserving, while the ($f\circ \tau_\sigma$)'s of 
Proposition~\ref{pro:XGautoplusGpitoYGauto} are. It turns out that being  
gadget-preserving is exactly a necessary and sufficient condition for an 
automorphism $g$ of $Y(G)$ to have the form $g= f\circ \tau_\sigma$ for some $f\in 
\Aut(X(G))$ and $\sigma\in \Aut(G)$. To prove this, we start by making a few 
observations on gadget-preserving automorphisms of $Y(G)$. The first one is 
trivial. 
\begin{lem}\label{lem:gadgetpreservingproperties}
Let $g, g_1, g_2\in \Aut(Y(G))$ be gadget-preserving. Then $g^{-1}$ and 
$g_2\circ g_1$ are gadget-preserving automorphisms of $Y(G)$, too. If  $g_1\in \Aut(Y(G))$ and $ g_2:Y(G)\cong \tilde Y(G) $ are gadget-preserving, then so is $g_2\circ g_1$.
\end{lem}

\medskip
\begin{lem}\label{lem:gadgetpreservingtoautomorphism}
  Let $g\in \Aut(Y(G))$ or $ g:Y(G)\cong \tilde Y(G) $ be gadget-preserving. Then, %
  $\sigma: V(G)\to V(G)$ defined by
\[
\mbox{for every $u\in V(G)$: \ \   }\sigma(u):= u^+, \mbox{ \ \ where $\bar g(V(Y(u)))= V(Y(u^+))$,}
\]
 is an automorphism of $G$. 
Moreover, for every edge $uv\in E(G)$,
\[
\big\{g(a(u,v)), g(b(u,v))\}= \big\{a(\sigma(u), \sigma(v)), b(\sigma(u), \sigma(v))\}.
\]
\end{lem}

\proof We first show $\sigma\in \Aut(G)$. The bijectivity of $u\mapsto u^+$ 
follows directly from the bijectivity of $g$. To see that it preserves edges 
and non-edges in $Y(G)$, consider any $u,v\in V(G)$ with $u\ne v$. Then, 
\begin{align*}
uv\in E(G) & \iff \text{there is an edge between $Y(u)$ and $Y(v)$} \\
 & \iff \text{there is an edge between $Y(u^+)$ and $Y(v^+)$} 
 & \text{(as $g$ is an isomorphism)} \\
 & \iff u^+v^+\in E(G). 
\end{align*}
Now consider an edge $uv\in E(G)$. Since $\bar g(V(Y(u)))= V(Y(u^+))= 
V(Y(\sigma(u)))$, it holds 
\[
\big\{g(a(u,v)), g(b(u,v))\}\subseteq V(Y(\sigma(u))).
\]
Similarly, we obtain
\[
\big\{g(a(v,u)), g(b(v,u))\}\subseteq V(Y(\sigma(v))).
\]
Recall that in $Y(G)$ there are two edges between 
$\big\{a(u,v), b(u,v)\}$ and $\big\{a(v,u), b(v,u)\}$\label{page:twist} and  $g\in \Aut(Y(G))$ or $ g:Y(G)\cong \tilde Y(G) $, so there must be two 
edges between 
\begin{eqnarray*}
\{g(a(u,v)), g(b(u,v))\}
 & \text{and} &
\{g(a(v,u)), g(b(v,u))\}.
\end{eqnarray*}
As the only edges between $  V(Y(\sigma(u)))$ and $ V(Y(\sigma(v)))$ are the edges between
\begin{eqnarray*}
\{a(\sigma(u),\sigma(v))), b(\sigma(u),\sigma(v))\}
 & \text{and} &
\{a(\sigma(v),\sigma(u))), b(\sigma(v),\sigma(u))\},\footnotemark
\end{eqnarray*}\footnotetext{Note, this 
property is also true for $\tilde Y(G)$, no matter which edge we twist.}
 in particular, we get 
 $\big\{g(a(u,v)), g(b(u,v))\}= 
\big\{a(\sigma(u),\sigma(v))), b(\sigma(u),\sigma(v))\}$. \proofend 
\medskip

\noindent
Now  we prove a converse of 
Proposition~\ref{pro:XGautoplusGpitoYGauto}. 

\begin{prop}\label{pro:YGtwinpreservingdecompose}
If $g\in \Aut(Y(G))$ (or $ g:Y(G)\cong \tilde Y(G)$) is gadget-preserving, then there are a unique $\sigma\in 
\Aut(G)$ and a unique $f\in \Aut(X(G))$ (or $ f:X(G)\cong \tilde X(G)$) such that for all $uv\in E(G)$ the 
equalities in~\eqref{eq:YGgtwinpreserving}, i.e., 
\begin{eqnarray*}
    g(a(u,v))= f(a(\sigma(u), \sigma(v)))
     & \text{and} &  
    g(b(u,v))= f(b(\sigma(u), \sigma(v)))
\end{eqnarray*}
hold. And in fact, if $g\in \Aut(Y(G))$, then $\sigma$ is precisely the automorphism of $G$ as stated in 
Lemma~\ref{lem:gadgetpreservingtoautomorphism}.
\end{prop}
As there is no $ f:X(G)\cong \tilde X(G)$ we get:
\begin{cor}\label{cor:YGtwinpreservingdecompose} There is no gadget-preserving $ g:Y(G)\cong \tilde Y(G)$.
\end{cor}

\noindent
\textit{Proof of Proposition~\ref{pro:YGtwinpreservingdecompose}.} Let $\sigma$ be the automorphism of $G$ as stated 
in~Lemma~\ref{lem:gadgetpreservingtoautomorphism}, i.e., with
\begin{equation}\label{eq:antimis}\bar g(V(Y(u)))=V(Y(\sigma(u)))
\end{equation}
for all $ u$ in $ G$.
By
Lemma~\ref{lem:GautotoYGauto} there is an automorphism $\tau_\sigma$ of $Y(G)$ such 
that (i) and (ii) hold for every $u\in V(G)$.  
\begin{itemize}
  \item[(i)] For every $v\in V(G)$ with $uv\in E(G)$, 
    \begin{eqnarray}\label{eq:rhopi}
    \tau_\sigma(a(u,v))= a(\sigma(u), \sigma(v))
    & \text{and} &  
    \tau_\sigma(b(u,v))= b(\sigma(u), \sigma(v)).
    \end{eqnarray}
\item[(ii)] $\overline{\tau_\sigma}(V(Y(u))= V(Y(\sigma(u)))$, i.e., $\tau_\sigma$ is 
    gadget-preserving. 
\end{itemize} 
Now we define a function 
\begin{equation*}%
f:= g\circ \tau_\sigma^{-1} \quad \ \text{as} \  
\begin{tikzcd}%
Y(G) \arrow[rd,"g"] & Y(G) \arrow[l, "\tau_\sigma^{-1}"swap] \arrow[d, "f"] \\
 & Y(G)
\end{tikzcd}\qquad \mbox{ or} \qquad\left( \ \text{as} \  
\begin{tikzcd}%[column sep=tiny]
Y(G) \arrow[rd,"g"] & Y(G) \arrow[l, "\tau_\sigma^{-1}"swap] \arrow[d, "f"] \\
 & \tilde Y(G)
\end{tikzcd}\right).
\end{equation*}
Lemma~\ref{lem:gadgetpreservingproperties} implies that $f\in \Aut(Y(G))$ (or $ f:Y(G)\cong \tilde Y(G)$) is gadget-preserving. We claim that $f\in \Aut(X(G))$  (or $ f:X(G)\cong \tilde X(G)$).

By (ii), for $ u\in V(G)$, $ \overline{\tau_\sigma^{-1}}(V(Y(u)))= V(Y(\sigma^{-1}(u)))$ and thus by \eqref{eq:antimis},
\[
\bar f(V(Y(u)))=\bar g(\overline{\tau_\sigma^{-1}}(V(Y(u))))=\bar g(V(Y(\sigma^{-1}(u))))= V(Y(u)).
\]
We still have to show that $ f$ preserves the colors of the link vertices.
This amounts to show that for every $uv\in E(G)$, 
\[
\big\{f(a(u,v)), f(b(u,v))\big\}
 = \big\{a(u,v), b(u,v)\big\}.
\] 
We deduce
\begin{align*}
\big\{f(a(u,v)), f(b(u,v))\big\} 
 & = \big\{g(\tau_\sigma^{-1}(a(u,v))), g(\tau_\sigma^{-1}(b(u,v)))\big\} \\
 & = \big\{g(a(\sigma^{-1}(u), \sigma^{-1}(v))), g(b(\sigma^{-1}(u), \sigma^{-1}(v)))\big\}
  & \text{(by~\eqref{eq:rhopi})} \\
 & = \big\{a(\sigma(\sigma^{-1}(u)), \sigma(\sigma^{-1}(v))), b(\sigma(\sigma^{-1}(u)), \sigma(\sigma^{-1}(v)))\big\}
  & \text{(by Lemma~\ref{lem:gadgetpreservingtoautomorphism})} \\ 
 & = \big\{a(u,v), b(u, v)\big\}. 
\end{align*}
Therefore, as already mentioned, there is no  gadget-preserving $ g:Y(G)\cong \tilde Y(G)$, since there is no isomorphism between $X(G)$  and $ \tilde X(G)$.
Hence, in the rest of this proof we can assume $ g\in \Aut(Y(G))$.

Using~\eqref{eq:rhopi} and $g= f\circ \tau_\sigma$, we obtain
\begin{align*}
g(a(u,v))= f(\tau_\sigma(a(u,v)))= f(a&(\sigma(u), \sigma(v)))\\
  &\text{and} \ \ \
g(b(u,v))= f(\tau_\sigma(b(u,v)))= f(b(\sigma(u), \sigma(v))).
\end{align*}
This proves  the equalities displayed in the statement of the proposition and shows that for $ uv\in E(G)$,
\begin{equation}\label{eq:gadrev}g(a(u,v))\in V(Y(\sigma(u))).
\end{equation}
It remains to show that $\sigma$ and $f$ are unique. So let us assume that some 
$\sigma_0\in \Aut(G)$ and $f_0\in \Aut(X(G))$ satisfy that for every edge $uv$ 
in $G$, 
\begin{eqnarray}\label{eq:gf0pi0}
    g(a(u,v))= f_0(a(\sigma_0(u), \sigma_0(v)))
     & \text{and} &  
    g(b(u,v))= f_0(b(\sigma_0(u), \sigma_0(v))). 
\end{eqnarray}
Since $ G$ is connected with more than one vertex, for every $ u$ in $ G$
there is a $ v$ with $ uv\in E(G)$.  As $f_0\in \Aut(X(G))$ we have
\[
\overline{f_0}(V(Y(\sigma_0(u))))= V(Y(\sigma_0(u))) 
\]
and hence by~\eqref{eq:gf0pi0}, $g(a(u,v))$ is in $Y(\sigma_0(u))$. Together with \eqref{eq:gadrev}, we see that
\[
\sigma_0(u)= \sigma(u).
\]
As $ u$ was an arbitrary vertex of $ G$, we get $\sigma_0= \sigma$. It remains to show that $ f=f_0$.
Now we invoke Proposition~\ref{pro:XGautoplusGpitoYGauto} on $f_0$ and $\sigma$. Thus 
the function
\[
g_0:= f_0\circ \tau_\sigma
\quad \text{i.e.,}\ \ 
\begin{tikzcd}
Y(d) \arrow[r, "\tau_\sigma"] \arrow[rd,"g_0"] 
 & Y(d) \arrow[d, "f_0"] \\%
 & Y(d) 
\end{tikzcd}.
\]
is the 
\emph{unique} automorphism of $ Y(G)$ that satisfies  for every $u\in V(G)$ 
 and for every $v\in V(G)$ with $uv\in E(G)$, 
    \begin{eqnarray*}%
    g_0(a(u,v))= f_0(a(\sigma(u), \sigma(v)))
    & \text{and} &  
    g_0(b(u,v))= f_0(b(\sigma(u), \sigma(v))).
    \end{eqnarray*}
 Hence, by \eqref{eq:gf0pi0}  we get $ g_0=g$. Thus, $f_0= g_0\circ \tau_\sigma^{-1}= g\circ \tau_\sigma^{-1}= f$, the desired equality.\proofend

\noindent
With Proposition 7.8 ready, the only missing piece for our proof of Theorem 7.4 is:

\begin{prop}\label{pro:deg3siblingpreserving} Assume $\deg(G)\ge 3$. Then every $g \in \Aut(Y(G))$ or $ g:Y(G)\cong \tilde Y(G)$ is gadget-preserving.
\end{prop}

\noindent
Therefore, by Corollary~\ref{cor:YGtwinpreservingdecompose}, we see that $Y(G)\not\cong \tilde Y(G) $ for graphs $ G$ with $\deg(G)\ge 3$. Together with Lemma~\ref{lem:1-2} we get the following corollary. 
\begin{cor}\label{cor:fifg} 
For all graphs $G$, \ \ $Y(G)\not\cong\tilde Y(G)$.
\end{cor}

\noindent
The following discussion will lead to a proof of Proposition~\ref{pro:deg3siblingpreserving}.
As Lemma~\ref{lem:1-2} and Proposition~\ref{pro:3conn} indicate, vertices 
$u\in V(G)$ of degree at least $3$ play a special role in $Y(G)$ and $\tilde 
Y(G)$. In fact, the vertex sets of their corresponding $Y(u)$'s can be easily 
recognized by the next lemma.

\begin{lem}\label{lem:deg3VYu}%
A vertex $x$ in $Y(G)$ is in a gadget $Y(u)$ for 
some $u\in V(G)$ with $\deg(u)\ge 3$ if and only if it is on a cycle of length 
at most $8$ in $Y(G)$. And in the yes case, 
\[
V(Y(u))= \big\{y\in V(Y(G)) \bigmid \text{$y$ and $x$ are on a cycle of length at most $8$}\big\}.
\]
The same properties hold for $\tilde Y(G)$ as well.
\end{lem} 

\proof We start by claiming that
every cycle in $Y(G)$ and every cycle in $\tilde Y(G)$ that contains vertices 
of at least two gadgets has length at least $9$. 

Assume first that the cycle $C$ contains vertices of exactly two gadgets, say, of 
$Y(u)$ and $Y(v)$. Then in $Y(G)$ the cycle $C$ must contain the edges 
of~\eqref{eq:eun} (possibly such a cycle in $\tilde Y(G)$ must instead 
contain the edges of~\eqref{eq:etw}) as these are the only edges between 
$Y(u)$ and $Y(v)$. It is not hard to see that in order to pass in $Y(u)$ from 
$a(u,v)$ to $b(u,v)$ (if it is possible) we need a path of length at least 
$4$. The same applies to paths of $Y(v)$ from $a(v,u)$ to $b(v,u)$. So in 
total such a cycle must have length at least $10$. 

If the cycle contains vertices from three gadgets, one easily verifies that 
the length has to be at least $9$ (in Figure~\ref{fig:gYC3} for $G:=C_3$ the 
graph $Y(G)$ is the disjoint union of two cycles of length $9$).
Thus, every cycle of length at most $8$ must be completely contained in a 
single gadget $Y(u)$. 

By Lemma~\ref{lem:cig} we know  that for $u\in V(G)$ with $\deg(u)\ge 3$ every two vertices of
$Y(u)$  are on a cycle of length at most $8$; and if 
$\deg(u)\le 2$, then there is no cycle within $Y(u)$. Hence, all the 
following are equivalent for every vertex $x$ of $Y(G)$ or of $\tilde Y(G)$.
\begin{itemize}
\item $x$ is on a cycle of length at most $8$.

\item $x$ is on a cycle of length at most $8$ which is completely contained 
    in a gadget $Y(u)$ for some $u\in V(G)$ with $\deg(u)\ge 3$. 

\item $x$ is in a gadget $Y(u)$ for some $u\in V(G)$ with $\deg(u)\ge 3$.

\item $x$ is in a gadget $Y(u)$ that contains exactly all the vertices $y$ 
    such that $y$ and $x$ are on a cycle of length at most $8$.
\end{itemize}
This finishes the proof. \proofend

\noindent

As discussed in Section~\ref{sec:autu}, in general, given a gadget  $Y(d)$ we cannot uniquely determine the structure of $Y(d)$. 
In particular, Example~\ref{exa:gY4} shows that there is an automorphism of $Y(4)$ sending 
link vertices onto middle vertices. However, a
whole CFI-graph $Y(G)$ is now available to us, from which we can obtain some 
 structural information about the gadgets $Y(u)$ provided $\deg(u)\ge 3$, e.g., the following whose proof is trivial.

\begin{lem}\label{lem:deg3Yustructure}
Let $u\in V(G)$ and $x\in V(Y(u))$. Then
 $x\in L(u)$ if and only if there is a vertex $y\in 
    V(Y(G))\setminus V(Y(u))$ such that $xy\in E(Y(G))$, i.e., $Y(G)$ 
    contains an edge between $x$ and a vertex outside $Y(u)$.
    The same is true for $\tilde Y(G)$.
  \end{lem}

  \noindent
For a further analysis we introduce two notions.
\begin{defn}\label{def:sigggadlink} Let $g \in \Aut(Y(G)) $ or $g: Y(G)\cong\tilde Y(G) $ and $ u\in V(G)$. Then $ g$ \emph{preserves the gadget $ Y(u)$} if there is a $ u^+\in V(G)$ with
  \[
\bar g(V(Y(u)))=V(Y(u^+)).
\]
And $ g$ \emph{preserves the gadget $ Y(u)$ and its twins} if $\bar g(V(Y(u)))=V(Y(u^+)) $ for some $ u^+$ in $ G$ and for all $ x\in L(u)$ we have $ g(x), g(x')\in L(u^+)$ and
\[
g(x')=g(x)'.
\]
\end{defn}

\noindent
The following example shows that a gadget of a CFI-graph can be preserved but not its twins.
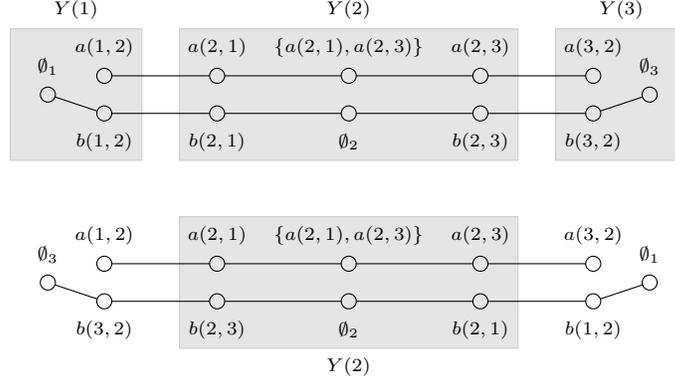
\begin{figure} 	
\centering
\begin{tikzpicture}[scale=0.25,vertex style/.style={draw,
                                   circle,
                                   minimum size=2mm,
                                   inner sep=0pt,
                                   outer sep=0pt,
                                   %shade
                                   }]

{\scriptsize                                  
\begin{scope}[xshift=-10cm,yshift=0cm]
% Y(P3)
% draw vertices

\def\cloudwidth{20}
\def\horwidth{7}

\def \j{1}
\pgfmathsetmacro\jaone{int(\j+1)}
\draw[fill=gray, opacity=.2] (\cloudwidth*\j-15+\horwidth+2,5.5) rectangle (\cloudwidth*\j-15+2*\horwidth+2,12.5);
\draw (\cloudwidth*\j-15+\horwidth*1.5+2,13.5) node {{\textcolor{black}{$Y(\j)$}}};

\draw (\cloudwidth*\j-15+\horwidth+4,9) coordinate[vertex style] (middledown\j);
\draw (\cloudwidth*\j-15+\horwidth+4,10.5) node {{\textcolor{black}{$\emptyset_1$}}};

\draw (\cloudwidth*\j-15+2*\horwidth,6.5) node {{\textcolor{black}{$b(1,2)$}}};
\draw (\cloudwidth*\j-15+2*\horwidth,8) coordinate[vertex style] (rightdown\j);
\draw (\cloudwidth*\j-15+2*\horwidth,10) coordinate[vertex style] (rightup\j);
\draw (\cloudwidth*\j-15+2*\horwidth,11.5) node {{\textcolor{black}{$a(1,2)$}}};

\foreach \j in {2} {
    \draw[fill=gray, opacity=.2] (\cloudwidth*\j-15-2,5.5) rectangle (\cloudwidth*\j-15+2*\horwidth+2,12.5);
    \draw (\cloudwidth*\j-15+\horwidth,13.5) node {{\textcolor{black}{$Y(\j)$}}};
    
    \pgfmathsetmacro\jmone{int(\j-1)}
    \pgfmathsetmacro\jaone{int(\j+1)}
    \draw (\cloudwidth*\j-15,6.5) node {{\textcolor{black}{$b(2,1)$}}};
    \draw (\cloudwidth*\j-15,8) coordinate[vertex style] (leftdown\j);
    \draw (\cloudwidth*\j-15,10) coordinate[vertex style] (leftup\j);
    \draw (\cloudwidth*\j-15,11.5) node {{\textcolor{black}{$a(2,1)$}}};

    \draw (\cloudwidth*\j-15+\horwidth,6.5) node {{\textcolor{black}{$\emptyset_2$}}};
    \draw (\cloudwidth*\j-15+\horwidth,8) coordinate[vertex style] (middledown\j);
    \draw (\cloudwidth*\j-15+\horwidth,10) coordinate[vertex style] (middleup\j);
    \draw (\cloudwidth*\j-15+\horwidth,11.5) node {{\textcolor{black}{$\{a(2,1),a(2,3)\}$}}};
    
    \draw (\cloudwidth*\j-15+2*\horwidth,6.5) node {{\textcolor{black}{$b(2,3)$}}};
    \draw (\cloudwidth*\j-15+2*\horwidth,8) coordinate[vertex style] (rightdown\j);
    \draw (\cloudwidth*\j-15+2*\horwidth,10) coordinate[vertex style] (rightup\j);
    \draw (\cloudwidth*\j-15+2*\horwidth,11.5) node {{\textcolor{black}{$a(2,3)$}}};
}

\def \j{3}
\pgfmathsetmacro\jmone{int(\j-1)}
\draw[fill=gray, opacity=.2] (\cloudwidth*\j-15-2,5.5) rectangle (\cloudwidth*\j-15+\horwidth-2,12.5);
\draw (\cloudwidth*\j-15+\horwidth*0.5-2,13.5) node {{\textcolor{black}{$Y(\j)$}}};

\draw (\cloudwidth*\j-15,6.5) node {{\textcolor{black}{$b(3,2)$}}};
\draw (\cloudwidth*\j-15,8) coordinate[vertex style] (leftdown\j);
\draw (\cloudwidth*\j-15,10) coordinate[vertex style] (leftup\j);
\draw (\cloudwidth*\j-15,11.5) node {{\textcolor{black}{$a(3,2)$}}};

\draw (\cloudwidth*\j-15+3,9) coordinate[vertex style] (middledown\j);
\draw (\cloudwidth*\j-15+3,10.5) node {{\textcolor{black}{$\emptyset_3$}}};

% draw lines

% lines in a cloud
\foreach \j in {1,...,3} {
    \ifthenelse{\not\equal{\j}{1}}{
        \draw (leftdown\j)--(middledown\j);
    }{};
    \ifthenelse{\not\equal{\j}{3}}{
        \draw (middledown\j)--(rightdown\j);
    }{};
    \ifthenelse{\not\equal{\j}{1} \and \not\equal{\j}{3}}{
        \draw (leftup\j)--(middleup\j);
        \draw (middleup\j)--(rightup\j);
    }{};
}

% lines between clouds
\foreach \j in {1,...,2} {
    \pgfmathsetmacro\jaone{int(\j+1)}
    \draw (rightup\j)--(leftup\jaone);
    \draw (rightdown\j)--(leftdown\jaone);
}
\end{scope}}

% {\scriptsize
% \draw (-18,5) node {{\textcolor{black}{$Y(P_3)$}}};
% \draw[dashed] (-20,4)--(25,4);
% \draw (-18,3) node {{\textcolor{black}{$\tilde Y(P_3)$}}};
% }

{\scriptsize                                  
\begin{scope}[xshift=-10cm,yshift=-10cm]
% Y(P3)
% draw vertices

\def\cloudwidth{20}
\def\horwidth{7}

\def \j{1}
\pgfmathsetmacro\jaone{int(\j+1)}
% \draw[dashed] (\cloudwidth*\j-15+\horwidth+2,5.5) rectangle (\cloudwidth*\j-15+2*\horwidth+2,12.5);
% \draw (\cloudwidth*\j-15+\horwidth*1.5+2,13.5) node {{\textcolor{black}{$Y(\j)$}}};

\draw (\cloudwidth*\j-15+\horwidth+4,9) coordinate[vertex style] (middledown\j);
\draw (\cloudwidth*\j-15+\horwidth+4,10.5) node {{\textcolor{black}{$\emptyset_3$}}};

\draw (\cloudwidth*\j-15+2*\horwidth,6.5) node {{\textcolor{black}{$b(3,2)$}}};
\draw (\cloudwidth*\j-15+2*\horwidth,8) coordinate[vertex style] (rightdown\j);
\draw (\cloudwidth*\j-15+2*\horwidth,10) coordinate[vertex style] (rightup\j);
\draw (\cloudwidth*\j-15+2*\horwidth,11.5) node {{\textcolor{black}{$a(1,2)$}}};

\foreach \j in {2} {
    \draw[fill=gray, opacity=.2] (\cloudwidth*\j-15-2,5.5) rectangle (\cloudwidth*\j-15+2*\horwidth+2,12.5);
    \draw (\cloudwidth*\j-15+\horwidth,4.5) node {{\textcolor{black}{$Y(\j)$}}};
    
    \pgfmathsetmacro\jmone{int(\j-1)}
    \pgfmathsetmacro\jaone{int(\j+1)}
    \draw (\cloudwidth*\j-15,6.5) node {{\textcolor{black}{$b(2,3)$}}};
    \draw (\cloudwidth*\j-15,8) coordinate[vertex style] (leftdown\j);
    \draw (\cloudwidth*\j-15,10) coordinate[vertex style] (leftup\j);
    \draw (\cloudwidth*\j-15,11.5) node {{\textcolor{black}{$a(2,1)$}}};

    \draw (\cloudwidth*\j-15+\horwidth,6.5) node {{\textcolor{black}{$\emptyset_2$}}};
    \draw (\cloudwidth*\j-15+\horwidth,8) coordinate[vertex style] (middledown\j);
    \draw (\cloudwidth*\j-15+\horwidth,10) coordinate[vertex style] (middleup\j);
    \draw (\cloudwidth*\j-15+\horwidth,11.5) node {{\textcolor{black}{$\{a(2,1),a(2,3)\}$}}};
    
    \draw (\cloudwidth*\j-15+2*\horwidth,6.5) node {{\textcolor{black}{$b(2,1)$}}};
    \draw (\cloudwidth*\j-15+2*\horwidth,8) coordinate[vertex style] (rightdown\j);
    \draw (\cloudwidth*\j-15+2*\horwidth,10) coordinate[vertex style] (rightup\j);
    \draw (\cloudwidth*\j-15+2*\horwidth,11.5) node {{\textcolor{black}{$a(2,3)$}}};
}

\def \j{3}
\pgfmathsetmacro\jmone{int(\j-1)}
% \draw[dashed] (\cloudwidth*\j-15-2,5.5) rectangle (\cloudwidth*\j-15+\horwidth-2,12.5);
% \draw (\cloudwidth*\j-15+\horwidth*0.5-2,13.5) node {{\textcolor{black}{$Y(\j)$}}};

\draw (\cloudwidth*\j-15,6.5) node {{\textcolor{black}{$b(1,2)$}}};
\draw (\cloudwidth*\j-15,8) coordinate[vertex style] (leftdown\j);
\draw (\cloudwidth*\j-15,10) coordinate[vertex style] (leftup\j);
\draw (\cloudwidth*\j-15,11.5) node {{\textcolor{black}{$a(3,2)$}}};

\draw (\cloudwidth*\j-15+3,9) coordinate[vertex style] (middledown\j);
\draw (\cloudwidth*\j-15+3,10.5) node {{\textcolor{black}{$\emptyset_1$}}};

% draw lines

% lines in a cloud
\foreach \j in {1,...,3} {
    \ifthenelse{\not\equal{\j}{1}}{
        \draw (leftdown\j)--(middledown\j);
    }{};
    \ifthenelse{\not\equal{\j}{3}}{
        \draw (middledown\j)--(rightdown\j);
    }{};
    \ifthenelse{\not\equal{\j}{1} \and \not\equal{\j}{3}}{
        \draw (leftup\j)--(middleup\j);
        \draw (middleup\j)--(rightup\j);
    }{};
}

% lines between clouds
\foreach \j in {1,...,2} {
    \pgfmathsetmacro\jaone{int(\j+1)}
    \draw (rightup\j)--(leftup\jaone);
    \draw (rightdown\j)--(leftdown\jaone);
}
\end{scope}}

\end{tikzpicture}
\caption{The automorphism $\id_{P_4}^{\ \ \frown}\,\textup{fl-end}_{P_6} $ of $ Y(P_2)$. The figure shows two copies of $ Y(P_2)$. The images of the points of the upper copy are located at the corresponding place in the lower copy.}\label{fig:YP4P6}
\end{figure}

\begin{exa}\label{exa:flends} Clearly, a path $ P_k$ with $ k$ edges has two automorphisms, the identity $ \id_{P_k}$ and the automorphism $ \textup{fl-end}_{P_k}$ that flips the ends of $ P_k$.

  Let $ G$ be the path $ P_2$ with  $ V(P_2)=\{1,2,3 \}$ and $ E(P_2)=\{12, 23 \}$. Then $ Y(G)$ is the disjoint union (of a copy) of $ P_4$ and (of a copy) of $ P_6$. Therefore, $\Aut(Y(P_2))$ consists of the following four automorphisms:
  \[
    \id_{P_4}^{\ \ \frown}\,\id_{P_6},\ \ \ \id_{P_4}^{\ \ \frown}\,\textup{fl-end}_{P_6}, \ \ \ \textup{fl-end}_{P_4}^{\ \ \frown}\,\id_{P_6}, \ \ \ \textup{fl-end}_{P_4}^{\ \ \frown}\,\textup{fl-end}_{P_6}.
  \]
Here for $ g_1\in \Aut(P_4)$ and $ g_2\in \Aut(P_6)$ by $ g_1^{\  \frown}\,g_2$ we denote the automorphism of $ Y(P_2)$ whose restriction to $ P_4$ and $ P_6$ are $ g_1$ and $ g_2$, respectively.

We leave it to the reader to verify that $ \id_{P_4}^{\ \ \frown}\,\id_{P_6} $ and $ \textup{fl-end}_{P_4}^{\ \ \frown}\,\textup{fl-end}_{P_6}$  preserve all gadgets of $ Y(P_2)$ and its twins, while the other two automorphisms only preserve the gadget $ Y(2)$ (even map $ Y(2)$ onto $ Y(2)$  but do not preserve the twins of $ Y(2)$ (for $\id_{P_4}^{\ \ \frown}\,\textup{fl-end}_{P_6} $ see Figure~\ref{fig:YP4P6}).\hfill{$ \dashv$}
\end{exa}

\noindent
With the next two lemmas we finish the proof of Proposition~\ref{pro:deg3siblingpreserving}.
\begin{lem}\label{lem:imgj}
  Let $ u$ be a vertex of $ G$ with $ \deg(u)\ge 3$. Then every $ g\in \Aut(Y(G))$ or $ g:Y(G)\cong \tilde Y(G)$ preserves the gadget $ Y(u)$ and its twins.
\end{lem}
\proof Every two vertices of 
$Y(u)$ are in $Y(u)$ on at least one cycle of length at most $8$ 
(cf.~Lemma~\ref{lem:cig}). As the same must hold for the images under $ g$ of these two vertices, by  Lemma~\ref{lem:deg3VYu} there must be a 
vertex $u^+\in V(G)$ such that $g$ maps $Y(u)$ into $Y(u^+)$. In particular, 
$\deg(u^+)\ge 3$. Looking at~$g^{-1}$, by the same reason we get~$g^{-1}$ maps $Y(u^+)$ into 
$Y(u)$. Hence, $g$ maps $Y(u)$ onto $Y(u^+)$.

Since for $ x$ in $ Y(G)$, the graph $Y(G)$ contains an edge between $x$ and a vertex outside $Y(u)$ if and only if  it  contains an edge between $g(x)$ and a vertex outside $Y(u^+)$, by Lemma~\ref{lem:deg3Yustructure} we get
$ \bar g(L(u))= L(u^+)$ and hence, $ \bar g(M(u))= M(u^+)$.

Finally, for every $ x\in L(u)$ we have
\[
xm\in E(Y(u)) \iff x'm\notin E(Y(u))
\]
for all $ m\in M(u)$ (see~\eqref{eq:cmeg}). As the  restriction of $ g$ to $ V(Y(u))$ is an isomorphism from $ Y(u)$ onto $Y(u^+) $, we obtain
\[
g(x)g(m)\in E(Y(u^+)) \iff g(x')g(m)\notin E(Y(u^+))
\]
for all $ m\in M(u)$. Since $\bar g(M(u))= M(u^+) $, we see that
\[
 g(x)\,m\in E(Y(u^+)) \iff g(x')\,m \notin E(Y(u^+))
\]
for all $ m\in M(u^+)$.
Hence, $g(x')=g(x)'$ by Lemma~\ref{lem:li'uni}. \proofend

\noindent
As $ G$ is connected,  the next lemma shows that an automorphism of $ Y(G)$ is gadget-preserving if it preserves at least one gadget and its twins. Therefore, together with the preceding lemma it yields the desired proof of Proposition~\ref{pro:deg3siblingpreserving}.

\begin{lem}\label{lem:prindstep} Let $ u$ be a vertex of $ G$ and assume that $ g\in \Aut(Y(G))$ or  $ g:Y(G)\cong \tilde Y(G)$ preserves the gadget $ Y(u)$ and its twins. Then $ g$ preserves the gadget $ Y(v)$ and its twins for every $ v$ in $ G$ with $ uv\in E(G)$.

\end{lem}
\proof So let $ v$ be a vertex of $ G$ with $ uv\in E(G)$. If $ \deg(v)\ge 3$, then we are done by the previous lemma. Hence, we can assume that $ \deg(v)\le 2$.

By assumption we 
know that for some vertex $u^+$ of $G$, \ $ \bar g(V(Y(u)))= V(Y(u^+))$ and that
\[
g(a(u,v)), g(b(u,v))\in L(u^+) \ \ 
 \text{ and } \ \ g(b(u,v))=g(a(u,v))'.
\]
Thus there is a $w\in V(G)$ with $u^+w\in E(G)$ such that 
\begin{equation}\label{eq:gba}\{g(a(u,v)), g(b(u,v))\}= \{a(u^+,w), b(u^+,w)\}.
\end{equation}
As $uv$ is an edge in~$G$, we know that the only edges with one end in $\{a(u,v), b(u,v)\}$ and one end not  in $Y(u)$ must have the second end in $\{a(v,u), b(v,u)\}$. Accordingly, the only edges with one end in $\{a(u^+,w), b(u^+,w)\}$ and one end not in $Y(u^+)$ have their second end in $\{a(w,u^+), b(w,u^+)\}$.

Therefore, by (\ref{eq:gba}) and since $g$ is an automorphism,
\begin{equation}\label{eq:arrdm}
    \{g(a(v,u)), g(b(v,u)\} = \{a(w,u^+), b(w,u^+)\}.
  \end{equation}

  \noindent
  Now we analyze the cases $ \deg(v)\in \{1,2  \}$.
  First we consider the case $\deg(v)= 1$. 
Then, $A(v)= \{a(v,u)\}$ and $B(v)= \{b(v,u)\}$. Hence $L(v)=A(v)\cup B(v)$ 
contains one vertex, namely $ a(v,u)$, that in $Y(G)$ (or $\tilde Y(G)$) is of degree $1$ and one vertex, namely $ b(v,u)$, that is 
of degree $2$ and whose neighbor not in $ Y(u)$ is the only vertex in $ M(v)$ (cf.~Figure~\ref{fig:X1X2X3}).
Hence one of the vertices in $g(a(v,u)),g(b(v,u))$, i.e., by 
\eqref{eq:arrdm} one of the vertices in $\{a(w,u^+), b(w,u^+)\}$, must have degree 1 and the other one, denote it by $ x$, has degree two. Hence, $ \deg(w)=1$ and thus, the other neighbor of $ x$ must be the only vertex in $ M(w)$. Therefore, $ \bar g(M(v)))= M(w)$, $L(w)= \{a(w,u^+), b(w,u^+)\}= \bar g(L(v))$ and thus $ \bar g(Y(v))= Y(w)$, i.e., $ g$ preserves the gadget $ Y(v)$. Since  by~\eqref{eq:arrdm}, $g(x')=g(x)'$ for 
$ x\in L(v)$, $ g$ preserves the twins of $ Y(v)$.

Finally, we consider the case $\deg(v)=2$ (we argue similarly to the case of degree 1). Then, $\deg(a(v,u))=\deg(b(v,u))=2$ 
and the neighbors of $a(v,u) $ and $b(v,u) $ not in $ Y(u)$ are the vertices of $ M(v)$  (cf.~Figure~\ref{fig:X1X2X3}). Again, by~\eqref{eq:arrdm}, $a(w,u^+)$  and $ b(w,u^+)$  must have  degree 2. Hence, $ \deg(w)=2$ and thus the neighbors of $a(w,u^+)$  and $ b(w,u^+)$ not in $ Y(u^+)$ are the vertices in $ M(w)$. Therefore, $ \bar g(M(v))=  M(w)$.  Thus, $ \bar g(L(v))=  L(w)$ (as the link vertices are the neighbors of the middle vertices) and therefore,  $ \bar g(Y(v))= Y(w)$, i.e., $ g$ preserves the gadget $ Y(v)$. As $g(x')=g(x)'$ holds for $x\in \{a(v,u), 
b(v,u)\}$ (by \eqref{eq:arrdm}), it holds for all $x\in L(v)$ (note 
that $|A(v)|=|B(v)|=2$). Hence,  $ g$ preserves the twins of $ Y(v)$.
\proofend

\medskip 
\begin{rem}
  We showed $Y(G)\not\cong \tilde Y(G)$ using $X(G)\not\cong \tilde X(G)$. We also
can show the result  with an argument similar to the one used in the 
colored case. In fact, let $G$ be a graph of degree at least $3$ and assume there is an isomorphism  
$g:Y(G)\cong \tilde Y(G)$. By the preceding results we have seen that $ g$ preserves all gadgets and its twins. In other words, there is a $ \sigma\in \Aut(G)$ such that
\begin{itemize}
\item[--] $\bar g(Y(u))= V(Y(\sigma(u)))$ \  for all $ u\in V(G)$,
\item[--] $ \{g(a(u,v)), g(b(u,v)) \}= \{a(\sigma(u),\sigma(v)), b(\sigma(u),\sigma(v))$ \ for all $ uv\in E(G)$.
\end{itemize}
 Now consider
\[
F_g:= \bigcup_{u\in V(G)}g(\emptyset_u). 
\]
As $g(\emptyset_u) \in M(\sigma(u))$, the set  $ F_g$ has even cardinality. Counting the number of vertices that 
each \emph{edge} of~$G$ contributes to this union, i.e.,
\begin{align*}
  F_g=\bigcup_{uv\in E(G)}=\{a(u,v) \mid g(a(u,v))=b(\sigma(u),\sigma(v))\} \cup \{a(v,u) \mid g(a(v,u))=b(\sigma(v),\sigma(u)) \},
\end{align*}
we see (similar to the proof of Theorem~\ref{thm:XGiso})
 that every edge  
contributes zero or two vertices except the twisted edge that only 
contributes one. We get the desired contradiction: the union must have even 
and odd cardinality. \hfill{$\dashv$}
\end{rem}

% Polynomial time algorithm accepting the $Y(G)'s$ and rejecting the $\tilde Y(G)'s$ 

\section{Polynomial time algorithm accepting the $Y(G)'s$ and rejecting the 
$\tilde Y(G)'s$}\label{sec:poly}

In  the Introduction we mentioned that the statements (b) and (c) 
(cf.~page \pageref{page:(b)(c)}) show that the Weisfeiler-Leman algorithm is 
not an optimal algorithm to distinguish the original graphs $ X^k$ from the  
$\tilde X^k$'s. The same holds for the uncolored graphs. We show here the 
corresponding  statement (c) and in Section~\ref{sec:tw1} the corresponding 
 statement (b). 
\begin{theo}\label{thm:polythm} 
There is a polynomial time algorithm accepting (all graphs isomorphic to) 
$Y(G)$ for all $G$ and rejecting the corresponding $\tilde Y(G)$.  
\end{theo}
\noindent Recall that $G$ always denotes a connected graph  with at least two 
vertices. Polynomial time refers to the size of the input graph. Note that the
algorithm does not know whether the input graph is a $Y(G)$ or a $\tilde 
Y(G)$ and in the positive case it does not have information on the corresponding $G$.  
We can require that for input graphs that are not isomorphic to one of the forms $Y(G)$ or $\tilde Y(G)$ the algorithm always stops, always accepts, or
always rejects. However, we don't care about this issue. 
\medskip

\noindent \textit{Proof of Theorem~\ref{thm:polythm}:} As $\deg(Y(G)), \deg(\tilde 
Y(G))\ge\deg(G)$, by Lemma~\ref{lem:1-2} it is clear how the desired 
algorithm can treat graphs~$Z$ with $\deg(Z)\in [2]$.

Now let $Z= Y(G)$ or 
$Z= \tilde Y(G)$ be a graph of degree at least three, then, again  by Lemma~\ref{lem:1-2},
\[
 V_{\ge 3}(G):=\{u\in V(G) \mid \deg(u)\ge 3 \}\ne\emptyset. 
\]
First, we show how we get the gadgets $ Y(u)$ with $ u\in  V_{\ge 3}(G)$.  
Consider the binary 
relation $\sim$ on $V(Z)$ given by 
\[
x\sim y \iff \text{$\deg(x)\ge 3$, $\deg(y)\ge 3$, and $x$ and $y$ are on a cycle of length at most eight.}
\]
By Lemma~\ref{lem:cig} and  Lemma~\ref{lem:deg3VYu},  $\sim$ is an equivalence relation with domain $\bigcup_{u\in V_{\ge 3}(G)}V(Y(u))$, whose equivalence 
classes are the gadgets $Y(u)$ with $u\in V_{\ge 3}(G)$. Note that the equivalence classes are the gadgets, however we do not have the corresponding vertices $ u\in G$, as in general, we do not know $ G$. What we have for $ u$ in $ G$ of degree at least 3 is
\begin{itemize} 
\item[--] the set $ V(Y(u))$ and $ \deg(u)$ \ (as $ |V(Y(u))|=2\cdot \deg(u)+2^{\deg(u)-1}$)
\item[--] (by Lemma~\ref{lem:deg3Yustructure}) the set $ L(u)$
\item[--]  (by Lemma~\ref{lem:imgj}) the partition of $ L(u)$ into the sets $ L(u,v) \  (=\{a(u,v), b(u,v) \})$, where $ v$ ranges over the neighbors of $ u$ in $ G$ (again we have the sets of the partition but neither the vertex $ u$ nor its neighbors).
\end{itemize}
We want to get this information for all $ u$ in $ G$. So assume that at some point, the set $V$ is the union of all $ V(Y(u))$ over the $ u$ for  which we  already have the information (i.e., at the beginning the set $ V$  is $\bigcup_{u\in V_{\ge 3}(G)}V(Y(u))$). If the set $ V(Z)\setminus V$ is nonempty, then it contains a vertex, say $ x$, that has a neighbor in some  of the already known $ L(u,v)$. In particular, $ x\in L(v,u)$ and $ \deg(v)\le 2$. Moreover, $ L(v,u)$ besides $ x$ contains the further neighbor not in $ V$ of a vertex in $ L(u,v)$. As $ \deg(v)\le 2$, the neighbors not in $ V$ of the vertices of $ L(v,u)$  are the vertices in $ M(v)$. If $ \deg(v)=1$, then $ V(Y(v))=L(v,u)\cup M(v)$. If $ \deg(v)=2 $, then $ V(Y(v))= L(v,u)\cup M(v) \cup L$ where $ L$ consists of the two vertices not in $  L(v,u)$ that have a neighbor in $ M(v)$. The sets   $L(v,u)$ and $ L$ constitute the desired partition of $ L(v)$ (note that $L=L(v,w)$, where $ w$ is the second neighbor of $ v$.

After finitely many steps we will have $V= V(Z)$ and thus we have the information for all gadgets of~$ Z$.  Hence, the parity of the number of twisted edges would 
allow us to decide whether $Z$ is $Y(G)$ or $\tilde Y(G)$. However, in general, for $uv\in E(G)$  we do not know which vertex 
of $L(u,v)$ is $a(u,v)$ and which is $b(u,v)$; we need this information 
about $Z$ (or about a copy isomorphic to $Z$) to count the twisted edges.

As for $ u,v\in V(G)$,
\[
uv\in E(G)\iff \mbox{$ u\ne v$ and  there is an edge between a vertex of $ Y(u)$ and a vertex of $ Y(v)$,}
\]
we get on the set of gadgets of $ Z$ a graph isomorphic to $ G$ by defining the edge relation $ E$ as follows:
\[
\{Y(u), Y(v) \} \in E \iff \text{$ u\ne v$ and there is an edge between a vertex of $Y(u)$ and a vertex of $Y(v)$}. 
\]

\noindent
In the following we identify this graph with $ G$ (clearly, if the graphs $ G_1$ and $ G_2$ are isomorphic, then so are $Y( G_1)$ and $Y( G_2)$ and  $\tilde Y( G_1)$ and $\tilde Y( G_2)$).

For the purpose of getting the parity of the number of twisted edges, we fix an arbitrary ordering 
$<$ of $V(Z)$. Consider a vertex $u\in V(G)$ of degree at least two. For $v$ 
with $uv\in E(G)$ we decide that the first vertex of $L(u,v)$ in the ordering 
$<$ should play the role of $a(u,v)$ and the second one the role of $b(u,v)$. 
We denote the first one by $\overline{a(u,v)}$ and the second one by 
$\overline{ b(u,v)}$.
We do this for all neighbors of $u$. Then we  set
\[
t:= \{\overline{a(u,v)}\mid v\in V(G), uv\in E(G), \overline{a(u,v)}= b(u,v)\}.
\]
 For all $m\in M(u)$, the set $m\,\Delta\, t$ has even cardinality (if 
$|t|$ is even) or   for all $m\in M(u)$ the set $m\,\Delta\, t$ has odd cardinality (if~$ |t|$ is odd). How can we find out whether $|t|$ is even or not?
Note that for $m\in M(u)$ we have
\[
m\,\Delta\, t \se \{\overline{a(u,v)} \mid v\in V(G) \mbox{ and\ } uv \in E(G)\}
.\]
In fact, assume $ x\in L(u)$ and $ x\in m\,\Delta\, t$. If $ x\in m$ and $x\notin t$, then $ x=a(u,v)$ for some $ v$ and $\overline{a(u,v)}=a(u,v)=x$. If $ x\in t$ (and $ x\notin m$), then $ x=\overline{a(u,v)}$ for some $ v$.
Moreover,
\begin{align*}
\overline{a(u,v)}\, m\in E(Z)\iff & \big(a(u,v)\in m 
 \text{ and\ } \overline{a(u,v)}=a(u,v)\big)\text{ or\ } \big(a(u,v)\notin m  
 \text{ and\ } \overline{a(u,v)}=b(u,v)\big)\\ 
\iff  & \ \overline{a(u,v)}\in m\,\Delta\, t.
\end{align*}
Thus, we get the information whether $ |t|$ is even, by counting,  for an arbitrary $ m\in M(u)$,  the $\overline{a(u,v)} $ with $ \overline{a(u,v)}\, m\in E(Z)$.

If  $ |t|$ is even, we set $ t(u):= t$. If  $ |t|$ is odd we take for example the last link vertex $ x$ for which we set  $\overline{a(u,w)} =x$ for some neighbor $ w$ of $u$ and redefine $\overline{a(u,w)}$ by $\overline{a(u,w)}=x'$, the twin of $ x$. Then we set
\[
t(u):= \{\overline{a(u,v)}\mid v\in V(G), uv\in E(G), \overline{a(u,v)}= b(u,v)\}.
\]
Note that $t(u)= t\setminus \{\overline{a(u,w)} \} $ (if $ x=b(u,w)$) or  $t(u)= t\cup \{\overline{a(u,w)} \} $ (if $ x=a(u,w)$).
Hence, $ t(u)$ has even cardinality.

If the vertex $u$ of $G$ has degree one and $v$ is its neighbor, then we know that $ a(u,v)$ is the vertex of degree 1 in $ L(u,v)$. We set $\overline{a(u,v)}= 
a(u,v)$ and $\overline{ b(u,v)}=b(u,v)$ and let $t(u)$ be the empty set in 
$M(u)$.

Hence for every $u$ of $G$ the mapping $f_{t(u)}$ 
(cf.~Proposition~\ref{pro:autcolgad}) is an automorphism of $Y(u)$ in $Z$. 
Let~$f$ be the union of the functions $f_{t(u)} $, i.e, the  function defined 
on $V(Y(G))$ such that 
\[
\text{for all $x\in Y(G)$, say $x\in Y(u)$:\quad $f(x)=f_{t(u)}(x)$.}
\]
Clearly $f$ is a bijection from $V(Z)$ onto $V(Z)$, whose restriction to 
every gadget $Y(u)$ is an automorphism of $Y(u)$. Denote by $Z^f$ the graph 
induced by this bijection, i.e., $f: Z\cong  Z^f$. Now counting the twisted  
edges of the form $\overline{a(u,v)}\,\overline{b(v,u)}$ for $uv\in E(G)$ 
tells us (whether $Z^f$ and hence) $Z$ are isomorphic to $Y(G)$ or to $\tilde 
Y(G)$. \proofend 

\noindent The following corollary is trivial for the colored CFI-graphs. 
\begin{cor}\label{cor:gdef}
If $Y(G)\cong Y(G')$ for graphs $G$ and $G'$, then $G\cong G'$. 
\end{cor}

\proof If $ \deg(G), \deg(G')\in [2]$, then the result follows from 
Lemma~\ref{lem:1-2}. If, say, $\deg(G)\ge 3$, then $\deg(Y(G))\ge 3$ by~\eqref{eq:degcfi}. Therefore,  $\deg(Y(G'))\ge 3$ and
thus, again by Lemma~\ref{lem:1-2}, $\deg(G')\ge 3$. In the previous proof we 
have seen that then isomorphic copies of $ G$ and $G'$ can be defined in 
$Y(G)$ and $Y(G')$, respectively. As $Y(G)\cong Y(G')$, these copies are isomorphic. \proofend 

\noindent Clearly, the previous corollary holds if we replace $Y$ by $\tilde 
Y$.

% Reinterpreting some results in terms of coloring and first-order definability 
\section{Reinterpreting some results in terms of coloring and first-order definability}\label{sec:fodef}

Various insights we obtained, mainly in the last two sections, are the basis for results concerning the question how far the colors of the colored CFI-graphs are implicitly present in the uncolored CFI graphs. The main result of this section, Theorem~\ref{thm:coldef}, shows that there is a first-order formula $ \varphi(x,y)$ expressing in the uncolored version $ Y(G)$ of $ X(G)$ (and in the uncolored version $\tilde Y(G)$ of $\tilde X(G)$) that vertices $ x$ and~$ y$ have the same color in $ X(G)$ (the same color in $\tilde X(G) $). This result holds for all graphs~$ G$ where every vertex is of degree at least 3.
\medskip

\noindent
We first consider \emph{rigid} graphs.
A graph  $H$ is 
\emph{rigid} (or, asymmetric) if the identity  is the only automorphism of 
$H$. 
\begin{prop}\label{pro:colrxgyg} If $ G$ is rigid, then  $\Aut(Y(G))= \Aut(X(G))$ and thus, automorphisms of $Y(G)$ \emph{preserves} the colors of $ X(G)$, i.e, $ x$ and $ g(x)$ 
have the same color in $X(G)$.
\end{prop}
\proof  Paths and cycles are not rigid. As $ G$ always denotes a connected graph with at least two vertices,  we see that a rigid graph $ G$ must contain a vertex of degree at least three.  Hence, by Theorem~\ref{thm:fifg},  $\Aut(Y(G))= \Aut(X(G))$ for every rigid graph.\proofend

\noindent
For arbitrary graphs $ G$ of degree 3 we get the following result.
\begin{prop}\label{pro:colxgyg} Assume $ \deg(G)\ge 3$. Every automorphism of $Y(G)$ is \emph{``same-color''-preserving}, i.e, if $g\in 
    \Aut(Y(G))$ and $x,y\in V(Y(G))$ have the same color in $X(G)$, then so 
    do $ g(x)$ and $ g(y)$. The same applies to $ \tilde Y(G)$.
  \end{prop}

\proof  So assume $g\in \Aut(Y(G))$.  Recall that $X(G)= 
(Y(G), c)$ where $ c=(c_1,c_2)$ is  
an appropriate coloring (cf.~\eqref{eq:defxg}). Assume $x,y\in V(Y(G))$ 
have the same color in $X(G)$, i.e., $c_1(x)= c_1(y)$ 
(or, $c_1(x)$ and $c_1(y)$ are undefined) and $c_2(x)=c_2 (y)$. By the second 
equality we know that $x,y\in X(u)$ for some $u\in V(G)$. Thus, by 
Lemma~\ref{lem:imgj}, for some $ u^+\in V(G)$ we have $g(x), g(y)\in X(u^+)$ and thus, $ 
c_2(g(x))=c_2(g(y))$. If $c_1(x)$ and $c_1(y)$ are undefined, then $ x,y\in M(u)$. Therefore, $ g(x), g(y)\in M(u^+)$ (again by Lemma~\ref{lem:imgj}) and thus,   $c_1(x)$ and $c_1(y)$ are undefined, too.
Otherwise, $c_1(x)=c_1(y)$ and therefore, $x,y\in L(u)$ and $ y\in 
\{x,x'\}$. Using once more  Lemma~\ref{lem:imgj}, we see that $g(x),g(y)\in L(u^+)$ and $ g(y)\in 
\{g(x),g(x)'\}$. Hence,  $c_1(g(x))= c_1(g(y))$. \proofend

\paragraph{First-order definability.}
In Section~\ref{sec:auts}, for graphs 
$G$ of degree at least $3$ we have seen that every
automorphism is
\begin{itemize}
\item gadget-preserving, i.e., maps vertices of a same gadget onto  vertices of a same gadget,
\item preserves twins, i.e, maps the graph of the function $'$  onto itself (recall that  $'$ maps every link vertex in $ Y(G)$ to its twin),
,\end{itemize}
and thus,
\begin{itemize}
  \item preserves links, i.e., maps link vertices onto link vertices.
 \end{itemize}
 In the  preceding proposition, we showed that such  automorphisms are
 \begin{itemize}
\item ``same-color''-preserving, i.e., map vertices of the same color onto vertices of the same color.
\end{itemize}
Here we address 
the question to what extent the  gadgets, the set of link 
vertices, the graph of the function~$'$, and the set of vertices of the same color  are first-order definable.

The answer is positive if we restrict to graphs $ G$ where \emph{every} vertex has degree at least 3 (a class of graphs that suffices to get the main results of~\cite{caifurimm92}). We state the corresponding result for vertices of the same color in the following theorem. In the course of its proof we derive the corresponding results for the other notions mentioned above.

\begin{theo}\label{thm:coldef}For the class consisting of all the graphs $ Y(G)$ and
$ \tilde Y(G)$ where every vertex of $ G$ has degree at least 3 there is a first-order formula $ \emph{same-color}(x,y)$ expressing (in $ Y(G)$ and
$ \tilde Y(G)$) that $ x$ and~$ y$ are  vertices of the same color in $ X(G)$ and
$ \tilde X(G)$, respectively.
\end{theo}
\proof %
Let $ G$ be an arbitrary graph  where every vertex has degree at least 3.
In the previous section, we considered the binary 
relation $\sim$ on $V(Y(G))$ (= $ V(\tilde Y(G))$) given by 
\[
x\sim y \iff \text{$ x$ and $y$ are on a cycle of length at most 8.}
\]
By Lemma~\ref{lem:deg3VYu} we get
\[
  x\sim y \iff \mbox{$ x$ and $ y$ are in the same gadget}.
\]
Hence, taking as $ \textit{gadget}(x,y)$ a first-order formula expressing that $ x$ and $ y$ are on a cycle of length at most~8, we see that $ \textit{gadget(x,y)}$  expresses that $ x$ and $ y$ are in the same gadget.

The following formula $\textit{link}(x)$ expresses that $ x$ is a link vertex, 
as a vertex is a link vertex if and only if  it has an edge to a vertex of  a distinct gadget.
\[
\textit{link}(x):=\exists y(\neg \textit{gadget}(x,y))\wedge Exy).
\]
Clearly, 
\[
\textit{middle}(x):=\neg\textit{link}(x)
\]
defines the set of all middle vertices.

By Lemma~\ref{lem:li'uni}, the pairs
$(a(u,v), b(u,v))$ and $(b(u,v), a(u,v))$ for $uv\in E(G)$, that is, the pairs on the graph of the function $ '$, are just the pairs that satisfy the  formula \textit{graph[$\ '$]}(x,y), where
\[
  \textit{graph[$\ '$]}(x,y):=\textit{link}(x)\wedge \textit{link}(y)\wedge\textit{gadget}(x,y)\wedge \forall z \big((\textit{gadget}(z,x)\wedge \textit{middle}(z))\to (Exz\leftrightarrow \neg Eyz)\big).
\]
This  set of pairs is precisely  the set of pairs of distinct link vertices of the same color in the corresponding  $ X (G) \ (= (Y(G), (c_1,c_2)))$ or  $ \tilde X (G)$.  Hence, we can set 
\[
\emph{same-color}(x,y):= \Big(x=y \vee \textit{graph[$\ '$]}(x,y) \vee \big(\textit{gadget}(x,y)\wedge\textit{middle}(x)\wedge \textit{middle}(y)\big)\Big).\benda
\]

\begin{rem}\label{rem:nondef}In  the class of graphs $ Y(G)$ where $ G$ is of degree at least 3 (i.e, with  one or more vertices of degree at least 3) we cannot express in first-order logic that vertices $ x$ and $ y$ of $ Y(G)$ have the same color in $ X(G)$. For a contradiction assume that the first-order 
 formula $\textit{sa-col}(x,y)$ does this job. Then there is an~$m$ (depending on the 
 quantifier rank of $\textit{sa-col}(x,y)$) with the following property. Assume that we add at some vertex $u_0$ of  a   graph $ G$ of degree at least 3   a path of length at least $3m$, thereby getting the graph $ G'$. We consider vertices $ u$ and $ v$ on this path that are neighbors, have a distance at least $m$ from~$u_0$, and also a  distance at 
least $m$ from the end of the path. Then in $ Y(G')$ the pairs $ (a(u,v), b(u,v))$ and $ (a(u,v), b(v,u))$ both satisfy $\textit{sa-col}(x,y)$ or both  satisfy $\neg \textit{sa-col}(x,y)$, a contradiction.\hfill{$ \dashv$}
\end{rem}
\medskip

\noindent
Erd\"os  and Rényi~\cite{erdren} showed that almost all graphs are rigid. By results of Fagin~\cite{fag}, we know that
\begin{itemize}
\item[--] almost all graphs are connected
\item[--] almost all graphs only contain vertices of degree at least 3. 
\end{itemize}
Therefore, we obtain the following corollaries from Proposition~\ref{pro:colxgyg} and Theorem~\ref{thm:coldef}.
\begin{cor}For almost all graphs $ G$,
  \[
\Aut(Y(G))=\Aut(X(G)).
\]
\end{cor}
\begin{cor}There is a first-order formula $ \emph{same-color}(x,y)$ that for almost all graphs $ G$ expresses in $ Y(G)$ and
$ \tilde Y(G)$ that $ x$ and~$ y$ are  vertices of the same color in $ X(G)$ and
$ \tilde X(G)$, respectively.
\end{cor}

% The relationship between the $L^k$-equivalence and the $C^k$-equivalence of CFI~graphs

\section{The relationship between the $L^k$-equivalence and the $C^k$-equivalence of CFI~graphs}\label{sec:ck}

One obtains the logic $C$, \emph{first-order logic with counting}, by adding 
to the formula formation rules of first-order logic, here denoted by $L$, the 
following rule: 
\begin{itemize}
\item If $\varphi$ is a formula and $\ell \ge 1$, then $\exists^{\ge 
    \ell}x\, \varphi$ is a formula. 
\end{itemize}
The meaning of $\exists^{\ge \ell}x\, \varphi$ is ``there are at least $\ell$ 
elements $x$ such that $\varphi$ holds.'' 

Of course, the logic $C$ is not more expressive than $L$ as we can express  
$\exists^{\ge \ell}x\, \varphi$ in first-order logic using variables 
$y_1,\ldots, y_\ell$ not occurring in $\varphi$ by 
\[
\exists y_1\ldots \exists y_\ell
 \Big(\bigwedge_{i,j\in[\ell],\, i<j}
  \neg y_i=y_j \wedge \forall x(\bigvee_{i\in[\ell]} x=y_i\to \varphi)\Big).
\]
Things change if we consider the fragments $L^k$ and $C^k$ where $k\ge 1$. 
Here by $L^k$ and $C^k$ we denote the fragment of first-order logic $L$ and the fragment
of first-order logic with counting $C$, respectively, whose formulas contain 
at most $k$ variables. More precisely, we fix a set $\{x_1, \ldots, x_k\}$ of 
$k$ variables and $L^k$ and~$C^k$ contain precisely the formulas of the 
corresponding logic ($L$ or $C$) whose free and bound variables are among 
$x_1,\ldots, x_k$. Note that we can express that a graph has exactly five 
vertices by the $C^1$-sentence $(\exists^{\ge 5}x_1\, x_1=x_1\, \wedge \, 
\neg \exists^{\ge 6}x_1\, x_1=x_1)$ but not by an $L^1$-sentence, not even by 
any $L^5$-sentence. 
\medskip

\noindent If $F$ is a set of formulas of a logic we write $H_1\equiv_{F}H_2$ 
for (colored or uncolored) graphs $H_1$ and $H_2$ and say that $H_1$ and 
$H_2$ are \emph{$F$-equivalent} if they satisfy the same sentences of $F$. The 
significance of the $C^k$-equivalence comes from the result
\[
\text{$H_1\equiv_{C^{k+1}}H_2$}  \iff   \text{the $k$-dimensional Weisfeiler-Leman 
 algorithm does not distinguish $H_1$ and $H_2$}.
\]
In the Introduction, we already mentioned that the Weisfeiler-Leman algorithm
has applications in various areas. In particular, for many classes of graphs 
there is a $k$ such that the isomorphism relation on such a class coincides 
with the indistinguishability in the  
$k$-dimensional Weisfeiler-Leman algorithm. 
\medskip

\noindent First we only consider uncolored graphs, hence only a binary 
relation symbol $E$ for the edge relation is present in the logics $L^k$ and 
$C^k$. Later we also consider the colored CFI-graphs and we  will explain what 
relation symbols we use for the colors.  

\begin{exa}\label{exa:ced2}  
We prove that $Y(P)\not\equiv_{C^2} \tilde Y(P)$ for every path $P$ with at 
least one edge (cf. Figure~\ref{fig:gYP3}). For a vertex $u$ in~$P$ let~$d_u$ 
be its shortest distance to an endpoint of $P$, i.e, its shortest distance to 
a vertex of~$P$ of degree one. In particular, $d_u=0$ for an endpoint $ u$ of $P$. 
By induction on $d$ we introduce a formula $\textit{end}_d(x)$ that in~$ P$ 
defines the set of points  of distance~$d$: 
\begin{eqnarray*}
\textit{end}_0(x):= \exists^{\le 1}y Eyx
 & \text{and} & 
\textit{end}_{d+1}(x)
 := \exists y (Eyx\wedge \textit{end}_{d}(y)) 
  \wedge \forall y( Eyx\to \neg \bigvee_{0\le i< d}\textit{end}_{i}(y))
\end{eqnarray*}
(of course, $\exists^{\le 1}y$ is an abbreviation for $\neg \exists^{\ge 
2}y$). Note that $P_1$ is the unique path satisfying $\forall x\, 
\textit{end}_0(x)$, and for $m\ge 1$, 
\begin{itemize}
\item $P_{2m}$ is the unique path satisfying $\exists^{=1}x \ 
    \textit{end}_m(x)\wedge \forall x \neg \textit{end}_{m+1}(x)$,
  
\item $P_{2m+1}$ is the unique path satisfying $\exists^{=2}x \ 
    \textit{end}_m(x)\wedge \forall x \neg \textit{end}_{m+1}(x)$. 
  \end{itemize}
So let $P:= P_m$ a path with $m\ge 1$ edges. In Lemma~\ref{lem:1-2} we saw 
that $Y(P_m)$ consists of a path $P_{3(m-1)+1}$ and of a path $P_{3(m-1)+3}$ 
and the graph $\tilde Y(P_m)$ consists of two copies of the path 
$P_{3(m-1)+2}$. Therefore, if for example $m$ is odd, then $Y(P_m)$ but not 
$\tilde Y(P_m)$  contains two vertices $u$ with $d_u=(3m-1)/2$, that is, 
$Y(P_m)$ but not $\tilde Y(P_m)$ satisfies the $C^2$-sentence $\exists^{=2}x 
\ \textit{end}_{(3m-1)/2}(x)$. \hfill{$ \dashv$}
\end{exa}

\noindent This ``technique based on the distance from endpoints'' can be used 
to prove a result due to Immerman and Lander~\cite{immlan} that 
$C^2$-equivalence and isomorphism coincide on the class of trees, i.e., we 
have for trees $T_1$ and $T_2$, 
\[
T_1\equiv_{C^2} T_2 \iff T_1\cong T_2.
\]

\medskip

\noindent Note that $P_1\not\equiv_{L^2} P_2$ as $P_2\models \exists x\exists 
y (x\ne y \wedge \neg Exy)$ and $P_1$ is not a model of this sentence. 
Similarly, $P_2\not\equiv_{L^2} P_3$ as $P_3\models \forall  x\exists y(x\ne 
y \wedge \neg Exy)$ and $P_2$ is not a model of this sentence. But are $P_i$ 
and $P_{i+j}$ for $i\ge 3$ and $j\ge 1$ $L^2$-equivalent and 
 are $Y(P_m)$ and $\tilde Y(P_m)$ $L^2$-equivalent?
\medskip

\noindent
 Before we present tools to decide these questions, we consider 
similar examples for the colored CFI-graphs $X(G)$ (and $\tilde X(G)$). For 
every $u\in V(G)$ in the logic we have a unary ``gadget'' relation symbol~$P_u$ interpreted by $X(u)$ and a unary ``link'' relation symbol $L_u$ 
interpreted by $\{a(v,u),b(v,u)\mid v\in V(G),\ uv\in E(G) 
\}$. That is, for $uv\in E(G)$ exactly the vertices in $\{a(u,v),b(u,v)\}$ 
satisfy the formula $P_ux\wedge L_vx$.

\begin{exa}\label{ex:colpat} 
For $m\ge 1$ we show that $X(P_m)\not\equiv_{L^2} \tilde X(P_m)$. By 
Lemma~\ref{lem:1-2} we know that $X(P_m)$ is the disjoint union of  
$P_{3(m-1)+1}$ and $P_{3(m-1)+3}$ (with their 
corresponding colors) and $\tilde X(P_m)$ the disjoint union of 
two copies of $P_{3(m-1)+2} $ (again, with their 
corresponding colors). No vertex of the path $P_{3(m-1)+3}$ has two neighbors 
of the same color. Hence by an $L^2$-sentence we can express that there is a 
path of length at least $3(m-1)+3 $. This sentence holds in $X(P_m)$ but not 
in $\tilde X(P_m)$. Let us be more precise for $m=3$~(see 
Figure~\ref{fig:gYP3}). Assume $V(P_3)= [4]$ and 
$E(P_3)=\{12, 23, 34\}$. Now $X(P_3)$ has a unique path of length 9, namely 
(the first line contains the vertices, the formulas in the second describe 
the colors of the vertices). 
{\small
\[
\begin{array}{c|c|c|c|c|c|c|c|c|c}
\emptyset_{1}  &b(1,2) &b(2,1) &\emptyset_{2}  &b(2,3) &b(3,2) &\emptyset_{3} &b(3,4) &b(4,3)&\emptyset_{4}\\\hline \\[-3mm] 
  P_1x& P_{1}x\! \wedge\! L_2x&P_{2}x\! \wedge\! L_1x & P_2x&P_{2}x\!\wedge\! L_3x &P_{3}x\!\wedge\! L_2x  &P_{3}x   &P_{3}x\!\wedge\! L_4x& P_{4}x\!\wedge\! L_3x&P_{4}x
  \end{array}.
\]}
Now the existence of this path of length (at least) 9 can be expressed by the $ L_2$-sentence
\[
\hspace{2cm}\exists x(P_1x\wedge \exists y(Exy \wedge  P_{1}y \wedge L_2y\wedge \exists x(Eyx\wedge P_{2}x \wedge L_1x\wedge \exists y(\ldots ) ) )). \hspace{2cm}\dashv
\] 
\end{exa}

\noindent So $X(P_m)\not\equiv_{L^2} \tilde X(P_m)$ but does 
$Y(P_m)\not\equiv_{L^2} \tilde Y(P_m)$ hold? We will see in~\eqref{eq:col2c2} that this is not the case for $ m\ge 3$. Moreover, so far we do not know whether there is a relationship between 
the $L^k$-equivalence and the $C^k$-equivalence of CFI-graphs. Among others, 
to answer these question, we use  the characterizations of the 
$L^k$-equivalence by the \emph{$k$-pebble game}  and of the $C^k$-equivalence 
by the \emph{bijective $k$-pebble game}. We recall the definitions of these 
games and the corresponding results. 
\medskip

\noindent
So let $k\ge 1$ and assume that $H$ and $H'$ are both  graphs or both colored 
graphs. The $k$-pebble game and the bijective $k$-pebble game on $H$ 
and~$H'$ are played by two players called the \emph{Spoiler} and the 
\emph{Duplicator}. Each of the two players has $k$ pebbles labelled 
$1,\ldots, k$. A  game consists of a (possibly infinite) sequence of 
\emph{rounds}. 

In each  \emph{round of the $k$-pebble game}, the Spoiler picks up one of his 
pebbles  and places it on an vertex of one of the graphs $H$ or $H'$ (the 
pebble he picks up may have been unused in the game so far or placed on some 
 vertex in an earlier round). The Duplicator answers by picking up her 
pebble with the same label and placing it on a vertex of the other graph. 
Therefore after each round, say the $r$th round, there is a subset 
$I\subseteq [k]$, consisting of the labels of the pebbles used so far. Hence, 
for $i\in I$ there are $u_i\in V(H)$ and $v_i\in V(H')$ on which a pebble with 
label $i$ is placed. We call the pair $(\bar u, \bar v)$ where $\bar u= 
(u_i\, |\, i\in I)$ and $\bar v= (v_i\, |\, i\in I)$ the \emph{position} 
after round $r$. 

The Duplicator \emph{wins} if after each round the mapping given by 
$u_i\mapsto v_i$  for $i\in I$ is a \emph{partial isomorphism from $H$ to $H'$}, that 
is, an isomorphism between the (colored) graphs induced by $H$ on $\{u_i \mid 
i\in I\}$ and induced by $H'$ on $\{v_i \mid i\in I\}$. 

The $k$-pebble game characterizes the $L^k$-equivalence in the following sense.
\begin{prop} 
$H\equiv_{L^k} H'$ if and only if  the Duplicator has a winning strategy for the 
$k$-pebble game on $H$ and~$H'$. 
\end{prop}
\noindent
We generalize the result of Example~\ref{ex:colpat} by showing the corresponding result for trees.
\begin{exa}\label{ex:coltr}Let $G$ be a tree $T$. We show $X(T)\not\equiv_{L^2} \tilde X(T)$.
We choose a vertex $t_1$ of degree 1 and let~$t_2$ be its unique neighbor. We 
can assume that $\tilde X(T)= X(T)^{t_1 t_2}$. We  consider $T$ as a rooted 
tree with root~$t_1$.

To get $X(T)\not\equiv_{L^2} \tilde X(T)$ we show that the Spoiler has a winning strategy in the $2$-pebble game on $X(T)$  and $ \tilde X(T)$. One feature of this winning strategy: the Spoiler places his pebbles always on vertices in $X(T)$ and hence, the Duplicator has to put her pebbles on vertices of $\tilde X(T) $. We do not mention this explicitly below.

The Spoiler in the first round places his first pebble on 
$\emptyset_{t_1} $, in the second round his second pebble on $b(t_1,t_2)$, 
and in the third round his first pebble on $b(t_2,t_1)$. If the Duplicator hasn't lost so far, then after the third round her 
first pebble is on $a(t_2,t_1)$  (in fact, after the first round her first pebble must be on $\emptyset_{t_1} $  and after the second round her second pebble on $b(t_1,t_2)$; as $a(t_2,t_1)$ is the only neighbor of $b(t_1,t_2)$ in the gadget $ X(t_2)$ of $\tilde X(T)$, after the third round her 
first pebble must be on $a(t_2,t_1)$).

Clearly, if $t_2$ is a leaf, then the Spoiler wins by setting his second pebble on 
$\emptyset_{t_2} $. So assume $t_2$ is not a leaf. Now the Spoiler can easily 
play in such a way that a path $t_1, t_2,t_3,...$ of $T$ is obtained where the Spoiler 
passes all $b(t_{i+1},t_{i})$ (in $ X(T)$) and at the same time the Duplicator  has 
to pass $a(t_{i+1},t_{i})$ (in $\tilde X(T)$). Thus the Spoiler wins once a leaf 
$t_{i+1}$ is reached.

More precisely, we already know that this holds for $i=1$. Now assume it 
holds for some $i\ge 1$. We show that it holds for $i+1$. Assume  
the Spoiler's has a pebble, say the second one, on $b(t_{i+1},t_{i})$  
and  the Duplicator's second pebble is on $a(t_{i+1},t_{i})$ . 
The Spoiler places his first pebble on~$\emptyset_{t_{i+1}} $. The Duplicator must 
answer by placing her first pebble on some $m$ in $M(t_{i+1})$ that contains $a(t_{i+1},t_{i})$  (as  
$\{b(t_{i+1},t_{i}),\emptyset_{t_{i+1}}\}\in E( X(T))$). As $|m|$ is even, $m$ must 
contain a  vertex of the form $a(t_{i+1},t)$ with $ t\ne t_{i}$. We set $t_{i+2}:=t$. Now 
the Spoiler puts his second pebble on $b(t_{i+1}, t_{i+2})$. The Duplicator must answer by putting her second pebble on 
$a(t_{i+1}, t_{i+2})$, the only neighbor of $ m$  of the same color as
$b(t_{i+1},t_{i}+2)$.
Finally,   the Spoiler moves his first pebble to $b(t_{i+2}, t_{i+1})$ and clearly 
the Duplicator has to move her first pebble to $a(t_{i+2}, t_{i+1})$. \hfill{$ \dashv$}
\end{exa}

\noindent We leave it to the reader to present a winning strategy for the 
Duplicator in  the 2-pebble game on $P_i$ and $P_{i+j}$ for $i\ge 3$ and $j\ge 
1$ and  a winning strategy for her in the 2-pebble game on $Y(P_m)$ and 
$\tilde Y(P_m)$. Together with Example~\ref{exa:ced2} we get for $m\ge 3$, 
\begin{equation}\label{eq:col2c2}
Y(P_m)\equiv_{L^2}\tilde Y(P_m) \ \ 
 \text{ but } \ \ Y(P_m)\not\equiv_{C^2}\tilde Y(P_m).
\end{equation}

\noindent So far we do not know whether there is a relationship between 
the $L^k$-equivalence and the $C^k$-equivalence of colored CFI-graphs. Among others, 
to answer this question, we use  the characterizations of the 
$L^k$-equivalence by the $k$-pebble game  and of the $C^k$-equivalence 
by the bijective $k$-pebble game. We recall the definition of the bijective $k$-pebble game.
\medskip

\noindent
 Let $k\ge 1$ and assume that $H$ and $H'$ are both  graphs or both colored 
graphs. We already mentioned that the  bijective $k$-pebble game on $H$ 
and~$H'$ is played by two players called the \emph{Spoiler} and the 
\emph{Duplicator} and that they have $k$ pebbles labelled 
$1,\ldots, k$. A  game consists of a (possibly infinite) sequence of 
\emph{rounds}.

In each \emph{round of the bijective $k$-pebble} game between  $ H$ and $ H'$  first the 
Spoiler picks up one of his pebbles, say the $ i$th pebble with $ i\in [k]$ (the 
 $ i$th pebble he picks up may have been unused in the game so far or placed on some 
vertex of $ H$ in an earlier round).
The Duplicator answers by presenting a bijection $ f:V(H)\to V(H')$ (if $ |V(H)|\ne |V(H')|$ the Spoiler wins).
Now the Spoiler places his $ i$th pebble on some vertex $ u_i$ of $ H$ and the Duplicator must place her  $ i$th pebble on $v_i:= f(u_i)$.

Note that after each round there is a subset $ I\se [k]$ consisting of the labels of the pebbles placed on a vertex. Hence, for $ i\in I$ there are a $ u_i\in V(H)$ and a $ v_i\in V(H')$ such that the $ i$th pebble of the Spoiler is on $ u_i$ and the $ i$th pebble of the Duplicator  is on $ v_i$. We call the pair $(\bar u, \bar v)$ where $\bar u= 
(u_i\, |\, i\in I)$ and $\bar v= (v_i\, |\, i\in I)$ the \emph{position} 
after this round. 

The Duplicator \emph{wins} if after each round the mapping given by 
$u_i\mapsto v_i$  for $i\in I$ is a partial isomorphism from $H$ to $H'$. %

The bijective $k$-pebble game characterizes the $C^k$-equivalence.
\begin{prop} 
$H\equiv_{C^k} H'$ if and only if the Duplicator has a winning strategy for the 
bijective $k$-pebble game on $H$ and~$H'$. 
\end{prop}

\noindent
We use the characterizations of the $ L^k$-equivalence and the $ C^k$-equivalence by their corresponding games to derive the following result (a result 
already mentioned at least implicitly in~\cite{caifurimm92}). 
\begin{prop}\label{pro:lkick}
For $k\ge 1$,
\[
\text{$X(G)\equiv_{L^k}\tilde X(G)$ \ implies $X(G)\equiv_{C^k}\tilde X(G)$}.
\]
\end{prop}
\noindent We have seen (cf.~\eqref{eq:col2c2}) that there are graphs $G$, 
namely paths, such that $Y(G)\equiv_{L^2}\tilde Y(G)$ and 
$Y(G)\not\equiv_{C^2}\tilde Y(G)$. However, is there for some $k\ge 3$ a graph $G$ 
 such that the implication of Proposition~\ref{pro:lkick} fails for 
the uncolored CFI-graphs? As far as we know this is still an open question. 
In other words, it is open whether for $ k\ge 3$ by a variant of the following proof of 
this proposition one can get the result for uncolored graphs.

\medskip 
\noindent \textit{Proof of Proposition~\ref{pro:lkick}.} For $k=1$ we have $X(G)\equiv_{C^1} \tilde X(G)$ for all $G$. In fact, in 
$C^1$ we essentially can only express how many vertices exists of a given 
color and these are the same in $X(G)$ and $\tilde X(G)$. Thus the 
implication of Proposition~\ref{pro:lkick} holds for $k= 1$. The case $ k=2$ will be treated in the next section.

So we assume $k\ge 3$ 
and let $G$ be a graph. We will make use of the following claims. 

\medskip
\noindent \textit{Claim 1}. Let $u,v\in V(G)$. If in a play of the $k$-pebble 
game on $X(G)$ and $\tilde X(G)$  the Duplicator applies her winning strategy   
and after some round a position is reached where the pebbles are on 
\[
\text{$\ldots a(u,v)\ldots$ \ in $X(G)$ \ \ and  
 $\ldots b(u,v)\ldots$ \ in $\tilde X(G)$},
\]
then also
\[
\text{$\ldots b(u,v)\ldots$ \ in $X(G)$ \ \ and 
 $\ldots a(u,v)\ldots$ \ in $\tilde X(G)$},
\]
is a winning position for the Duplicator. 

\medskip
\noindent \textit{Proof of Claim 1}. By~the definition of $X(d)$,  $ X(G)$, and  $\tilde X(G)$ this 
replacement always leads from a partial isomorphism to a partial isomorphism 
(cf.~\eqref{eq:m-edges}, \eqref{eq:ccg}, \eqref{eq:eun}, and~\eqref{eq:etw}).  \hfill$\dashv$  
\medskip

\noindent \textit{Claim 2}. Assume that at some point of a $k$-pebble game a 
position is reached where the pebbles are on  (for a vertex $ v$ of $ G$ we denote by $ m(v), m'(v), \ldots $ vertices in $ M(v)$, the set of middle vertices of the gadget $ X(v)$)
\begin{equation}\label{eq:posit}
\text{$\ldots m(v)\ldots  m'(v)\ldots$ \ in $X(G)$ \ \ 
 and $\ldots m_1(v)\ldots  m_1'(v)\ldots$ \ in $\tilde X(G)$}
\end{equation}
with $m(v) \Delta m_1(v)\ne m'(v) \Delta m'_1(v)$. Then  the Spoiler can win the 
game.  
\medskip

\noindent \textit{Proof of Claim 2}: If the positions of the pebbles 
in~\eqref{eq:posit} do not yield a partial isomorphism, the Spoiler already wins. 
So assume they correspond to a partial isomorphism. Set $m_0(v)= m(v) 
\Delta m_1(v)$ and $m'_0(v)= m'(v) \Delta m'_1(v)$. By assumption, $m_0(v)\ne  
m'_0(v)$, say $a(v,w)\in m_0(v)\setminus m'_0(v)$. 

Assume the pebbles labelled $s$ are on $m(v)$ and $m_1(v)$ and the pebbles 
labelled $t$ are on $m'(v)$ and $m'_1(v)$. The Spoiler chooses a further 
pebble and puts it on $a(v,w)$ in $ X(G)$. Due to the color of $a(v,w)$ the Duplicator has 
to put her corresponding pebble on $a(v,w)$ or on $b(v,w)$ in $\tilde X(G)$. We show that in 
both cases her response does not lead to a partial isomorphism. 

First we consider the case where the Duplicator puts her pebble on $a(v,w)$. As 
$a(v,w)\in m_0(v)$, we know that $(a(v,w)\in m(v)\iff a(v,w)\notin m_1(v))$, 
or equivalently, 
\begin{equation*}\{a(v,w), m(v)\} \in E(X(G))\iff \{a(v,w), m_1(v)\}\notin 
E(\tilde X(G)),
\end{equation*}
which shows that the Duplicator's answer doesn't lead to a partial isomorphism. 
Now assume the Duplicator puts her pebble on $b(v,w)$. As $a(v,w)\notin m'_0(v)$, 
we know that $(a(v,w)\in m'(v)\iff  a(v,w)\in m'_1(v))$. Hence, $(\{a(v,w), 
m'(v)\} \in E(X(G))\iff \{a(v,w), m'_1(v)\}\in E(\tilde X(G))$ or 
equivalently,    
\begin{equation*}\{a(v,w), m'(v)\} \in E(X(G))\iff \{b(v,w), 
m'_1(v)\}\notin E(\tilde X(G),
\end{equation*}
which again shows that Duplicator's answer doesn't lead to a partial 
isomorphism. \hfill$\dashv$
\medskip

\noindent By this  claim  we have shown that in case that the position in \eqref{eq:posit} is a winning position for the Duplicator in the $ k$-pebble game, then we have $ m_1(v)=f_{m(v)\Delta m_1(v)}(m(v))$ (by the definition of $f_{m(v)\Delta m_1(v)} $, see Corollary~\ref{cor:uniquefm}) and $ m_1'(v)=f_{m'(v)\Delta m'_1(v)}(m'(v))=f_{m(v)\Delta m_1(v)}(m'(v))$ by Claim 2. We generalize this insight in the next claim.
\medskip

\noindent \textit{Claim 3.} If in a play of the $k$-pebble 
game on $X(G)$ and $\tilde X(G)$  the Duplicator applies her winning strategy   
and after some round a position is reached where for some $ m(v), x,m_1(v),y\in M(v)$ the pebbles are on 
\begin{equation*}
\text{$\ldots m(v)\ldots  x\ldots$ \ in $X(G)$ \ \ 
 and $\ldots m_1(v)\ldots  y\ldots$ \ in $\tilde X(G)$},
\end{equation*}
then $ y=f_{m(v)\Delta m_1(v)}(x)$.
\medskip

\noindent \textit{Proof of Claim 3}: To simplify the presentation let us assume that $ m(v)=\emptyset_{v}$ and thus, $f_{m(v)\Delta m_1(v)} = f_{m_1(v)}$.
If $ x\in M(v)$, say $ x=m'(v)$, then we  just saw in Claim 2 that $ y= f_{m_1(v)}(m'(v))=f_{m_1(v)}(x)$.

Now, assume that $ x\in \{a(v,w) ,b(v,w) \}$ for some $ w$ with $ vw\in E(G)$. As the above position is a partial isomorhism, we get  $ y\in \{a(v,w) ,b(v,w) \}$ (by the colors) and
\[
\{x, \emptyset_{v}\}\in E(X(G)) \iff \{y,m_1(v)\}\in E(\tilde X(G)).
\]
As $f_{m_1(v)} $ is an isomorphism of the gadget $ X(v)$, we know that $ f_{m_1(v)}(x)\in \{a(v,w) ,b(v,w) \}$ and
\[
\{x, \emptyset_{v}\}\in E(X(G)) \iff \{f_{m_1(v)}(x),m_1(v)\}\in E(\tilde X(G)).
\]
Hence by~\eqref{eq:m-edges}, $ f_{m_1(v)}(x)=y$. \hfill{$\dashv$}
\medskip

\noindent \textit{Claim 4}. If in a play of the $k$-pebble 
game on $X(G)$ and $\tilde X(G)$  the Duplicator applies her winning strategy   
and after some round a position 
\begin{equation}\label{eq:posit1}
\text{$\ldots \ldots  \emptyset_{v} \ldots \ldots$ \ in $X(G)$ \ \ 
 and $\ldots \ldots \  m_1(v) \ \ldots\ldots $ \ in $\tilde X(G)$}
\end{equation}
 for some $ v$ in $ G$ and some $ m_1(v)\in M(v)$ is reached, then for each $ m(v)\in M(v)$
\begin{equation}\label{eq:posit2}
\text{$\ldots \ldots \  m(v) \ \ldots \ldots$ \ \ \  in $X(G)$ \ \ 
 \   and \ \ $\ldots \ldots m(v)\Delta m_1(v)\ldots\ldots $ \ in $\tilde X(G)$}
\end{equation}
is a winning poisition for the Duplicator in the $k$-pebble 
game.
\medskip

\noindent \textit{Proof of Claim 4}: As $ f_{ m(v)\Delta(m(v)\Delta m_1(v)) }=f_{m_1(v)}$, by the previous claim we see that \eqref{eq:posit2} represents a partial isomorphism as \eqref{eq:posit1} represents one.

Assume that the relevant pebble is the  $ s$th pebble (e.g., the $ s$th pebbles are  on $\emptyset_{v} $ and $ m_1(v)$ in \eqref{eq:posit1}).
As long as the $ s$th pebble is not moved by the Spoiler in \eqref{eq:posit2},  moves of the Spoiler in \eqref{eq:posit2} are answered by the Duplicator as he would do in \eqref{eq:posit1}. Again by the previous claim the position obtained from  \eqref{eq:posit2}  represents a partial isomorphism.  If the Spoiler moves the 
$ s$th pebble in \eqref{eq:posit2}, again the Duplicator answers moving the $ s$th pebble as he would do in \eqref{eq:posit1}. Then the positions obtained from \eqref{eq:posit1} and \eqref{eq:posit2} are identical. \hfill{$ \dashv$}

\medskip

\noindent Now we present a winning strategy for the Duplicator in the bijective  
$k$-pebble game. It suffices to show that the Duplicator can play in such a way 
that after every round of a bijective $k$-pebble play the pebbles are in a 
position that is a winning  position for the Duplicator in the $k$-pebble game. 

We fix an arbitrary ordering of $ V(X(G))$ with the property that there is an initial segment consisting exactly of all vertices $ \emptyset_{v}$ for $ v$ in $ G$. For $ n\le |V(X(G))|$ we let $ A_n$ be the set of the 
first~$n$ vertices in this ordering. In particular, $A_0= \emptyset$ and $ A_{|V(X(G))|}= |V(X(G))|$.

We assume that after a certain round in the bijective   
$k$-pebble game we have a position $ P$ that is a winning  position for the Duplicator in the $k$-pebble game. 

Suppose the Spoiler  in the bijective  
$k$-pebble game
chooses the pebble labelled~$s$. The Duplicator must answer with a bijection between $ V(X(G)) $ and $V(\tilde X(G))) $.
We 
stepwise define sets $B_0, B_1,\ldots \ (\se  V(\tilde X(G)))$ with $\emptyset= B_0\subset B_1\subset 
\ldots$ and where each $B_{n+1}\setminus B_n$ is a singleton. Clearly, $ B_{|V(X(G))|}=V(\tilde X(G))$. The bijection offered by the Duplicator to the Spoiler will be $ x_n\mapsto y_n$ where $A_{n+1}\setminus A_n=\{x_n \}$ and $B_{n+1}\setminus B_n=\{y_n \}$. The general idea: if $ x\in A_{n+1}\setminus A_n$, then  $B_{n+1}=B_n\cup \{y  \}$ where $ y$ is the answer of the Duplicator in position $ P$ if the Spoiler sets the $ s$th pebble on $ x$. By Claim 1 -- Claim 4 the new position will be a winning position  for the Duplicator in the $k$-pebble game. Let us see the details.

If $\emptyset_{v}$ is in $A_{n+1}\setminus A_n$, then $B_{n+1}\setminus B_n=\{ m(v) \}$, where $ m(v)$ is the answer in the $ k$-pebble game of Duplicator according to its winning strategy if the Spoiler in the position $ P$
puts his $ s$th pebble on $ \emptyset_{v}$.

If $m_1(v)\in M(v)$ with $ m_1(v)\ne  \emptyset_{v}$ is in $A_{n+1}\setminus A_n$, then $B_{n+1}\setminus B_n=\{f _{m(v)}(m_1(v)) \}$, where $ m(v)$ was the Duplicator's response to $\emptyset_{v}$ (note that $\emptyset_{v}$ preceeds $m_1(v)$ in the ordering).

Assume for some  $ uv\in E(G)$ the vertex $a(u,v)$ is in $A_{n+1}\setminus A_n$ and $b(u,v)\notin 
A_n$.  If  in the $k$-pebble game the Spoiler in position $ P$ puts his $ s$th pebble on  $a(u,v)$, then $B_{n+1}$ contains Duplicator's answer, i.e., $B_{n+1}$ contains this answer and  the vertices of $B_n$. 
If $b(u,v)$ is already in $A_n$, then $B_{n+1}$ contains the 
vertex in $\{a(u,v), b(u,v)\}$ not already in $B_n$. The
case ``$b(u,v)$ is in $A_{n+1}\setminus A_n$'' is treated similarly.
\medskip

\noindent
By construction we see that  $ B_{|V(X(G))|}=V(\tilde X(G))$. As already mentioned, in position $ P$ of the bijective $ k$-game the Duplicator answers, when the Spoiler chooses the $ s$th pebble, with the bijection  $ x_n\mapsto y_n$ where $A_{n+1}\setminus A_n=\{x_n \}$ and $B_{n+1}\setminus B_n=\{y_n \}$. If now Spoiler puts the $ s$th pebble on $ x_i$, then the Duplicator puts it on~$ y_i$. By Claim 1 -- Claim 4 the resulting position is a winning position for the Duplicator in the $ k$-pebble game. 
\medskip

\noindent
We will address the case $k= 2$ at the end of the next section. \proofend

% Tree-width I and II

\section{The role of tree-width. I}\label{sec:tw1}

In this section we derive a first relationship between the tree-width of a 
graph $G$ and the $C^k$-equivalence of $X(G)$ and $\tilde X(G)$ (and of 
$Y(G)$ and $\tilde Y(G)$).

First we recall the definition of the tree-width of a (colored) graph $ H$.  A \emph{tree decomposition} of $H$ is a pair $(T,B)$, 
where $B= \{B_t\}_{t\in V(T)}$ is a family of nonempty  subsets (called \emph{bags}) of 
vertices of $H$ indexed by the nodes $t$ of the tree~$T$. Moreover, $(T,B)$ 
must satisfy the following two properties. 
\begin{itemize}
\item For every edge $uv$ of $ H$ there is a node $t\in V(T)$ such that
    $u,v\in B_t$.

\item For every vertex $v$ of $ H$ the set of nodes $t\in V(T)$ with $v\in 
    B_t$ is nonempty and connected in $T$. 
\end{itemize}
The \emph{tree-width of the tree decomposition $(T,B)$} is equal to $(\max_{t\in 
V(T)}|B_t|)- 1$ and the \emph{tree-width $\tw(H)$ of  $H$} is the 
minimum tree-width of all tree decompositions of $H$.
\smallskip

\noindent One easily verifies that $\tw(H)\le |V(H)|-1$ and that the tree-width 
of a tree (with at least one edge)  is $1$. 
\medskip

\noindent In this section we show the following result. 
\begin{theo}\label{thm:crtw} 
Let $k\ge 1$ and $G$ a graph. If $\tw(G)\ge k$, then $X(G)\equiv_{C^k}\tilde 
X(G)$. 
\end{theo}%
\noindent Dawar and Richerby~\cite{dawric} proved this result under the additional assumption that every vertex of $ G$ has at least degree 2  
(previously  Cai et al.~\cite{caifurimm92} showed the result   for the class 
of graphs with an appropriate separator). % Therefore, 

 \paragraph{The cops-and-robber game.}
We introduce the cops-and-robber game that characterizes the tree-width of graphs.
Fix some integer $k\ge 1$ and a graph $G$. \emph{The $k$ cops and the robber 
game  on $G$}, is played by two players, the Cops (consisting of  $ k$ cops) and the Robber. A 
\emph{position} of the game is a pair $((v_i\, |\,  i\in I),u)$, where $I\se 
[k]$, $v_i\in V(G)$ for $i\in I $, and $u\in V(G)$. Intuitively, it tells us 
that the $i$th cop is on the vertex~$v_i$ if $i\in I$, and not  on the graph 
if $i\in ([k]\setminus I)$, and that the Robber is on the vertex $u$.

The play proceeds in rounds. In the first round the Cops choose an $i\in [k]$ and a vertex $v_i$ of $G$. 
Then the Robber chooses some vertex $u \in V(G)$ with $v_i\ne u$ (recall that $ G$ has at least two vertices). The position 
after the first round  is then $((v_j\, |\, j\in I), u)$ with $ I:=\{i \}$. 

Every further round consists of the following two steps. Assume the current 
position is $((v_i \, |\, i\in I ),u)$. First the Cops choose some $j\in [k]$ 
and $v_j'\in V(G)$. Then the Robber chooses a vertex~$ u'$ such that there 
exists a path  from~$u$ to $u'$ in $G\setminus \{v_i \mid i\in (I\setminus 
\{j\})\}$ (the path the Robber uses to arrive at $ u'$ from~$ u$ after the $ j$th cop left his actual position and before he arrives at $ v_j'$). After this round the new  position is $((v_i' \, |\, i\in I\cup\{j \} 
),u')$, where $v_i'=v_i$ for $i\in (I\setminus \{j\})$. 
If some cop is on the vertex $u'$ of the Robber, i.e., if $u'\in \{v_i' \mid 
i\in I\cup\{j \}\}$, then the Cops win.

If there is no position of the play 
such that the Cops win, then the Robber wins.

\medskip
\noindent The following result is due to Seymour and Thomas~\cite{seytho93}. 
\begin{theo}\label{thm:crst} 
The Robber has a winning strategy for the $k$ cops and robber game on $G$ if 
and only if $\tw(G)\ge k$. 
\end{theo}
\noindent We make use of this theorem in the following proof. 

\medskip

\noindent \textit {Proof of Theorem~\ref{thm:crtw}.} Let $u_0u_0'\in 
E(G)$ and $\tilde X(G)=X^{u_0u_0'}(G)$. For $x\in V(X(G))\ (=V(\tilde X(G)))$ 
let $p(x)$ be the vertex of $G$ such that $x\in  X(p(x))$. 

As $\tw(G)\ge k$, the Robber has a winning strategy in the $k$ cops and 
robber game on $G$. We show that the Duplicator has a 
winning strategy for the  bijective $k$-pebble game based on information from the 
Robber's winning strategy. 

Assume that in the first round of a play of the bijective $k$-pebble game on $X(G) $ and $ \tilde X(G)$, the Spoiler  
chooses his $i$th pebble for some $ i\in [k]$. Now the Duplicator has to answer with a bijection  between $ V(X(G))$ and $ V(\tilde X(G))$.
Of course, he has no idea on what vertex $ x$ of $ X(G)$ the Spoiler will put his $ i$th pebble. To get this bijection, for every $ v\in V(G)$ the Duplicator looks at the answer of the Robber according to his  winning strategy if in the first round  of a play of the $k$ cops and the robber game on $G$ the $ i$th cop moves to $ v$. This will provide the Duplicator with the values of the required bijection for all $ x$  in the gadget $ X(v)$.

So let $ v$ be an arbitrary vertex of $ G$. We start  a play of the $k$ cops and robber game where in the first round the Cops
 choose the $ i$th cop and the vertex $v$ (the vertex to which the  $ i$th cop wants to go). According to its winning 
strategy  the Robber answers by going to a vertex $u=u(v)$; in particular, we know 
that
\begin{equation}\label{eq:crsd2}u\ne v
\end{equation}
  If $u\ne u_0$, we choose a path from $u$ to $u_0$, say $u= 
  u_1,u_2,u_3,u_4=u_0$  (we leave the general case to the reader). By repeatedly applying, first Lemma~\ref{lem:twist2adjacentedges} and then~\eqref{eq:zwgl},
 we get (arguing as in the proof of  Lemma~\ref{lem:twist2nonadjacentedges}) 
\begin{align*}
X^{uu_2}(G) \cong X^{uu_2\ uu_2\ u_2u_3\ u_2u_3\ u_3u_4\ u_3u_4\ u_0u_0'}(G)= X^{u_0u_0'}(G) = \tilde X(G).
\end{align*}
Moreover, by Lemma~\ref{lem:twist2adjacentedges} there is an
isomorphism $f^v$,
\begin{equation}\label{eq:crsd1}f^v: X^{uu_2}(G)\cong X^{u_0u_0'}(G),
\end{equation}
that is the identity 
on all gadgets $X(w)$ for $w\notin\{u_2,u_3,u_4\}$.
If $ u=u_0$, we choose $ u_2:=u_0'$ and as
$f^v$ the identity. Then, again, $f^v$ is an isomorphism  from $X^{uu_2}(G) \ (= X^{u_0u_0'}(G))$ onto $X^{u_0u_0'}(G)$.
 \medskip

\noindent
As $ v\in V(G)$ was arbitrary, we define $ h$ on $ V(X(G))$ by 
\begin{equation}\label{eq:crsd4}\mbox{ for $ x\in V(X(G))$}, \ \ h(x):=f^{p(x)}(x)
\end{equation}
(recall that $ p(x)$ is the vertex of $ G$ with $x \in X(p(x))$).

Then  $ h$ is a bijection between $ V(X(G))$ and $ V(\tilde X(G))$ as  by~\eqref{eq:crsd1},  $f^{p(x)}$ maps $ X(p(x))$ onto $ X(p(x))$.
\medskip

\noindent
Recall that we considered a play of  the bijective $k$-pebble game on $X(G) $ and $ \tilde X(G)$ where in the first round  the Spoiler  
chooses his $i$th pebble for some $ i\in [k]$.
Now the Duplicator answers  with the bijection~$ h$. Then the Spoiler puts his $ i$th pebble on some vertex $ x_i$ of $ X(G)$ and the Duplicator must put her $ i$th pebble on
\begin{equation}\label{eq:crsd8}
h(x_i) \ (= f^{p(x_i)} (x_i))   
\end{equation}
We consider now the play of the $ k$ cops and the robber game where in the first round the $ i$th cop moves to $v:= p(x_i)$ and the Robber  according to his winning strategy moves to $ u=u(v)$.
\medskip

\noindent 
Now, it suffices to show that throughout the play of the bijective $ k$-pebble game, the Duplicator is able to 
maintain the following invariant. Let $(\bar x, \bar y)$, where $\bar x= 
(x_i\, |\, i\in I)$ and $\bar y= (y_i\, |\, i\in I)$, be the current position 
in the bijective $k$-pebble game. Then (a) and (b) hold.
\begin{itemize}
\item[(a)] In the corresponding $k$ cops and the robber game, which the Robber plays 
  according to his winning strategy, the actual position is $ ((p(x_i)\, | \, i\in I),u)$, i.e, for $ i\in I$ the $ i$th cop is on $p(x_i) $ and the Robber on $ u$. 

\item[(b)] For some $u'$ with $uu'\in E(G)$  there is an isomorphism  
  $f$,
\begin{equation}\label{eq:crsd5}f:X^{uu'}(G)\cong X^{u_0u_0'}(G) %
\end{equation}
and
\begin{equation}\label{eq:crsd3}
  y_i=f(x_i) \mbox{ \ for all $ i\in I$}.
\end{equation}
\end{itemize}
The statements in (a) and (b) guarantee that $x_{i}\mapsto y_{i}$ for $i\in I$ is a partial isomorphism from 
  $X(G)$ to $\tilde X(G)$ (and thus the Duplicator wins the bijective $ k$-pebble game). In fact,
from~(b) we get (as $X^{uu'}(G)$ is obtained from $ X(G)$ by adding and deleting some edges between the gadgets $ X(u)$ and $ X(u')$)
\[
f: X(G)\setminus V(X(u))\cong X^{u_0u_0'}(G)\setminus V(X(u)). 
\]
As the Robber applies its winning strategy, by (a) we know that $ u\notin \{p(x_i) \mid i\in I \}$.
Therefore, by~\eqref{eq:crsd3}  $x_{i}\mapsto y_i$ for $i\in I$ is a partial isomorphism from 
  $X(G)$ to $\tilde X(G)$.
\medskip

\noindent We  already have seen that the invariant is satisfied after the 
first round of the play of the bijective $k$-pebble game by \eqref{eq:crsd1} and \eqref{eq:crsd8}  taking $ u_2$ as $ u'$ and $ f^{p(x_i)}$ as $ f$.
\medskip

\noindent
Now assume that the actual 
position $(\bar x, \bar y)$ of the cardinality $k$-pebble game, where $\bar x= (x_i\, |\, 
i\in I)$ and $\bar y= (y_i\, |\, i\in I)$, the actual position $((p(x_i) \, |\,  i\in I),u)$ of the corresponding $k$ cops and the robber game, together 
with some vertex $u'$ and function $f$ satisfy the invariant, i.e., (a) and (b) hold. 

In the next round of the bijective $k$-pebble play let the Spoiler  choose his $j$th 
pebble, where $j\in [k]$. The Duplicator proceeds as above to get the desired bijection.
That is, let $ v$ be an arbitrary vertex of~$ G$. In the actual position $((p(x_i) \, |\,  i\in I),u)$ of the $ k$ cops and the robber game   let the Cops choose the vertex~$ v$ for the $j$th cop. The Robber, using his winning strategy answers by getting from his actual positioin $ u$ to a vertex $ w= w(v)$ of $ G$ where $ w\notin \{p(x_i) \mid i\in I\setminus \{j \}\}\cup\{v  \}$.
 \begin{itemize}
\item If $w\ne u$, we know  that there is a path from $w$ to 
  $u$ that avoids the vertices in $\{p(x_i) \mid i\in I\setminus \{j \}\}$, say $w=w_1,w_2,\ldots, w_{m-1}, w_m =u$. As above (cf.~\eqref{eq:crsd1}) we see that there is an isomorphism $g^v$,
\[
g^v: X^{w w_2}(G)\cong X^{uu'}(G),
\]
that is the identity on all gadgets $X(v')$ for 
$v'\notin\{w_2,\ldots, w_{m}\}$; in particular, it is the identity on the gadgets $ X(p(x_i))$ for $i\in I\setminus \{j \}\} $. By (b), we get
\begin{equation}\label{eq:crsd6}
f\circ g^v: X^{ww_2}(G)\cong 
    X^{u_0u_0'}(G).
  \end{equation}
  Hence, (for all $ v$ in $ G$)
  \begin{equation}\label{eq:crsd7}
    (f\circ g^v)(x_i)= f(x_i) \ \ \mbox{for $i\in I\setminus \{j \}$}.
    \end{equation}
 \item If $w= u$, we set $ w_2=u'$  and take  as $g^v$ the 
    identity on $X^{uu'}(G)$. Again, \eqref{eq:crsd6} and \eqref{eq:crsd7} hold.
  \end{itemize}
  Finally we define $ h$ on $ V(X(G))$ by
 \[
\mbox{ for $ x\in V(X(G))$}, \ \ h(x):=(f\circ g^{p(x)})(x).
\]
Again $ h$ is a bijection between $ V(X(G))$ and $ V(\tilde X(G))$. Now Duplicator answers in the bijective $ k$-pebble game with the bijection $ h$.
Next the Spoiler puts his $ j$th pebble on some vertex $ x_j'$ of $ X(G)$.  Then  Duplicator has to put her $ j$th pebble on $h(x_j') \ (=f\circ g^{p(x_j')})(x_j')$.

Hence, the new position of the cardinality $k$-pebble game is $((x_i' \, |\, i\in 
 I'), (y_i'\, |\, i\in I'))$. Here  $I'= I\cup \{j\}$,
 \[
\mbox{for $ i\in (I\setminus \{j \})$, \ \  $ x_i':=x_i$ and  $y_i':=y_i=f(x_i)=(f\circ g^{p(x_j')})(x_i)$ \ \ (by (b) and \eqref{eq:crsd7}),}
\]
and $y_j':= (f\circ g^{p(x_j')})(x_{j}')$.
\medskip

\noindent
In the  corresponidng $k$ cops and the
robber game the $ j$th cop addresses the vertex $v:=p(x_j')$ and according to its winning strategy the Robber moves to the vertex $ w=w(v)$. That, is  $((p(x_i') \, | \, i\in I'), w)$ is the new position in the $k$ cops and the
robber game.

 Together with $ww_2$ for $uu'$ and $f\circ g^{p(x_j')}$ for $f$ they 
satisfy the invariant, i.e., (a) and (b).\proofend
\bigskip

\noindent
Recall that  Proposition~\ref{pro:lkick} of the previous  section claims that for $ k\ge 1$,
\[
X(G)\equiv_{L^k}\tilde X(G) \ \ \mbox{ implies} \ \ X(G)\equiv_{C^k}\tilde X(G).
\]
There we proved it for $ k\ne 2$. For $ k=2$ we can argue as follows. If $ \tw(G)\ge 2$, then $  X(G)\equiv_{C^2}\tilde X(G)$ by Theorem~\ref{thm:crtw}. Therefore, the implication holds.
If 
 $ \tw(G)= 1$, then $ G$ is a tree. In Example~\ref{ex:coltr} we have seen that
$ X(G)\not\equiv_{L^2}\tilde X(G)$ for trees $ G$, so again the implication holds.

\section{The role of tree-width. II}\label{sec:tw2} 
By~Theorem~\ref{thm:crtw} we know that for $k\ge 1$, 
\begin{equation*}\tw(G)\ge k \  \text{ implies} \ \ X(G)\equiv_{C^k} \tilde X(G).
\end{equation*}
Clearly, $X(G)\equiv_{C^k} \tilde X(G) \  \text{ implies} \  Y(G)\equiv_{C^k} 
\tilde Y(G)$. Hence, 
\begin{equation}\label{eq:driy}\tw(G)\ge k \  \text{ implies} \  Y(G)\equiv_{C^k} \tilde Y(G).
\end{equation}
Roberson~\cite{Roberson} showed for a further uncolored variant $Z(G)$ and 
$\tilde Z(G)$ of the graphs $X(G)$ and $\tilde X(G)$ that the condition 
$\tw(G)\ge k$ is not only sufficient but also necessary for the 
$C^k$-equivalence, i.e., 
\begin{eqnarray*}
\tw(G)\ge k & \iff & Z(G)\equiv_{C^k} \tilde Z(G).
\end{eqnarray*}
We show how Roberson's proof can be adapted to get the equivalence 
\[
\tw(G)\ge k\iff Y(G)\equiv_{C^k} \tilde Y(G).
\]
Therefore, by~\eqref{eq:driy}  we must show that 
\begin{equation}\label{eq:twkk} 
\tw(G)< k \ \ \text{ implies} \ \ Y(G)\not\equiv_{C^k} \tilde Y(G).
\end{equation}

\noindent The proof of~\eqref{eq:twkk} is based on the following result of 
Dvo\v{r}\'{a}k~\cite[Theorem 6]{Dvorak} that relates $C^k$-equivalence to 
homomorphism counts. 
\begin{theo}\label{thm:Dvorak} 
Assume $k\ge 1$. For graphs $H_1$ and $H_2$ the following two statements are 
equivalent. 
\begin{itemize}
\item $H_1\not\equiv_{C^k} H_2$. 

\item There is a graph $F$ of tree-width less than $k$ such that 
    $\hom(F,H_1)\neq \hom(F,H_2)$. 
\end{itemize}
\end{theo}

\noindent Here $\hom(F,H)$ denotes the number of homomorphisms from $F$ to 
$H$. Recall that a homomorphism from~$F$ to $H$ is a mapping $f:V(F)\to V(H)$ 
that preserves edges, i.e., if $uv\in E(F)$, then $f(u)f(v)\in E(H)$. Later 
we will use the notation $\Hom(F,H)$ to denote the set of all homomorphisms 
from~$F$ to $H$. Hence, $\hom(F,H)= |\Hom(F,H)|$. 

\medskip
\noindent Therefore, to get \eqref{eq:twkk} for $G$ with $\tw(G)< k$ it 
suffices to find a graph $F$ with $\tw(F)< k$ such that $\hom(F, Y(G))\neq 
\hom(F,\tilde Y(G))$. As already mentioned to get such a graph $F$ we adapt 
the proof of~\cite[Theorem 3.14]{Roberson} to uncolored \CFI-graphs.

\begin{figure} 	
\centering
\begin{tikzpicture}[scale=0.3,vertex style/.style={draw,
                                   circle,
                                   minimum size=3mm,
                                   inner sep=0pt,
                                   outer sep=0pt,
                                   %shade
                                   }]
{\small                                   
\begin{scope}%[scale=1]

  \path \foreach \i in {1,...,2}{
    \foreach \j in {1,...,2}{ 
       (6*\i,6*\j) coordinate[vertex style, fill=white] (t\i\j)}};
    
    \draw[left] (5,6) node{$v_4$};  
    \draw[right] (13,6) node{$v_3$};  
    \draw[left] (5,12) node{$v_1$};  
    \draw[right] (13,12) node{$v_2$};  

    \draw[thick] (t11)--(t21)--(t22)--(t12)--(t11);

\end{scope}}

{\small
\begin{scope}[xshift=22cm,yshift=0cm]

  \path \foreach \i in {1,...,2}{
    \foreach \j in {1,...,2}{ 
       (9*\i-3,9*\j-3) coordinate[vertex style, fill=white] (t\i\j)}};
    
    \draw[left] (5,6) node{$v_4$};  
    \draw[right] (16,6) node{$v_3$};  
    \draw[left] (5,15) node{$v_1$};  
    \draw[right] (16,15) node{$v_2$};  
    
    \draw[thick] (t11)--(t21)--(t22)--(t12)--(t11);

  \path \foreach \i in {1,...,2}{
    \foreach \j in {1,...,2}{ 
       (3*\i+6,9*\j-3) coordinate[vertex style, minimum size= 1mm, fill=black] ()}};

  \draw[above] (9,15.5) node{$w_{v_1,v_2}$};  
  \draw[above] (12,15.5) node{$w_{v_2,v_1}$};  

  \draw[below] (9,5.5) node{$w_{v_4,v_3}$};  
  \draw[below] (12,5.5) node{$w_{v_3,v_4}$};  
   
  \path \foreach \i in {1,...,2}{
    \foreach \j in {1,...,2}{ 
       (9*\i-3,3*\j+6) coordinate[vertex style, minimum size= 1mm, fill=black] ()}};

  \draw[left] (5.5,9) node{$w_{v_4,v_1}$};  
  \draw[left] (5.5,12) node{$w_{v_1,v_4}$};  

  \draw[right] (15.5,9) node{$w_{v_3,v_2}$};  
  \draw[right] (15.5,12) node{$w_{v_2,v_3}$};

\end{scope}}

r\end{tikzpicture}
\caption{A cycle $G= C_4$ and its corresponding $G_2$.}\label{fig:C4division}
\end{figure}
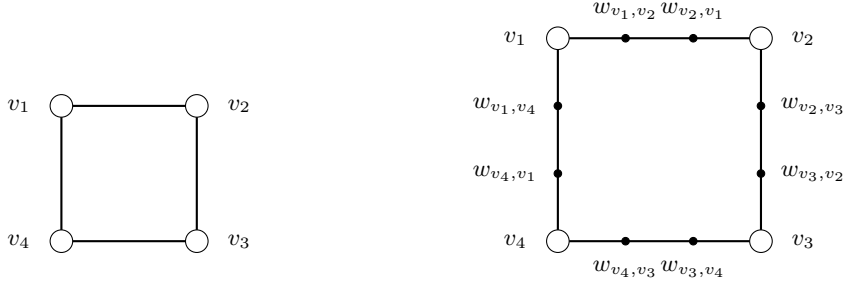

We fix $G$ with $\tw(G)< k$. Theorem~\ref{thm:homcfi} below shows that as $F$ 
we can take the $2$-subdivision $G_2$ of~$G$. The graph~$G_2$ is obtained 
from $G$ by adding for every edge $uv$ of  $G$ two new vertices $w_{u,v}$ and $w_{v,u}$  and replacing the edge $uv$ of  $G$ by the path $u,w_{u,v},w_{v,u},v$ (cf.\ Figure~\ref{fig:C4division}). Then $\tw(G_2)=\tw(G)$. In fact, if 
$\tw(G)= 1$, then~$G$ is a tree (recall that we always assume that $G$ is a 
connected graph). Then $G_2$ is a tree too and thus, $\tw(G_2)= \tw(G)$. 
Assume $\tw(G)\ge 2$. Fix a tree decomposition of $G$ of tree-with $\tw(G)$. 
For every edge $uv\in E(G)$ choose a node $t$ of the corresponding tree such 
that the bag $B_t$ contains $u$ and~$v$. Add to $t$ a path of length $2$ with 
new bags $B_1$ and $B_2$ where $B_1:= \{u, w_{u,v}, v\}$ and $B_2:=\{w_{u,v}, 
w_{v,u}, v\}$. In this way we obtain a tree decomposition of $G_2$ of 
tree-width $\tw(G)$.

\begin{theo}\label{thm:homcfi}
Let $G$ be a connected graph. Then 
\[
\hom(G_2,Y(G))> \hom(G_2,\tilde Y(G)).
\]
\end{theo}

\noindent By Theorem~\ref{thm:Dvorak} we get the desired result~\eqref{eq:twkk} (as $G$ is 
connected and has at least two vertices, we know that $\tw(G)\ge 1$). 
\begin{cor}\label{cor:twkyg} 
Assume $k\ge 2$. If $\tw(G)< k$, then $Y(G)\not\equiv_{C^k} \tilde Y(G)$. 
\end{cor}
\noindent Hence, by the results mentioned at the beginning of this section, 
for $k\ge 1$ we get the following equivalences. 
\begin{equation}\label{eq:twkygxg}
\begin{array}{rcl} 
 X(G)\equiv_{C^k} \tilde X(G) & \iff & \tw(G)\ge k\\
  & \iff & Y(G)\equiv_{C^k} \tilde Y(G).
\end{array}
\end{equation}

\noindent The following considerations will ultimately lead to a proof of 
Theorem~\ref{thm:homcfi}. We fix an edge $u_0v_0$ of $G$ and set 
\begin{eqnarray*}
Y^0(G):= Y(G) & \text{and} & Y^1(G):= Y^{u_0v_0}(G),
\end{eqnarray*}
i.e., $Y^1(G)= \tilde Y(G)$. We consider the function $p: V(Y^0(G)) 
 \ \big(= V(Y^1(G))\big) \to V(G_2)$ given by
\begin{equation}\label{eq:defp}
\text{for $uv\in E(G)$, \ \ $p(a(u,v))= p(b(u,v))= w_{u,v}$
 \quad and \quad for $m\in M(u)$, \ \ $p(m)=u$.}
\end{equation}
Note that for $i= 0, 1$ the function $p$ is a homomorphism from $Y^i(G)$ to 
$G_2$, we call it the \emph{projection from $Y^i(G)$ to $G_2$}. For $ uv\in 
E(G)$ it sends the link vertices $ a(u,v)$ and $b(u,v)$ of $Y^i(u)$ to $w_{u,v} 
$ and all the middle vertices of $Y^i(u)$ to $u$. 

For all homomorphisms $f\in \Hom(G_2, Y^i(G))$, the composition $g:= p \circ 
f$ is in $\Hom(G_2, G_2)$. Clearly, if $f_1, f_2\in \Hom(G_2, Y^i(G))$ and 
 $g_1:=  p \circ f_1$ and $g_2:=  p \circ f_2$  are distinct, then so are $f_1$ and $f_2$. 
Hence, 
\[
\Hom(G_2, Y^i(G)) 
 = \bigcup_{g\in \Hom(G_2, G_2)} \{f\in \Hom(G_2, Y^i(G))\mid p \circ f=g\}
\]
and this union is a disjoint union. For $g\in \Hom(G_2, G_2)$ and $i\in \{0, 
1\}$ we set 
\[
\Hom^i_{g}:= \{f\in \Hom(G_2,Y^i(G))\mid p\circ f= g\}.
\]
\begin{exa}\label{exa:g20}
To become familiar with these terms we show that for $i= 0$ and $g\in 
\Hom(G_2, G_2)$ the set $\Hom^i_{g}$ is nonempty. Thus we have to present an 
$f\in \Hom(G_2,Y(G))$ with $p\circ f= g$. For $\alpha\in V(G_2)$ (we use 
$\alpha, \beta, \ldots$ to denote vertices of $G_2$) 
\begin{itemize}
\item if $g(\alpha)=u\in V(G)$, then we set $f(\alpha)=\emptyset_{u}$ 
    ($\emptyset_{u}$ denotes the empty set in $M(u)$), 

\item if $g(\alpha)=w_{u,v}$, then we set $f(\alpha)=b(u,v)$. 
\end{itemize}
One easily verifies that $f$ is a homomorphism from $G_2$ to $Y(G)$ with 
$p\circ f= g$. \hfill{$ \dashv$}
\end{exa}
\noindent To obtain Theorem~\ref{thm:homcfi} it suffices to show that 
\begin{itemize}
\item $|\Hom^0_{g}|\ge |\Hom^1_{g}|$ for all $g\in \Hom(G_2, G_2)$ \qquad 
    and 
  
\item $|\Hom^0_{g}|> |\Hom^1_{g}|$ for at least one $g\in \Hom(G_2, G_2)$. 
\end{itemize}

\noindent To get these inequalities for $g\in \Hom(G_2, G_2)$ and $i\in \{0, 
1\}$, we introduce a system of linear equtions whose solutions correspond to 
the functions in $\Hom^i_{g} $.

We fix $g\in \Hom(G_2, G_2)$ and $i\in \{0, 1\}$. We introduce variables that 
will range over the field $\mathbb F_2$, namely 
\begin{align*}
x_{\alpha,u} & \quad \text{for all $\alpha\in V(G_2)$ with $g(\alpha)\in V(G)$ and $u\in V(G)$ 
 with $g(\alpha)u\in E(G)$,} \\
x_{\alpha } & \quad \text{for all $\alpha\in V(G_2)$ with $g(\alpha)=w_{u,v}$
 for some $uv\in E(G)$}.
\end{align*}
We consider the following system Eq$^i_g$ of linear equations (over $\mathbb 
F_2$). 
\begin{align*}
\textup{Eq$_g$-1}: \quad & \sum_{\textup{$u\in V(G)$ with $g(\alpha)u\in E(G)$}} x_{\alpha,u} = 0 
 & & \quad \text{for $\alpha\in V(G_2)$ with $g(\alpha)\in V(G)$}, \\
\textup{Eq$_g$-2}: \quad & x_{\alpha,u}+ x_{\beta}=0 
 & & \quad \text {for $\alpha, \beta\in V(G_2)$ and $u\in V(G)$ with $\alpha \beta\in E(G_2)$, } \\
 & & &\quad \hspace{0.2cm}\text{$g(\alpha)\in V(G)$, $g(\alpha)u\in E(G)$, and $g(\beta)=w_{g(\alpha),u}$}, \\
\textup{Eq$^i_g$-3}: \quad & x_{\alpha} + x_{\beta} = c(i;uv)
 & & \quad \text{for $uv\in E(G)$ and $\alpha\beta\in E(G_2)$ with $g(\alpha)=w_{u,v}$} \\
 & & & \hspace{5.5cm}\text{and $g(\beta)=w_{v,u}$}.
\end{align*}
By definition, $c(i;uv)=1$ if $i=1$ \big(i.e, we consider $Y^1(G)$\big) and 
$uv= u_0v_0$ \big(recall that $u_0v_0$ is the edge twisted to obtain 
$Y^1(G)$ from $Y^0(G)$\big); otherwise, $c(i;uv)= 0$. 

Note that in general Eq$_g$-2 for $\alpha, \beta$ is distinct from Eq$_g$-2 
for $\alpha:= \beta$ and $\beta:=\alpha$. On the other hand, Eq$^i_g$-3 is 
determined by the edge $uv$ \ ($=vu$).

\begin{lem}\label{lem:eqsize}
The number of solutions of the system \textup{Eq$^i_g$} is $|\Hom^i_{g}|$. 
\end{lem}

\proof We present a bijection between $\Hom^i_{g}$ and the set of solutions 
of \textup{Eq$^i_g$}. First for $f\in \Hom^i_{g}$ we define the values of the 
variables in such a way that the system Eq$^i_g$ is satisfied. Let $\alpha\in 
V(G_2)$. 
 
\begin{itemize}
\item[(a)] If $g(\alpha)=w_{u,v}$, i.e., $p(f(\alpha))= w_{u,v}$, then, 
    by~\eqref{eq:defp}, $f(\alpha)\in \{a(u,v),b(u,v)\}$. We set 
    \[
    x_{\alpha}=
    \begin{cases}
        1 & \textup{if $f(\alpha)=a(u,v)$} \\
        0 & \textup{if $f(\alpha)=b(u,v)$.}
    \end{cases}
    \]

\item[(b)] If $g(\alpha)\in V(G)$, i.e., $p(f(\alpha))\in V(G)$, then, 
    by~\eqref{eq:defp}, $f(\alpha)\in M(g(\alpha))$. For $u\in V(G)$ with 
    $g(\alpha)u\in E(G)$ we set 
    \[
    x_{\alpha,u}=
    \begin{cases}
        1 & \textup{if } a(g(\alpha),u)\in f(\alpha)\\
        0 & \textup{if }a(g(\alpha),u) \not\in f(\alpha).
    \end{cases}
    \]
\end{itemize}
By (b) we see that the equation Eq$_g$-1 is satisfied as the sets in 
$M(g(\alpha))$ have even cardinality. We turn to Eq$_g$-2. Then we know that 
\[
\text{$\alpha, \beta\in V(G_2)$ and $u\in V(G)$ with $\alpha \beta\in E(G_2)$, $g(\alpha)\in V(G)$, $g(\alpha)u\in E(G)$, and $g(\beta)=w_{g(\alpha),u}$}.
\]
As $g(\beta)= w_{g(\alpha),u}$, we know that $f(\beta)\in 
\{a(g(\alpha),u),b(g(\alpha),u)\}$ and thus by~(a), 
\begin{equation*}
x_\beta =1 \iff f(\beta)= a(g(\alpha),u).
\end{equation*}
As $g(\alpha)\in V(G)$, we know that $f(\alpha)\in M(g(\alpha))$ and thus by 
(b) we have for $v\in V(G)$ with $g(\alpha)v\in E(G)$, 
\begin{equation*}
x_{\alpha,v}= 1 \iff a(g(\alpha),v)\in f(\alpha).
\end{equation*}
 Thus, if $x_\beta= 1$, 
then $f(\beta)= a(g(\alpha),u)$. Since $f(\alpha)f(\beta)\in E(Y^i(G))$ as 
$\alpha\beta\in E(G_2)$, we get 
 $a(g(\alpha),u)\in f(\alpha)$, that is, $x_{\alpha,u}= 1$ 
and therefore, $x_{\alpha,u}+ x_{\beta}= 0$. Similarly we argue if $x_\beta = 
0$: Then $f(\beta)= b(g(\alpha),u)$. Since $f(\alpha)b(g(\alpha),u) \in 
E(Y^i(G))$, we see that $a(g(\alpha),u)\notin f(\alpha)$, that is, 
$x_{\alpha,u}= 0$ and therefore, $x_{\alpha,u}+x_\beta= 0$. Hence, Eq$_g$-2 
holds. 

Finally we show that Eq$^i_g$-3 is satisfied. Assume $uv\in E(G)$ and 
$\alpha\beta\in E(G_2)$ with $g(\alpha)=w_{u,v}$ and $g(\beta)=w_{v,u}$. 
Hence, $f(\alpha)\in \{a(u,v),b(u,v)\}$ and $f(\beta)\in \{a(v,u),b(v,u)\}$. 
Since $\alpha\beta\in E(G_2)$, we get $f(\alpha)f(\beta)\in E(Y^i(G))$. 

If $c(i;uv)=0$, then $i= 0$ or $uv\ne u_0v_0$. Then 
\[
\text{$f(\alpha)f(\beta)=a(u,v)a(v,u)$ \ \ \ or \ \ \ $f(\alpha)f(\beta)=b(u,v)b(v,u)$ }
\]
By (a) in both cases we get $x_\alpha+x_\beta= 0= c(i;uv)$. 

If $c(i;uv)=1$, we have $i= 1$ and $uv= u_0v_0$. As the edge $u_0v_0$ has 
been twisted and $f(\alpha)f(\beta)\in E(Y^i(G))$, we see that 
\[
\text{$f(\alpha)f(\beta)=a(u,v)b(v,u)$ \ \ \ or \ \ \ $f(\alpha)f(\beta)=b(v,u)a(v,u)$. }
\]
Now in both cases we get $x_\alpha+x_\beta= 1= c(i;uv)$. Therefore, 
Eq$^i_g$-3 holds. 

\medskip
\noindent We denote by $\bar x(f)$ the solution of the system of equations 
Eq$^i_g$ just described for $f\in \Hom^i_g$. It is not hard to see that for 
$f_1, f_2\in \Hom^i_g$ with $f_1\ne f_2$ we have $\bar x(f_1)\ne \bar 
x(f_2)$. In fact, assume $f_1(\alpha)\ne f_2(\alpha)$ for some $\alpha\in 
V(G_2)$. 

If $g(\alpha)\in V(G)$, then $f_1(\alpha),f_2(\alpha)\in M(g(\alpha))$. By 
$f_1(\alpha)\ne f_2(\alpha)$ there is a $u$ with $g(\alpha)u\in E(G)$ such 
that ($a(g(\alpha),u)\in f_1(\alpha)\iff a(g(\alpha),u)\notin f_2(\alpha)$). 
Hence, by~(b), 
\[
x_{\alpha,u}= 1 \text{ in} \ \bar x(f_1) \iff x_{\alpha,u}=0 \text{ in} \ \bar x(f_2).
\]
If $g(\alpha)\notin V(G)$, then $g(\alpha)=w_{u,v}$ for some $uv\in E(G)$. 
Then, $f_1(\alpha), f_2(\alpha)\in \{a(u,v), b(u,v)\}$. Since $f_1(\alpha)\ne 
f_2(\alpha)$, we have ($f_1(\alpha)=a(u,v)\iff f_2(\alpha)\ne a(u,v))$. By 
(a) this shows that the value of $ x_\alpha$ in $\bar x(f_1)$ is distinct 
from that of $x_\alpha$ in $\bar x(f_2)$. 

\medskip
\noindent Now we finish the proof of the lemma by presenting for every 
solution $\bar x$ of Eq$^i_g$ an $h\in \Hom^i_g$ with $\bar x(h)= \bar x$. 

So suppose $\bar x$ is a solution of Eq$^i_g$. We define a mapping $h:G_2\to 
Y^i(G)$. For $\alpha\in G_2$: 
\begin{itemize}
\item[(c)] If $g(\alpha)=w_{u,v}$ with $uv\in E(G)$, we set $h(\alpha):= 
    \begin{cases} 
        a(u,v) & \textup{if $x_{\alpha}= 1$} \\
        b(u,v) & \textup{if $x_{\alpha}= 0$}. 
    \end{cases}$

\item[(d)] If $g(\alpha)\in V(G)$, we set $h(\alpha):= \{a(g(\alpha),u)\mid 
    u\in V(G), \ g(\alpha)u\in E(G), \ x_{\alpha,u}=1\}$. The equation 
    Eq$_g$-1 guarantees that $h(\alpha)\in M(u)$. 
\end{itemize}
(In a certain sense (c) and (d) are the inverse of (a) and (b), 
respectively). 

By definition of $h$, it is easy to see $p\circ h= g$. We still have to 
verify that $h$ is a homomorphism from $G_2$ to $Y^i(G)$. Let $\alpha\beta\in 
E(G_2)$. We know that $g(\alpha)g(\beta)\in E(G_2)$ as $g\in\Hom(G_2, G_2)$. 
That is, for some $uv\in E(G)$ we have $g(\alpha)g(\beta)= w_{u,v}w_{v,u}$ or 
$g(\alpha)g(\beta)=v w_{v,u}$. 
\begin{itemize}
\item Let $g(\alpha)g(\beta)= w_{u,v}w_{v,u}$, then by (c), $h(\alpha)\in 
    \{a(u,v), b(u,v)\}$ and $h(\beta)\in \{a(v,u),b(v,u)\}$. 

    If $c(i;uv)= 0$, then $x_\alpha+ x_\beta= 0$. Hence, 
    $h(\alpha)h(\beta)= a(u,v)a(v,u)$ (if $x_\alpha= x_\beta= 1$) or 
    $h(\alpha)h(\beta)= b(u,v)b(v,u)$ (if $x_\alpha= x_\beta= 0$). In both 
    cases, $h(\alpha)h(\beta)\in E(Y^i(G))$ as $uv$ is not twisted in 
    $Y^i(G))$ (by $c(i;uv)= 0$). 

    If $c(i;uv)= 1$, then $i= 1$, $uv= u_0v_0$, and $x_\alpha+ x_\beta= 1$. 
    Hence, $h(\alpha)h(\beta)= a(u_0,v_0)b(v_0,u_0)$ (if $x_\alpha= 1$ and 
    $x_\beta= 0$) or $h(\alpha)h(\beta)= b(u_0,v_0)a(v_0,u_0)$ (if 
    $x_\alpha= 0$ and $x_\beta=1$). In both cases, $h(\alpha)h(\beta)\in 
    E(Y^1(G))$ as the edge $u_0v_0$ is twisted in $Y^1(G)$. 

\item Let $g(\alpha)g(\beta)=v w_{v,u}$. From $g(\alpha)\in V(G)$ we get 
    by~(d), 
    \[
    h(\alpha)=\{a(g(\alpha),u') \mid u'\in V(G), \ g(\alpha)u'\in E(G), \ x_{\alpha,u'}=1 \}.
    \]
    From $g(\beta)= w_{v,u}$ we get by (c), ($h(\beta)= a(v,u)\iff 
    x_\beta=1$), that is, ($h(\beta)= a(g(\alpha),u)\iff x_\beta=1$). 
    Hence, by the equation Eq$_g$-2, $a(g(\alpha),u)\in h(\alpha)$ and 
    $h(\beta)= a(g(\alpha),u)$ (if $x_{\alpha,u}= x_\beta= 1$) or 
    $a(g(\alpha),u)\notin h(\alpha)$ and $h(\beta)= b(g(\alpha),u)$ (if 
    $x_{\alpha,u}= x_\beta= 0$). Again in both cases 
    $\{h(\alpha),h(\beta)\}\in E(Y^i(G))$. 
   \end{itemize}
Hence, $h\in \Hom^i_g$. \proofend 

\noindent \emph{Proof of Theorem~\ref{thm:homcfi}.} Note that for every $g\in 
\Hom(G_2, G_2)$, the system of linear equations Eq$^0_g$ is the homogeneous 
system corresponding to Eq$^1_g$ (possibly Eq$^0_g=$ \, Eq$^1_g$). Hence, 
\[
\big|\Hom^1_g\big| =
 \begin{cases}
  0, & \text{if Eq$^1_g$ has no solution} \\
  \big|\Hom^0_g\big|, & \text{otherwise}.
\end{cases}
\]
Hence to get the statement claimed by the theorem it suffices to show that 
there is a $g\in \Hom(G_2, G_2)$ such that Eq$^1_g$ has no solution. Note 
that Eq$^0_g$ has always a solution, namely the trivial one where all 
variables get the value zero. 

We take as $g$ the identity $\textup{id}:G_2\to G_2$. For better reading we 
repeat the equalities of Eq$^1_{\textup{id}}$. 
\begin{align*}
\textup{Eq$_{\textup{id}}$-1}: \quad & \sum_{\textup{$u$ with $vu\in E(G)$}} x_{v,u} = 0 
 & & \text{for $v\in V(G)$}, \\
 \textup{Eq$_{\textup{id}}$-2}: \quad & x_{v,u}+x_\beta= 0 
  & &\text{for $v,u\in V(G)$, $\beta \in V(G_2)$ with $v\beta \in E(G_2)$ and}\\
  & & & \hspace{2.9cm}\text{$ vu\in E(G)$, and $\beta= w_{v,u}$,} \\ 
 \textup{Eq$^1_{\textup{id}}$-3}: \quad & x_{\alpha} + x_{\beta} = c(1;uv)  
  & & \text{for $\alpha\beta\in E(G_2)$ and $uv\in E(G)$ with $g(\alpha)=w_{u,v}$} \\
  & & &\hspace{5.5cm}\text{and $g(\beta)=w_{v,u}$}.
\end{align*}
Every variable occurs twice. The variable $x_{v,u}$ only occurs if $vu\in 
E(G)$ and then it occurs twice, namely in Eq$_{\textup{id}}$-1 and in 
Eq$_{\textup{id}}$-2. 

If a variable $x_{w_{u,v}}$ with $uv\in E(G)$ occurs as $x_\alpha$ in 
Eq$^1_{\textup{id}}$-3, then for $v:= u$ and $\beta:=w_{u,v}$ it occurs in 
Eq$_{\textup{id}}$-2.

As the edge $u_0v_0$ is twisted, we have the equality
\begin{equation}\label{eq:0=1} x_{w_{u_0,v_0}}+ x_{w_{v_0,u_0}}=1.
\end{equation}
We assume Eq$^1_{\textup{id}}$ has a solution. For this solution we add all 
the left sides of Eq$^1_{\textup{id}}$. This sum is zero as every variable 
occurs twice. The sum of the right side is one (by~\eqref{eq:0=1}), thus we 
get the contradiction $0= 1$. \proofend

% The computational complexity of the $L^k$-equivalence of graphs

\section{The computational complexity of the $L^k$-equivalence of graphs}\label{sec:cclke}
The following result was shown in \cite[Theorem 34]{licrassch25} and 
\cite[Theorem~5]{sepp24}.
\begin{theo}\label{thm:sepp}
The problem \textup{EQ-C} of deciding given graphs~$H_1$ and~$H_2$ and an 
integer $k\in \mathbb N$ whether~$H_1$ and~$H_2$ are $C^{k}$-equivalent is 
\textup{coNP}-hard under polynomial-time many-one reductions. 
\end{theo}

\noindent In this section we show that this already holds for the 
$L^{k}$-equivalence of graphs, i.e., we show: 
\begin{theo}\label{thm:seppmin}
The problem \textup{EQ-L} of deciding given graphs~$H_1$ and~$H_2$ and an 
integer $k\in \mathbb N$ whether~$H_1$ and~$H_2$ are $L^{k}$-equivalent is 
\textup{coNP}-hard under polynomial-time many-one reductions.  
\end{theo}

\medskip
\noindent The proof in~\cite{sepp24} of Theorem~\ref{thm:sepp} can be adopted 
literally in order to get the following statement. 
\begin{quote}
{\em The problem of deciding given a graph $G$, where each vertex has degree at least 2, and an integer $k\in \mathbb 
N$ whether $Y(G)$ and $\tilde Y(G)$ are $C^{k}$-equivalent is 
\textup{coNP}-hard under polynomial-time many-one reductions.} 
\end{quote}
Strictly speaking, we have to show why we can restrict to graphs of degree at least 2. One can  verify that we can add this restriction to the input of the problem \textsc{BoundedDegreeTreewidth} considered in \cite{sepp24} (in fact, if the input graph $ G$  is a forest, then $ \tw(G)\le 1$; if $ G$ is not a forest, then consider the graph $ G^*$ obtained from $ G$ by adding to each vertex $ u$ of degree 1 vertices $ u_1$ and $ u_2$ and the edges $ uu_1,u_1u_2 $, and $ u_2u$; note that $ \deg(G)=\deg(G^*)$).
\medskip

\noindent
Thus, if the equivalence 
\begin{eqnarray}\label{eqn:lkck}
Y(G)\equiv_{L^k} \tilde Y(G)
 & \iff & 
Y(G)\equiv_{C^k} \tilde Y(G) 
\end{eqnarray}
would hold (for sufficiently large $k$), then in the previous result we can 
replace the $C^{k}$-equivalence by the $L^{k}$-equivalence and thus get 
Theorem~\ref{thm:seppmin}. 
However, just after  Proposition~\ref{pro:lkick} we mentioned that the direction from left to 
right in~\eqref{eqn:lkck} is still open (also for other uncolored variants of 
the CFI~graphs).
Nevertheless we know that (cf.~\eqref{eq:twkygxg}) 
\[
Y(G)\equiv_{C^k} \tilde Y(G)\iff X(G)\equiv_{C^k} \tilde X(G)
\]
and that (cf.~Propopsition~\ref{pro:lkick}), 
\begin{eqnarray*}
X(G)\equiv_{L^k}\tilde X(G) 
 & \iff &
X(G)\equiv_{C^k}\tilde X(G). 
\end{eqnarray*}
Hence, ($Y(G)\equiv_{C^k} \tilde Y(G)\iff X(G)\equiv_{L^k}\tilde X(G)$). 
Therefore, from the result mentioned above we get: 
\begin{quote}
{\em The problem \P\ of deciding given a graph $G$, where every vertex has degree at least 2, and an integer $k\in 
\mathbb N$ whether $X(G)$ and $\tilde X(G)$ are $L^{k}$-equivalent is 
\textup{coNP}-hard under polynomial-time many-one reductions.} 
\end{quote}

\medskip

\noindent \emph{Proof of Theorem~\ref{thm:seppmin}.} It suffices to present a 
polynomial time  many-one reduction from the problem \P\ just mentioned to the problem 
\textup{EQ-L} of Theorem~\ref{thm:seppmin}. 

 So let $G$ be a 
connected graph,  where every vertex has degree at least 2, and $k\in \mathbb N$.  As for fixed $k$ the $L^k$-equivalence 
of graphs can be decided in polynomial time, we can assume that $k\ge 4$. We
define \emph{uncolored} graphs $X^{\textup{path}}(G)$ and $\tilde 
X^{\textup{path}}(G)$ such that 
\begin{equation}\label{eq:red}
X(G)\equiv_{L^k}\tilde X(G) \ \ \iff \ \ X^{\textup{path}}(G)\equiv_{L^k}\tilde X^{\textup{path}}(G). 
\end{equation}
This yields the desired reduction.

Let $n:=|V(G)|$, $d:=\deg(G)$, and fix an arbitrary order of $V(G)$. We can 
view the colors in $X(G)$ (and in $\tilde X(G)$) as pairs $(\ell, i)\in 
[n]\times [d+1]$. For a vertex $x$ of $X(G)$ the number $\ell$ tells us that 
$x \in X(u)$ where $u$ is the $\ell\/$th element in the order of $V(G)$. Furthermore, 
$i=d+1$ tells us that $x$ is in $M(u)$, and $i\in [d]$ says that the vertex 
$x$ is $a(u,v)$ or $b(u,v)$, where $v$ is the ``$i$th neighbor of $u$.'' 

Recall that $P_m$ denotes a path with $m$ edges. 
We obtain the graphs $X^{\textup{path}}(G)$ and $\tilde X^{\textup{path}}(G)$ 
from $X(G)$ and $\tilde X(G)$ by deleting all colors and by adding to every 
vertex $x$ a path $P(x)$. If the vertex $x$ had the color $(\ell, i)$, then 
$P(x)$ is of length $(\ell-1)\cdot(d+1)+i$. We identify one end of this path 
with $x$ and we say that~$x$ is the \emph{base point of $P(x)$}. We set 
$\base(y):=x$ for every vertex $y$ of $P(x)$; in particular, $\base(x)= x$. 
\medskip

\noindent

\medskip
We prove \eqref{eq:red}. First assume that $X^{\textup{path}}(G) 
\equiv_{L^k}\tilde X^{\textup{path}}(G)$. Therefore, the Duplicator has a winning strategy in the $ k$-pebble game on $X^{\textup{path}}(G) 
$  and $\tilde X^{\textup{path}}(G)$.

Assume that in a $k$-pebble play on $X(G) 
$  and $\tilde X(G)$ the Spoiler puts a pebble on a vertex $ x$, say in $ X(G)$. Then the Duplicator looks on her response $ y$ according to her winning strategy  if in a  $ k$-pebble play on $X^{\textup{path}}(G) $ and $\tilde X^{\textup{path}}(G) $ the Spoiler puts his pebble on $ x$ in $X^{\textup{path}}(G) $. As $\base(x)=x$ in $X^{\textup{path}}(G) $, then so is $ \base(y)=y$.
 Otherwise, in the game on  $X^{\textup{path}}(G) $ and $\tilde X^{\textup{path}}(G) $  the Spoiler in the next three rounds  puts  pebbles (distinct from the one on $ x$) onto three neighbors of $ x$ in $ X(G)$ and he would win (as only base points have degree at least three), a contradiction.
 This shows that  Duplicator's winning strategy in the $ k$-pebble game on $X^{\textup{path}}(G) 
$  and $\tilde X^{\textup{path}}(G)$  yields the desired winning strategy of the Duplicator for the $ k$-pebble game on $X(G)$  and $\tilde X(G)$.

\medskip
 Now we show the other direction of~\eqref{eq:red}. So assume 
$X(G)\equiv_{L^k}\tilde X(G)$. Hence, the Duplicator has a winning strategy in the $ k$-pebble game on $X(G)$  and $\tilde X(G)$. Using this winning strategy she gets the following winning strategy for the  $ k$-pebble game on  $X^{\textup{path}}(G) 
$ and $\tilde X^{\textup{path}}(G)$. Let in a play of this game the Spoiler put a pebble on a vertex on $ x$ of $X^{\textup{path}}(G)$, say, on the thirteenth vertex of the path  $P(base(x))$. Then the Duplicator looks at her  answer  $y$ (in $\tilde X(G)$) according to her winning strategy for  the game  on $X(G)$  and $\tilde X(G)$
if the Spoiler puts its pebble on $\base(x)$. In particular, $ base(x)$ and $ y$ have the same color. Then the Duplicator answers in the game on $X^{\textup{path}}(G) 
$ and $\tilde X^{\textup{path}}(G)$ with the thirteenth vertex of the 
path $P(y)$ in $\tilde X^{\textup{path}}(G) $. \proofend

\paragraph{Acknowledgment.} We thank Tim Seppelt for bringing~\cite{Roberson} to our attention.

\bibliographystyle{plain}
\bibliography{refer}

\end{document}